\documentstyle[12pt,aasms4]{article}

\def\linebreak{\hfil\break}

\def\
{\hfil\linebreak}

\def\singlespace{\baselineskip=14pt}

\def\singlespace{%
    \lineskip                .15ex
    \baselineskip            3.0ex
   \lineskiplimit              0ex
   \parskip                0.60ex plus .30ex minus .15ex
   }%

\def\etal{{\it et al}. }
\def\degree{\ifmmode {^\circ}\else {$^\circ$}\fi}
\def\mum{\ifmmode {\rm \mu {\rm m}}\else $\rm \mu {\rm m}$\fi}
\def\arcsec{\ifmmode ^{\prime \prime}\else $^{\prime \prime}$\fi}

\def\inch{\ifmmode ^{\prime \prime}\else $^{\prime \prime}$\fi}
\def\arcmin{\ifmmode ^{\prime}\else $^{\prime}$\fi}

\def\msun{\ifmmode {\rm M_{\odot}}\else $\rm M_{\odot}$\fi}
\def\mearth{\ifmmode {\rm M_{+\mskip-14.6muO\,}}\else $\rm M_{+\mskip-14.6muO\,}$\fi}
\def\mearth{\ifmmode {\rm M_{\earth}}\else $\rm M_{\earth}$\fi}
\newbox\grsign \setbox\grsign=\hbox{$>$} \newdimen\grdimen \grdimen=\ht\grsign
\newbox\simlessbox \newbox\simgreatbox
\setbox\simgreatbox=\hbox{\raise.5ex\hbox{$>$}\llap
     {\lower.5ex\hbox{$\sim$}}}\ht1=\grdimen\dp1=0pt
\setbox\simlessbox=\hbox{\raise.5ex\hbox{$<$}\llap
     {\lower.5ex\hbox{$\sim$}}}\ht2=\grdimen\dp2=0pt
\def\simgreat{\mathrel{\copy\simgreatbox}}
\def\simless{\mathrel{\copy\simlessbox}}

\begin{document}

\singlespace

\centerline{\Large {\bf Accretion in the Early Kuiper Belt }}
\vskip 2ex
\centerline{\Large {\bf I. Coagulation and Velocity Evolution }}
\vskip 7ex
\centerline{Scott J. Kenyon}
\centerline{Harvard-Smithsonian Center for Astrophysics}
\centerline{60 Garden Street, Cambridge, MA 02138} 
\centerline{e-mail: skenyon@cfa.harvard.edu}
\vskip 3ex
\centerline{and}
\vskip 3ex
\centerline{Jane X. Luu}
\centerline{Department of Astronomy}
\centerline{Harvard University }
\centerline{60 Garden Street, Cambridge, MA 02138} 
\centerline{e-mail: jluu@cfa.harvard..edu}
\vskip 7ex
\centerline{to appear in}
\centerline{{\it The Astronomical Journal}}
\centerline{May 1998}
\vskip 7ex
\received{5 November 1997}
\accepted{2 February 1998}

%
%
\singlespace

\begin{abstract}

We describe planetesimal accretion calculations in the Kuiper Belt.  
Our evolution code simulates planetesimal growth in a single annulus 
and includes velocity evolution but not fragmentation.  Test results 
match analytic solutions and duplicate previous simulations at 1 AU.

In the Kuiper Belt,
simulations without velocity evolution produce a single runaway 
body with a radius $r_i \gtrsim$ 1000 km on a time scale 
$\tau_r \propto M_0^{-1} e_0^{x}$, where $M_0$ is the initial mass 
in the annulus, $e_0$ is the initial eccentricity of the planetesimals, 
and $x \approx$ 1--2.  
Runaway growth occurs in 100 Myr for $M_0 \approx$ 10 \mearth~and
$e_0 \approx 10^{-3}$ in a 6 AU annulus centered at 35 AU.  
This mass is close to the amount of dusty material expected in 
a minimum mass solar nebula extrapolated into the Kuiper Belt.

Simulations with velocity evolution produce runaway growth on a 
wide range of time scales.  Dynamical friction and viscous stirring
increase particle velocities in models with large (8 km radius) initial bodies.
This velocity increase delays runaway growth by a factor of two 
compared to models without velocity evolution.  In contrast, 
collisional damping dominates over dynamical friction and viscous stirring 
in models with small (80--800 m radius) initial bodies.  Collisional damping
decreases the time scale to runaway growth by factors of 4--10 relative to constant 
velocity calculations.  Simulations with minimum mass solar nebulae,
$M_0 \sim$ 10 \mearth, and small eccentricities, $e \approx 10^{-3}$, 
reach runaway growth on time scales of 20--40 Myr with 80 m initial 
bodies, 50--100 Myr with 800 m bodies, and 75--250 Myr for 8 km initial 
bodies.  These growth times vary linearly with the mass of the annulus, 
$\tau_r \propto M_0^{-1}$, but are less sensitive to the 
initial eccentricity than constant velocity models. 

In both sets of models, the time scales to produce 1000+ km objects
are comparable to estimated formation time scales for Neptune. 
Thus, Pluto-sized objects can form in the outer solar system 
in parallel with the condensation of the outermost large planets.

\end{abstract}

\section{INTRODUCTION}

Current models for planet formation involve aggregation of solid 
planetesimals and gas accretion in a circumstellar disk (see,
for example, \cite{hay85}; \cite{bos93}; and references therein).  
Large dust grains within the disk first settle to the midplane.  
These grains may then coagulate into successively larger grains (e.g., 
\cite{wei80}; \cite{wei93}) or continue to settle in a very thin layer 
that eventually becomes gravitationally unstable (e.g., \cite{gol73}).  
Both paths produce km-sized planetesimals that collide and merge to
produce large bodies such as planets.  Despite the complex and sometimes
unknown physics, many simulations produce objects resembling known 
planets on time scales roughly comparable to the expected lifetime 
of the protosolar nebula (see, for example, \cite{saf69}; 
\cite{gre78}, 1984; \cite{nak83}; \cite{wet89}; 1993; \cite{spa91}; 
\cite{kol92}; \cite{wei92}; \cite{pol96}).

Recent observations of slow-moving objects in the outer solar system 
offer a new challenge to planet formation models.  The trans-Neptunian
region now contains several dozen Kuiper Belt objects (KBOs) with
estimated radii of 100--300 km (\cite{jew96}).  The orbits of known 
KBOs suggest a division into at least three dynamical components 
with an inner radius of 30 AU and an unknown outer radius:
(i) the classical KBOs, objects with roughly circular orbits,
(ii) the resonant KBOs, objects in orbital resonance with Neptune 
(\cite{jew96}), and
(iii) the scattered KBOs, objects with large, eccentric orbits (\cite{luu97}).
Although the known population is still small, Jewitt \etal (1996)
estimate that the region between 30 and 50 AU contains $\approx$ 
70,000 objects larger than 100 km.  The total mass in the classical
Kuiper Belt is thus at least 0.1 \mearth.  This mass probably represents
a small fraction of the initial mass, because dynamical interactions 
with Neptune reduce the number of KBOs on short time scales compared 
to the age of the solar system (\cite{lev93}; \cite{mal96}).

Despite these new observations, the origin of KBOs remains uncertain.
Edgeworth (1949) and Kuiper (1951) first suggested that the Kuiper Belt 
was a natural extension of the original solar nebula.  
Holman \& Wisdom (1993) later showed that small KBOs, 
once formed, can survive at 30--50 AU for times approaching 
the age of the solar system.  More recent dynamical studies confirm this
conclusion and explain the observed distribution of KBOs in a general
way (\cite{lev93}).  The formation process and time scale for KBOs, 
however, is still
controversial.  Planetesimal simulations for plausible protosolar nebulae
at 25--30 AU show that Neptune can grow to its present size in 10--100
Myr (\cite{fer81}, 1984; \cite{ip89}; \cite{pol96}).  These results suggest
that Pluto might form on a similar time scale at $\sim$ 40 AU, because 
growth times are not a steep function of semi-major axis (see \cite{aar93};
\cite{pol96}; and references therein).  Nevertheless, Stern (1995, 1996)
and Stern \& Colwell (1997) conclude that KBO formation requires 
100--1000 Myr for the conditions expected in the outer solar system.  
These limits far exceed the time scale required to produce Neptune,
whose accretion time is constrained by the 10--100 Myr lifetime of
the protosolar nebula (see \cite{pol96} and references therein).

In this paper, we attempt to resolve the uncertainties surrounding
KBO production with a new planetesimal simulation at 35 AU.  We have
developed an evolution code to follow the growth and velocity evolution
of planetesimals with a wide range of initial masses.  The code matches
analytic models and duplicates Wetherill \& Stewart's (1993; hereafter WS93)
simulation of planetesimal evolution at 1 AU.  Our numerical results 
demonstrate that small bodies with initial radii of 80 m to 8 km can 
produce 1000+ km objects on time scales of 10--100 Myr.  We confirm 
these calculations with a simple analytic estimate of the growth time 
as a function of semi-major axis.  This analysis supports previous
estimates for a short growth phase for Neptune, 10--100 Myr, and 
indicates that Pluto-Charon can form just outside the current orbit
of Neptune on a similar time scale.

We outline the accretion model in Sec. 2, describe our calculations
in Sec. 3, and conclude with a discussion and summary in Sec. 4.
The Appendix contains a complete description of the algorithms
and detailed comparisons with analytic models.

\section{The Accretion Model}

For our simulations of accretion in the Kuiper Belt, we adopt 
Safronov's (1969) particle-in-a-box method, where planetesimals are 
treated as a statistical ensemble of masses with a distribution of 
horizontal and vertical velocities about a Keplerian orbit.  
Our simulations begin with a differential mass distribution, $n(m_i$), 
in a single accumulation zone centered at a heliocentric distance,
$a$, with an inner radius at $a - \Delta a/2$ and an outer radius 
at $a + \Delta a/2$.  We approximate the continuous distribution 
of particle masses with discrete batches having particle populations
$n_i(t)$ and total masses $M_i(t)$ (WS93).  The average mass of 
a batch, $m_i(t)$ = $M_i(t) / n_i(t)$, changes with time as collisions 
add and remove bodies from the batch.  This procedure naturally 
conserves mass and allows a coarser grid than simulations with 
fixed mass bins (\cite{wet90}, and references therein; \cite{ws93}).

To evolve the mass and velocity distributions in time, we solve the
coagulation and energy conservation equations for an ensemble of
objects with masses ranging from $\sim 10^{12}$~g to $\sim 10^{26}$ g.  
The Appendix describes our model in detail and compares our numerical
results with analytic solutions for standard test cases.  We adopt 
analytic cross-sections to derive collision rates and compute
velocity changes from gas drag and collective interactions such as
dynamical friction and viscous stirring.  Our initial approach to this
problem ignores fragmentation, which we will consider in a later paper.
In this study, we focus on developing a good understanding of 
planetesimal growth as a function of initial conditions in the 
Kuiper Belt.  

To test our numerical procedures in detail, we attempt to duplicate 
WS93's simulations of planetary embryo formation at 1 AU.  
WS93 (see also \cite{wet89}; \cite{bar90}, 1991, 1993; 
\cite{spa91}; \cite{aar93}) demonstrate that an ensemble of
8 km objects can produce $10^{26}$ g (Moon-sized) objects on a $10^5$ 
yr time scale.  WS93's model begins with 8.33 $\times~10^8$ planetesimals
having radii of 8 km and a velocity dispersion of 4.7 m s$^{-1}$ relative
to a Keplerian orbit (Table 1; see also Table 1 of \cite{ws93}).  
Tables 2--3 summarize our results using the WS93 initial conditions with 
a mass spacing factor of $\delta \equiv m_{i+1}/m_i$ = 1.25 and 1.4 
between successive mass batches
and two different analytic cross-sections.  Figure 1 shows our 
reproduction of the WS93 results without fragmentation for $\delta = 1.25$.
This simulation produces 14 3--9 $\times~10^{25}$ g objects in 
1.5 $\times~10^5$ yr, which agrees with the results in WS93 (see 
their Figure 12).  Our simulation confirms the broad ``plateau'' in 
the cumulative number, $N_C$, at log $m_i$ = 24--26 and the rough
power law dependence, $N_C \propto m_i^{-1}$, at log $m_i$ = 21--23.  
The broad plateau extends across a smaller mass range and becomes more
rounded as $\delta$ increases (Figure 2).  The maximum planetesimal mass,
$m_{max}$, at the conclusion of the calculation depends on both $\delta$ 
and the cross-sections.  We find marginally larger $m_{max}$ for the 
Spaute \etal (1991) cross-sections.  In general, $m_{max}$ increases 
as $\delta$ increases.  

The evolution of particle velocities in our simulations agrees with the 
WS93 results (Figure 1b).  
All of the velocities increase monotonically with time due 
to viscous stirring.  The velocities of the larger bodies increase 
very slowly, because dynamical friction transfers their kinetic 
energy to the smaller bodies.  The simulation maintains a nearly constant
ratio of vertical to horizontal velocity, $v_i/h_i \approx$ 0.53, for
all but the most massive bodies, which have $v_i/h_i < 0.5$.  
The equilibrium ratio of $v_i/h_i$ yields $<i>/<e> \approx $ 0.6, 
in agreement with Barge \& Pellat (1990, 1991; see also \cite{hor85}).
At the conclusion of the simulation,
our velocities for small bodies, $h_i \approx$ 500 m s$^{-1}$ at 
$m_i \sim 10^{19}$ g, are roughly 50\% larger than those of WS93.  
Our velocities for large bodies, $h_i \approx$ 10 m s$^{-1}$ at 
$m_i \sim 10^{26}$ g, are roughly 50\% smaller than those of WS93.
We also fail to reproduce WS93's abrupt drop in $h_i$ at log $m_i$ = 24.
However, these differences -- which are independent of $\delta$ -- have a 
negligible effect on the final mass distribution and probably result 
from slightly different algorithms for low velocity collisions.

Gas drag is included in our simulations but has a negligible 
impact on the evolution.  All of the 1 AU models lose $\sim$ 
0.01\% of their initial mass over 1.5 $\times ~ 10^5$ yr.  
Velocity changes due to gas drag are essentially zero, 
because the particle masses are so large.

To understand the sensitivity of these results to initial conditions,
we consider the growth time of planetesimals from the coagulation equation,
equation A4.  For most cases of interest, the growth time for bodies 
with $m_i$ is approximately 
$ \tau \approx n_0/(dn/dt) \propto H a \Delta a / n_j V F_g (r_i + r_j)^2 $,
where $n_j$ is the number of lower mass bodies,
$V$ is the relative velocity, $F_g$ is the gravitational focusing factor, 
$r_i$ and $r_j$ are the radii of particles $i$ and $j$, and 
$H$ is the vertical scale height.
Collisions between low mass objects are in the high velocity regime, 
where the gravitational focusing factor is $F_g \approx$ 1 and 
$ \tau \propto a^{5/2}~\Delta a~n_j^{-1}~(r_i + r_j)^{-2}$.  
This growth time is independent of the initial $e$ and $i$.
Gravitational focusing becomes effective in low-velocity 
collisions of massive objects; the growth time then depends 
on the initial velocity and is 
$ \tau \propto a^{5/2}~\Delta a~n_j^{-1}~(r_i + r_j)^{-1}~V^2$.
The extreme sensitivity of the growth time to velocity is the reason
why low-velocity planetesimals experience runaway growth in our
1 AU simulations (\cite{wet89}; \cite{id92a}, 1992b; \cite{kok96}; 
see also \cite{ws93} and references therein).
We adopt 1000 km as a useful reference radius and write the time for 
8 km objects to produce 1000 km objects at 1 AU as (see also \cite{bar90}):

\begin{equation}
\tau \approx \tau_0 \left ( \frac{a}{\rm 1~AU} \right )^{5/2}
                    \left ( \frac{\Delta a}{\rm 0.17~AU} \right )
                    \left ( \frac{\rho_0}{\rm 3~g~cm^{-3}} \right )^{1/3}
                    \left ( \frac{V_0}{\rm 450~cm~s^{-1}} \right )^2
                    \left ( \frac{\rm 6 \times 10^{27}{\rm g}}{M_0} \right ) ~ .
\end{equation}

\noindent
Using our simulations with $\delta$ = 1.4, we derive the proportionality 
constant for this standard case with velocity evolution, 
$\tau_{0,v}$ = 10700 yr, and for a model with 
no velocity evolution, $\tau_{0,nv}$ = 3750 yr.  
Additional simulations confirm the mass, velocity, and volume dependence 
of this relation for factor of two variations in $a$, $\delta a$,
$\rho_0$, $V_0$, and $M_0$ about the values in Eqn 1
(see also \cite{aar93}; \cite{pol96}).

\section{Kuiper Belt Calculations}

\subsection{Starting Conditions}

To choose appropriate constraints on planetesimal simulations in 
the outer solar system, we rely on observations of other stellar
systems and models of the protosolar nebula.  First, current data 
indicate lifetimes of $\sim$ 5--10 Myr for {\it typical} gaseous disks 
surrounding nearby pre-main sequence stars and for the solar nebula
(\cite{sar93}; \cite{str93}; \cite{rus96}).  
We adopt this estimate as a rough lower limit to the formation
time scale of KBOs and assume that interactions between gas and
planetesimals disappear on a similar time scale, $\tau_g \approx$ 10 Myr
(see the Appendix, equation A29).
Neptune formation places an upper limit on the KBO growth time, 
because Neptune excites KBOs through gravitational perturbations.
Recent calculations suggest Neptune can form in 5--100 Myr 
(Ip 1989; Lissauer \& Stewart 1993; Lissauer \etal 1995; 
Lissauer \etal 1996).  Once formed, Neptune inhibits KBO
formation at 30--40 AU by increasing particle random velocities
on time scales of 20--100 Myr (\cite{hol93}; \cite{dun95}).  
We thus adopt 100 Myr as a rough upper limit to the KBO formation
time scale at 30--40 AU.

We assume a wide range of starting conditions for KBO simulations.
Our model annulus is centered outside the orbit of Neptune at 35 AU 
and has a width of 6 AU. This annulus can accommodate at least 
10--100 isolated bodies\footnote{``Isolated bodies'' are planetesimals
that cannot collide with one another, as defined in the Appendix
following Eqn (A5)}
with $m_i \simgreat 10^{24}$ g for $e \le 0.01$. 
The simulations begin with $N_0$ bodies of radius $r_0$, with $r_0$ = 
80 m, 800 m, and 8 km.  These bodies have small initial eccentricities,
$e \sim 10^{-3}$ to $10^{-2}$ (\cite{mal95}), and an equilibrium ratio 
of inclination to eccentricity, $\beta = <i>/<e>$ = 0.6 (\cite{bar90}, 
1991, 1993).  The mass density of each body is fixed at 1.5 g cm$^{-3}$.
To set $N_0$, we extend the minimum mass solar nebula to the Kuiper Belt 
and integrate the surface density distribution for solid particles,
$\Sigma = \Sigma_0 (a/a_0)^{-3/2}$, across the 6 AU annulus.  The 
dust mass is then $M_{min} \approx 0.25~\Sigma_0$ \mearth~at
32--38 AU ($a_0$ = 1~AU).  Most minimum mass solar nebula models 
have $\Sigma_0$ = 30--60 g cm$^{-2}$, which sets $M_{min} \approx$ 
7--15 \mearth~(\cite{wei77}; \cite{hay81}; \cite{bai94}).  
We thus consider models with initial masses of $M_0$ = 1--100 \mearth~to 
allow for additional uncertainty in $\Sigma_0$.  Table 1 compares 
input parameters for all Kuiper Belt models with initial conditions 
at 1 AU (see also \cite{ws93}).  Tables 4--5 summarize other initial
conditions and results for the Kuiper Belt simulations summarized
below.

Our success criteria are based on direct observations of KBOs.
The present day Kuiper Belt contains at least 70,000 objects 
with diameters exceeding 100 km at 30--50 AU (\cite{jew95}, \cite{jew96}).
This population is some fraction of the initial Kuiper Belt population, 
because Neptune has eroded the Kuiper Belt over time (\cite{hol93};
\cite{dun95}).  Thus, a successful KBO simulation must achieve 
$r_5 \simgreat$ 50 km in $\simless$ 100 Myr, where $r_5$ is the 
radius where the cumulative number of objects exceeds $N_C \ge 10^5$.
Pluto formation is our second success criterion: plausible models must 
produce 1 or more objects with maximum radius $r_{max} \ge$ 1000 km.  
In models with 
velocity evolution, we end simulations at 100--200 Myr or when $r_{max}$
exceeds $\sim$ 1000 km.  To evaluate the dependence of runaway growth 
on the initial conditions, we extend simulations without velocity evolution 
to 5000 Myr or to when $r_{max}$ exceeds $\sim$ 2000 km.

\subsection{Models Without Velocity Evolution}

To isolate important processes in trans-Neptunian planetesimal
evolution, we begin with constant velocity solutions to the
coagulation equation.  We ignore fragmentation and fix the
velocities for all masses at $h_i = 4.0~e_{init}$ km s$^{-1}$ 
and $v_i$ = 3.6 ${\rm sin}~i_{init}$ km s$^{-1}$. The initial 
eccentricity and inclination are set at $i_{init}$ = 0.60 $e_{init}$.
The total mass and kinetic energy remain constant throughout the 
calculation.  We also adopt a coarse mass spacing factor, $\delta$ = 1.4.
This choice limits our ability to follow runaway growth with high accuracy
during the late stages of the simulation but allows us to investigate 
a wide range of initial masses and velocities with a modest investment 
of computer time.  Finally, we adopt simple formulae for gravitational 
focusing to speed our calculations (see equation A18; \cite{spa91}).
Table 4 summarizes the initial conditions and results for models 
with $M_0$ = 1--100 \mearth~and $e_{init}$ = $10^{-3}$ and $10^{-2}$.

Figure 3 shows how $N_C$ evolves with time for a model with
an initial planetesimal radius $r_0$ = 8 km, 
total mass $M_0$ = 10 \mearth, and $e_{init} = 10^{-3}$.
This model begins with $1.87 \times 10^{10}$ initial bodies
and produces $\sim$ 42,500 objects with twice the initial mass
after $\tau$ = 100 yr. Roughly half of the original population 
experiences at least one collision by $\tau \approx$ 16~Myr,
when the 17 largest bodies have $m_i \approx 10^{20}$~g. 
Slow growth continues until $\tau \approx$ 59 Myr when 
the 3 largest objects have sizes comparable to large KBOs, 
$r_i \approx$ 100~km and $m_i \approx 10^{22}$~g.
The growth rate of the large masses then increases considerably 
due to gravitational focusing.  Runaway growth ensues.  The cumulative
mass distribution then follows a power law, $N_C \propto r_i^{-2.75}$,
at low masses and develops a high mass shoulder that extends to
larger and larger masses as the simulation proceeds.   This shoulder 
resembles the runaway plateau observed in 1 AU models but does not 
evolve into a true plateau with $N_C \approx$ constant as in Figs. 1--2.
The largest planetesimals reach $r_{max} \approx$ 200~km at 
$\tau \approx$ 66~Myr; $r_{max}$ exceeds 1000~km only 9~Myr later.
A single runaway body with $r_{max} \approx$ 4000~km begins to 
sweep up lower mass planetesimals at $\tau \approx$ 80~Myr; by 
$\tau \approx$ 85~Myr, it contains essentially all of the mass 
in the annulus.

Simulations with $r_0$ = 8 km produce runaway growth independent of 
the initial mass in the annulus.  Figure 4(a) indicates that each model 
experiences a long, linear growth phase until $r_{max} \approx$ 100--200 km.
The largest objects then begin a short, rapid growth phase that produces
several isolated, runaway bodies with $r_{max} \approx$ 1000 km.  
These runaway bodies accumulate all of the lower mass bodies and
may merge to form a single runaway body if the isolation criterion
permits.  The time to produce runaway bodies with $r_i$ = 1000 km
scales with the mass in the annulus,
$\tau_r \approx$ 753 Myr ($M_0/{\rm 1 ~ \mearth}$)$^{-1}$.
For comparison, our scaling relation in Eqn (1) predicts 
$\tau_r \approx$ 775 Myr ($M_0/{\rm 1 ~ \mearth}$)$^{-1}$ for 
$\rho_0$ = 1.5 g cm$^{-3}$ in a 6 AU annulus centered at 35 AU.

Runaway growth also occurs independent of the initial radius, $r_0$
(Figures 4(a)--4(c)).  Due to smaller initial cross-sections, models 
with $r_0$ = 80--800 m take longer to reach the rapid growth phase.
These models make the transition from rapid growth to runaway growth 
more quickly, because it is easier for 100+ km objects to sweep up small
objects with $r <$ 1 km.  In all cases, a single runaway body with 
$r >$ 1000 km eventually accumulates all of the mass in each simulation, 
although the time scale is quite long, $\tau_r \approx$ 2700 Myr, 
for simulations with $M_0$ = 1 \mearth~and $r_0$ = 80 m.  Again,
the runaway growth time scales with mass:
$\tau_r \approx$ 2340 Myr ($M_0/{\rm 1 ~ \mearth}$)$^{-1}$ for 
$r_0$ = 800 m and
$\tau_r \approx$ 2700 Myr ($M_0/{\rm 1 ~ \mearth}$)$^{-1}$ for $r_0$ = 80 m.
The small increase in $\tau_r$ with initial radius for $r_0 \simless$ 800 m
suggests that models with $r_0 < $ 80 m will reach runaway growth on
time scales of $\sim$ 3000 Myr, which is $\sim$ 40 times slower than 
models with $r_0$ = 8 km.

Our results also confirm the velocity dependence derived in Eqn (1).  
Low eccentricity simulations with 50\% smaller initial velocities reach 
runaway growth in 25\% of the time for our standard model; simulations 
with twice the initial velocity require four times as long to achieve 
runaway growth.  This simple relation begins to break down as the
eccentricity increases to $e \approx 10^{-2}$, as outlined below.
The runaway time also scales with the width of the annulus, $\Delta a$, 
and the semi-major axis, $a$, as indicated in equation (1).

High eccentricity models also achieve runaway growth but do not 
follow precisely the velocity scaling in Eqn (1).  Figure 5 shows the 
radius evolution for models with various $r_0$ and $m_0$ for
$e = 10^{-2}$ (see also Table 4).  The growth time for 1000+ km 
objects is $\tau_r \approx$ 20--25 Gyr ($M_0/{\rm 1 ~ \mearth}$)$^{-1}$
nearly independent of the initial radius and velocity.  This relation
contrasts with the low eccentricity results, where the growth time is 
very sensitive to the initial conditions.  In all of our simulations,
planetesimal growth is orderly until gravitational focusing becomes 
important and runaway growth occurs.  However, the radius where 
gravitational focusing becomes important increases from $r_i \sim$ 
10 km at $e = 10^{-3}$ to $r_i \sim$ 100 km at $e = 10^{-2}$.  For
models with small initial bodies, $r_0 \simless$ 800 m, the time scale
to reach runaway growth is directly proportional to $e$.  For models
with large initial eccentricity, the long orderly phase also ``erases'' 
memory of the initial radius.  Thus, $\tau_r$ is nearly independent of 
$r_0$ for large $e$.  The relatively short orderly growth phase of
low $e$ models does not erase memory of $r_0$; $\tau_r$ decreases
with increasing $r_0$ for $e_{init} \lesssim$ 0.05.  For models with
$r_0 \sim$ 8 km, gravitational focusing accelerates growth immediately
at low $e$.  These simulations do not have an orderly growth phase; 
instead, they follow the 1 AU simulations and satisfy the scaling 
relation in Eqn (1).

Before we consider Kuiper Belt simulations with velocity evolution,
our basic result that constant velocity models achieve runaway growth 
deserves some comment.  First, previous simulations at 1 AU show that 
runaway growth requires dynamical friction to decrease the velocities 
of the largest bodies to the regime where gravitational focusing 
becomes important (\cite{wet89}, 1993; \cite{bar90}, 1991; \cite{ida90};
\cite{id92a}, 1992b; \cite{oht92}; \cite{kok96}).  In Kuiper Belt models 
with $e_{init} = 10^{-3}$, gravitational focusing factors become 
very large at planetesimal masses of $10^{23}$ to $10^{24}$ g.  
Further growth of these bodies only enhances gravitational focusing,
because the escape velocity increases while the impact velocities 
remain low.  More massive objects thus ``run away'' from their 
lower mass counterparts.  This response occurs in {\it any}
constant velocity simulation as long as bodies can reach masses where 
the escape velocity is large compared to the relative impact velocity,
$V_{e,ij}/V_{ij} \gg 1$.  Models with $e_{init} \simgreat$ 0.1 never reach 
this limit for plausible $M_0$ and thus do not experience runaway growth.
Models with $e_{init} \simless$ 0.05 always produce runaway bodies, albeit 
at much later stages than models with $e_{init} \sim 10^{-3}$ (Figure 5).

Our final comment on runaway growth concerns the shape of the 
cumulative number distribution near the end of the simulation.
During runaway growth, models at 1 AU develop a plateau in the 
cumulative number distribution that extends from $m_i = 10^{23}$~g 
to $m_i = 10^{25}$ to $10^{26}$~g (see Figures 1--2).  This plateau 
separates runaway bodies from the lower mass objects which are in
the orderly growth regime and have a power law size distribution, 
$N_C \propto r_i^{-3}$ for log $m_i$ = 21--24 (\cite{ws93}; 
see Figures 1--2).  The Kuiper Belt simulations also produce a 
power law size distribution, $N_C \propto r_i^{-2.7}$ for 
$m_i \simless 10^{25}$ g, but they develop a high mass ``shoulder''
instead of a marked plateau at runaway growth (see Figure 3).  
To test if this feature is a function of the mass resolution 
as in 1 AU models, we simulate evolution at 35 AU with 
$\delta$ = 1.1 and 1.25 for $M_0$ = 10 \mearth~and $r_0$ = 8 km.
As the mass resolution in the simulation increases from $\delta$ = 1.4
to $\delta$ = 1.1, the high mass shoulder follows a very 
shallow power law, $N_C \propto r_i^{-1.7}$ (Figure 6).
This power law becomes better defined as the mass resolution increases
further, but it never develops into the ``runaway plateau'' produced 
in the 1 AU models (Figures 1--2).  This result suggests that the 
broad plateau in 1 AU models is the result of velocity evolution, 
which reduces the velocity of the most massive objects and enhances 
gravitational focusing (\cite{ws93}).  We will now test this hypothesis 
by considering Kuiper Belt models with velocity evolution.

\subsection{Models with Velocity Evolution}

To understand the importance of velocity evolution in the Kuiper Belt,
we add several physical processes to the calculation:
(i) gas drag, 
(ii) dynamical friction and viscous stirring from 
long-range (elastic) collisions, and
(iii) dynamical friction and viscous stirring from 
short-range (inelastic) collisions.
As in our constant velocity models, we begin with $N_0$ bodies 
at radii, $r_0$ = 80 m, 800 m, and 8 km.  
We adopt $\delta$ = 1.4 for the mass spacing factor and use 
our simple expression for gravitational focusing, Eqn A18. 
The initial velocities are 
$h_i$ = 4.0 ($e_{init}/10^{-3}$) m s$^{-1}$ and 
$v_i$ = 2.1 ($e_{init}/10^{-3}$) m s$^{-1}$.
The eccentricity and inclination evolve separately due to
collisions and collective interactions (see section A.3 of the Appendix).  
Table 5 summarizes the initial conditions and results for models
with $M_0$ = 1--30 \mearth~and $e_{init}$ = $10^{-3}$ to $10^{-2}$.

Before describing the results of our simulations, it is useful 
to compare various time scales for velocity evolution at 35 AU.
First, gas drag is negligible in models that ignore fragmentation.
A typical simulation at 35 AU loses $\sim 10^{-5}$\% of its total
mass due to gas drag in 100 Myr.  Velocity changes due to gas drag
are also insignificant; the time scale for gas drag to modify the 
velocity exceeds 10 Gyr for all masses in our simulation.

Velocity changes due to elastic and inelastic collisions,
however, are significant.  Figure 7(a) compares time scales,
$\tau_{v,h} = h_i / (d h_i/dt) $, for horizontal velocity 
evolution as a function of particle mass at 35 AU.  
The two curves show $\tau_{v,h}$ for interactions between 
particles of the same mass with a power-law size distribution,
$N_C \propto r^{-2.7}$, and constant velocity.
These time scales are {\it not} integrated 
over the size distribution and are relevant only when a simulation
has a small range of masses.
Viscous stirring -- which tends to increase particle velocities -- 
is ineffective for $m_i \lesssim 10^{13}$ kg, 
$\tau_{v,h} \approx 10^{16}$ yr, but very effective, 
$\tau_{v,h} \simless 10^{6}$ yr, at $m_i \gtrsim 10^{17}$ kg.
Collisional damping is also more effective at large masses, 
but the time scale is much less mass-sensitive than viscous stirring.
Collisional damping balances viscous stirring for an initial particle
mass, $m_0 \approx 10^{15}$ kg, which corresponds to $r_0 \approx$ 5 km.

To illustrate these points in more detail, Figures 7(b) and 7(c)
plot the {\it integrated} time scales, 
$\tau_{v,h} = \sum_{i=1}^{N} h_i / (d h_i/dt) $,
for the horizontal velocity at two stages of a model with 
velocity evolution.  In Figure 7(b), the maximum mass has 
$m_{max} \approx 10^{14}$ kg.  Collisional damping still 
dominates viscous stirring for the lowest masses, but 
the mass where the two processes balance has moved from 
$m_i \approx 10^{15}$ kg to $m_i \approx 5 \times 10^{11}$ kg.
Once $m_{max} \approx 10^{19}$ kg, viscous stirring dominates
collisional damping for all masses.  Particle velocities thus
increase once massive objects with $r_i \simgreat$ 100 km are
produced.  The time scale for viscous stirring is quite short,
$\simless 10^6$ yr, at the lowest masses considered in our models,
so the velocity increases can be large during the 100 Myr time scale
of a typical simulation.  

Figure 8 shows how $N_C$ and $h_i$ evolve with time in a model
with $r_0$ = 8 km, $M_0$ = 10 \mearth, and $e_{init} = 10^{-3}$.
The simulation begins with $N_0 = 1.87~\times~10^{10}$ and 
produces 762 objects having eight times the initial mass in 1 Myr.  
Roughly half of the initial bodies experience at least one collision 
by 24 Myr, when the largest object has $r_{max} \approx$ 37~km.  
The horizontal velocities then range from $h_i$ = 1 m s$^{-1}$ 
at $r_i$ = 37~km up to $h_i$ = 7 m s$^{-1}$ at $r_i$ = 8~km.  
Orderly growth produces $r_i$ = 100~km objects at 56 Myr,
and this population reaches $N_C \approx$ 100 at 61 Myr.  This phase 
continues until $\sim$ 100 Myr, when Charon-sized objects with 
$r_i \approx$ 500~km begin to grow rapidly. There are 
10 ``Charons'' at 110 Myr, 47 ``Charons'' at 125 Myr, 
107 ``Charons'' at 150 Myr, and 202 ``Charons'' at 180 Myr 
when we ended the simulation. At 180 Myr, 20 Pluto-sized objects 
with $r_i \approx$ 1000 km are isolated bodies about to run away 
from the rest of the mass distribution.

In contrast to the constant velocity simulations with low $e_{init}$,
this model does not immediately enter a rapid growth phase once 
objects with $r_i \approx$ 100~km are first produced.   Viscous stirring 
and dynamical friction slowly increase the velocities of low mass 
bodies throughout the simulation: the horizontal velocity increases 
from $h_{init}$ = 4 m~s$^{-1}$ to $h_i \approx$ 65 m~s$^{-1}$ at 180 Myr
(Figure 8).
This twentyfold increase in the eccentricity reduces gravitational 
focusing by a factor of 400 and retards the growth of the most 
massive objects.  Evolution thus proceeds at a pace between the 
constant velocity simulations with $e_{init}$ = $10^{-3}$ and
$e_{init}$ = $10^{-2}$.

The evolution for simulations with smaller initial masses is different,
because collisional damping then dominates the velocity evolution (Figure 7). 
Figure 9 shows the time evolution of $N_C$ and $h_i$ for 
$r_0$ = 800 m, $M_0$ = 10 \mearth, and $e_{init} = 10^{-3}$.
This simulation begins with $N_0 = 1.87~\times~10^{13}$ and produces 
five objects having eight times the initial mass in 1 Myr.
It takes only 7.7 Myr for half of the initial objects to collide
at least once.  The maximum radius is then $r_{max} \approx$ 2.5~km.
Velocity damping from inelastic collisions overcomes viscous stirring,
so the particle velocities remain low and do not change significantly 
with mass.  As evolution proceeds, dynamical friction efficiently damps
the velocities of the largest bodies, but collisional damping still
maintains modest velocities at low masses.  
Bodies with $r_i$ = 8--10 km begin to form at 32--33 Myr, when 
only 11\% of the original objects remain.  The evolution soon 
overtakes the $r_0$ = 8 km model.  Orderly growth produces 50~km 
objects at 45 Myr and 100 km objects at 48 Myr.  Runaway growth 
begins shortly thereafter.  Charon-sized objects form at 60 Myr 
and reach Pluto-size at 80--81 Myr.

Simulations with $r_0$ = 80 m reach runaway growth on even faster
time scales.  Figure 10 shows the time evolution of $N_C$ and $h_i$ 
for $M_0$ = 10 \mearth~and $e_{init} = 10^{-3}$.  At 1 Myr, only 
43\% of the initial bodies have yet to experience a collision;
33 objects already have $r_i \approx$ 270 m.  The maximum radius
reaches $r_{max}$ = 800~m in 9 Myr and $r_{max}$ = 8~km in 17 Myr.
The low mass bodies first lose $\sim$ 50\% of their initial velocity,
$h_{init}$ = 4 m s$^{-1}$, and begin to increase in velocity at 
17--18 Myr when viscous stirring from long-range collisions finally 
overcomes damping from inelastic collisions.  At this time, the 
high mass bodies have low velocities due to dynamical friction, 
$h_i \approx$ 0.01 m s$^{-1}$, and begin to grow rapidly.
A runaway plateau in the $N_C$ distribution develops at 24 Myr 
and extends to Charon-sized objects at 25 Myr.  At the conclusion 
of this simulation at 33 Myr, five Pluto-sized objects have
$r_i \approx$ 900--1000 km.  The velocities of these high mass 
objects are then $h_i \approx$ 0.03--0.05 m s$^{-1}$.

With their long runaway growth times, models with $r_0$ = 8 km 
cannot meet both of our success criteria unless the initial mass 
is very large, $M_0 \gtrsim$ 15--20 \mearth.  Viscous stirring and 
dynamical friction increase the velocities of the small objects 
throughout these simulations, which reduces gravitational focusing 
and delays runaway growth compared to models with smaller $r_0$.  
The long approach to runaway growth allows the production of many 
large KBO's; simulations with $M_0 \approx$ 6--20 \mearth~have 
$r_5 \approx$ 50--100 km and thus reach our first success criterion.
However, the time scale to produce 1000+ km objects is 4--5 times 
longer than models with $r_0$ = 80 m, i.e.,
$\tau_r \approx$ 130 Myr ($M_0 / 10~\mearth$)$^{-1}$.
Most of these models thus fail to make Pluto on a reasonable time scale.

Models starting with lower mass objects, $r_0$ = 80 m and 800 m, meet 
both success criteria. Although viscous stirring and dynamical friction 
stir up the velocities of the lowest mass objects, the time scale 
for the velocity to increase is large compared to models with 
$r_0$ = 8 km (see Figure 7). The $r_0$ = 800 m models reach runaway 
growth faster than models with $r_0$ = 8 km and produce Pluto-sized 
objects in $\tau_r \approx$ 83 Myr ($M_0 / 10~\mearth$)$^{-1}$
for $e_{init}$ = $10^{-3}$.  The combination of smaller particle 
velocities and a shorter time to runaway growth results in fewer KBOs 
compared to models with $r_0$ = 8 km.  Nevertheless, these models 
achieve $r_5 \approx$ 50--90 km during the runaway growth phase.

Models with $r_0$ = 80 m and $e_{init} = 10^{-3}$ have the shortest
runaway growth times and produce the fewest numbers of KBOs.  
The time scale to produce Pluto-sized planets is $\tau_r \approx$ 32 Myr
($M_0 / 10~\mearth$)$^{-1}$ for $e_{init}$ = $10^{-3}$, which easily 
allows Pluto formation in a minimum mass solar nebula as Neptune
forms at a smaller semi-major axis.  These models, however, struggle 
to build a population of $10^5$ KBOs during the runaway growth phase.
With relatively low particle velocities at all masses (see Figure 10b),
objects with $r_i \approx$ 10--20 km do not grow as rapidly as larger
bodies.  This evolution tends to concentrate material in the more massive
KBOs and reduces the number of lower mass KBOs with $r_i \approx$ 50--100 km.
This smaller $r_5$ in the calculations leads to partially-successful
models that yield several Pluto-sized objects and $r_5 \lesssim$ 50 km
(Table 5).

At large initial eccentricity, planetesimal growth follows the evolution 
of low eccentricity models but on longer time scales (Figure 12).  
In simulations with $r_0$ = 80 m and 800 m, there are enough small
bodies for inelastic collisions to damp the particle velocities 
substantially.  Dynamical friction further decreases the 
velocities of the most massive bodies and allows runaway 
growth to occur on reasonable time scales.  In simulations 
with $r_0$ = 80 m and $e_{init} = 10^{-2}$, runaway growth 
is delayed by a factor of $\sim$ 2.5 compared to simulations
with $e_{init} = 10^{-3}$.  This delay increases to a factor of 
$\sim$ 3 for $r_0$ = 800 m.  At $r_0$ = 8 km, collisional damping 
initially reduces particle velocities but is overcome by viscous 
stirring when $r_{max} \approx$ 50 km.
At this time, the velocities are large, $V_i \approx 30$ m s$^{-1}$, 
and growth is slow.
None of these models reach runaway growth on a 100 Myr time scale.
Runaway growth, if it occurs, is delayed by a factor of $\sim$ 6
in large $e_{init}$ models compared to that in low $e_{init}$ models.

Our results for large $e_{init}$ thus favor low mass initial bodies.
Simulations with $r_0$ = 80 m and $e_{init} = 10^{-2}$ produce more KBOs
with $r_i \approx$ 50--100 km than their low $e_{init}$ counterparts.
The longer orderly growth phase and somewhat larger particle velocities
at the onset of runaway growth favor the growth of 50--100 km objects.
The $r_5$ for these mass distributions is 10\% to 20\% larger than the 
$r_5$ for low $e_{init}$ ($r_5 \approx$ 50--90 km for $e_{init} = 10^{-2}$
compared to $r_5 \approx$ 50--75 km for $e_{init} = 10^{-3}$).
Models with larger $r_0$ are less successful.  For $r_0$ = 800 m, 
runaway growth begins well after 100 Myr unless $M_0 \simgreat$ 
20--30 \mearth.  Runaway growth always occurs on a very long time scale,
$\gtrsim$ 150--200 Myr, for $r_0$ = 8 km and $M_0 \simless$ 30 \mearth.

Unlike the constant velocity models, simulations with velocity evolution 
begin to develop a ``runaway plateau'' in the cumulative mass distribution 
when $r_{max} \simgreat$ 500 km (see Figures 8--10).  In constant
velocity simulations, we found two power law cumulative mass 
distributions, $N_C \propto r^{-2.75}$ at low masses and
$N_C \propto r^{-1.75}$ at large masses.  The total mass per mass batch
is then roughly constant at low masses and increases slowly with mass at 
large masses.  In models {\it with} velocity evolution, dynamical friction 
reduces the velocities of the largest bodies to $V_i \simless$ 0.1 m s$^{-1}$
and maintains velocities of $V_i \sim$ 1--10 m s$^{-1}$ for 
smaller objects with $r_i \simless$ 10 km.  As noted by WS93, the 
increase in the escape velocity with mass coupled with the decrease 
in $V_i$ produce substantial increases in the collisional cross-sections
(see also \cite{wet89}; \cite{bar90}, 1991, 1993; \cite{id92a}, 1992b, 
\cite{oht92}; \cite{kok96}).
In our models, the velocity distribution resembles a step function and 
produces a step-like increase in the gravitational focusing factors,
from $F_g \sim$ 10--100 at $r_i \sim$ 10--100 km up to $F_g \sim 10^4$ 
at $r_i \sim $ 300--1000 km.  Runaway growth then converts the 
$N_C \propto r^{-1.75}$ mass distribution into $N_C \approx$ constant,
because objects with $r_i \sim$ 100--200 km grow too slowly to fill
in the power law as objects with $r_i \simgreat $ 500 km run away.
At low masses, the size distribution remains a power law, 
$N_C \propto r^{-2.75}$, because runaway growth does not change
the size distribution significantly.

To conclude this section, Figure 13 summarizes results for accretional 
evolution in the Kuiper Belt with velocity evolution and no fragmentation.  
Successful simulations that produce $\sim 10^5$ KBOs and a few Pluto-sized 
objects on time scales of 100 Myr or less have initial masses somewhat larger
than that predicted for a minimum mass solar nebula extrapolated into the 
Kuiper Belt, $M_0 \gtrsim$ 10 \mearth, and bodies with initial radii of 
$r_0 \approx$ 80 m to 8 km.  Simulations with small initial bodies, 
$r_0$ = 80 m, tend to produce fewer KBOs than simulations with larger
initial bodies, $r_0 $ = 800 m and 8 km.  This feature of our calculations
results in partially-successful models that produce Pluto-sized objects 
but too few KBOs in 100 Myr.  In models with $M_0 \approx$ 6 \mearth
and $r_0$ = 80 m, runaway growth removes KBOs from the mass distribution 
more rapidly than they are produced from lower mass objects.  This 
evolution does not occur in models with larger initial bodies, because 
collisional damping is less effective at ``circularizing'' the orbits 
during the orderly growth phase.  The higher particle velocities in 
these models allows formation of more KBOs during the runaway growth 
of Charon-sized objects.

Collisional evolution often fails to produce 100+ km objects on any
useful time scale.  Simulations with $M_0 \approx$ 1--6 \mearth~produce 
neither Pluto-sized objects nor a substantial number of 100+ km KBOs in 100 Myr.
Large initial eccentricities exacerbate this problem for models with
$r_0 \gtrsim$ 800 m, because collisional damping cannot reduce the
particle velocities before 100+ km objects form.  These simulations 
can produce KBOs and Plutos on longer time scales, 100--1000 Myr, in 
systems where a Neptune-sized object does not constrain the formation time.
Extrapolating our results to smaller initial masses, simulations with 
$M_0 \simless$ 0.1 \mearth~fail to produce KBOs during the age of the 
solar system, $\sim$ 5 Gyr.

\subsection{Limitations of the Models}

Statistical simulations of planetesimal growth have
well-documented approximations and uncertainties.  
The model assumes a homogeneous spatial distribution of
planetesimals whose velocities are small compared to the
orbital velocity.  These assumptions are good during the
early stages of planetesimal evolution.  As planetesimals
grow, dynamical friction can reduce the velocities of high 
mass objects below limits where the statistical approach is
valid (\cite{bar90}).  Once this limit is reached, runaway
growth produces a few large bodies that are not distributed
homogeneously in space (\cite{ws93}; \cite{kok96}).  These large 
bodies can then pump up the velocities of the smallest bodies
on short time scales through viscous stirring (Figure 7).  
We end the simulations with velocity evolution during 
the runaway growth stage when the basic assumption of
a homogeneous distribution of planetesimals begins to
break down.  The velocities of low mass bodies remain small 
compared to the Keplerian velocity, but the most massive objects 
often have velocities below the low velocity limit of the kinetic
approximation.  We will discuss this problem below.

The remaining limitations of the statistical approach involve 
our implementation of standard algorithms.  We adopt a single
accumulation zone and thus cannot follow the evolution in
semi-major axis of a planetesimal swarm (see \cite{spa91}).
We use a coarser grid than some simulations, but this choice
has little impact on the results at 35 AU.  At 1 AU, the lag
of simulations with $\delta >$ 1.1 relative to a simulation
with $\delta$ = 1.1 increases with increasing $\delta$;
we find a 12\% lag for $\delta$ = 1.4 but only a 2--3\% lag
for $\delta$ = 1.25.  At 35 AU, the lag in runaway growth 
relative to a $\delta$ = 1.1 model increases from 4\%--5\% 
for $\delta$ = 1.4 to 10\%--15\% at $\delta$ = 2.  
Our $\delta$ = 1.4 simulations thus overestimate the runaway
growth time only by 4--5\% (see also \cite{wet90}; \cite{kol92}).
This error is small compared to other uncertainties in the calculation.

Our choice of the initial mass distribution has a modest 
impact on our results.  We calculated the evolution of several
size distributions with equal mass per mass batch for 
$ r_{min} \simless r_i \simless r_{max}$.  Simulations with 
$r_{min} \sim$ 100--1000 m and $r_{max} \simless$ a few km are 
nearly indistinguishable from simulations with a single starting 
radius, $r_0 \approx r_{max}$.
In these models, collisional damping effectively reduces all
particle velocities as described above and allows runaway growth
to occur.  Simulations with large $r_{max} \approx$ 8 km are similar
to those with a single starting mass, unless $r_{min}$ is small.
For $r_{min} \simless$ 800 m, collisional damping keeps the 
particle velocities small compared to models with a single 
starting mass.  Runaway growth occurs in these models, but 
the time scale to runaway is sensitive to $r_{min}$.  We plan
to explore this sensitivity in more detail when we include 
fragmentation in the calculation.

The most uncertain approximation in our calculations is the
treatment of low velocity collisions.  During the late stages
of most simulations, the massive bodies have very low velocities
and very large gravitational ranges.  The velocities are often
smaller than the Hill velocity, $V_H$, which invalidates the
basic assumptions for our velocity evolution calculations
(\cite{bar90}, 1991, 1993).  Barge \& Pellat (1990) and WS93 have
developed different approximations to low velocity collisions based
on Ida's (1990) $n$-body simulations (see also \cite{oht92}).
We note that these two approximations produce different velocities
for high mass planetesimals, which we plan to examine in more 
detail when we include fragmentation in our calculations.  
The mass evolution of runaway bodies is not affected by our 
treatment of velocity evolution in the low velocity regime.

The large gravitational ranges of the most massive bodies 
also invalidate the standard treatment of collisions.  
We use the WS93 prescription for isolating the largest 
bodies from collisions with one another
and adopt the Greenberg \etal (1992) approach to low velocity
collisions in the two-dimensional regime.  Removing the isolation 
criteria allows the largest body to grow more rapidly than 
the isolated bodies and reduces the time scale to runaway growth 
by 5\%--10\%.  Removing the Greenberg \etal (1992) two-dimensional 
cross-sections has no substantial effect on our results.

Aside from fragmentation, we have included all important physical
processes in planetesimal evolution.  Our neglect of fragmentation,
however, is a serious limitation.  
In previous simulations, fragmentation of relatively strong bodies
with $r_i \simgreat$ 1--10 km produces a significant amount of cratering
debris that can be accumulated later by runaway bodies (e.g., 
\cite{ws93}; \cite{bar93}).  This process usually becomes important 
only in the late stages of calculations at 1 AU: it slows growth during 
early stages but speeds up runaway growth later in the evolution 
(\cite{bar93}; \cite{ws93}).  However, collisions between very weak
bodies can disrupt and thereby prevent {\it any} growth of icy 
planetesimals at modest velocities.  The importance of fragmentation 
at $a \approx$ 35 AU thus depends on the unknown strength of KBOs.

We can estimate the importance of fragmentation in Kuiper Belt simulations
using Barge \& Pellat's (1993) results for a reasonable fragmentation
model.  They adopt the Housen \etal (1991) energy prescription for
planetesimal disruption and derive collisional outcomes for several
test cases.  These results are most appropriate for rocky asteroids, 
but it is straightforward to scale them to the weaker, icy bodies 
that might exist at 35 AU.  We consider the two cases recently adopted
by Stern \& Colwell (1997): strong, rocky KBOs and weak, icy KBOs.

Fragmentation does not significantly change our main conclusions if KBOs 
are strong objects.  According to Figure 5 of Barge \& Pellat (1993),
fragmentation modifies the growth of 10 km bodies only when 
$e \simgreat e_{crit} \approx$ 0.025.  The critical eccentricity for 
fragmentation decreases to $e_{crit} \approx 10^{-2}$ for 1 km objects 
and $e_{crit} \approx 2~\times~10^{-3}$ for 0.1 km objects
(see also Fig. 1 of \cite{ste96}).  Our low $e$ simulations never 
reach these critical values.  Fragmentation is important in large 
$e$ simulations, but most of these models do not produce KBOs on 
a reasonable time scale. 

The growth of icy KBOs is probably very sensitive to the time scale 
for velocity evolution.
We expect fragmentation to dominate the early evolution of all 
simulations considered above, because only objects with 
$r_i \simgreat$ 20--30 km can survive collisions
and produce larger bodies (see also Stern \& Colwell 1997).  
As the evolution proceeds, however, inelastic collisions should
damp the velocities of bodies with $r_i \simless$ 1 km, while
dynamical friction damps the velocities of the most massive
objects (see Figure 7).  These damping time scales are short compared 
to the viscous stirring time scales, so the particle velocities 
decrease on relatively short time scales, $\sim$ 10 Myr.  This
damping is probably sufficient to allow the growth of 1--10 km
objects on time scales similar to those found in our models without
fragmentation.  Smaller bodies may not grow unless dynamical friction 
is very efficient.  We will explore this possibility in our second paper.

\section{DISCUSSION AND SUMMARY}

We have developed a time-dependent planetesimal evolution
program similar to the WS93 code used to simulate the
formation of terrestrial embryos from small bodies.
The program incorporates coagulation with realistic 
cross-sections and velocity evolution using the 
statistical formulation of Barge \& Pellat (1990, 1991; 
see also Hornung \etal 1985).  Our numerical solutions to the
coagulation equation agree with analytical solutions for 
three standard test cases.  Our results also agree with WS93's 
simulation of the formation of the Earth at 1 AU.  The present 
models do not incorporate fragmentation of bodies during collisions.
We will include fragmentation in a separate paper.

We have considered two simple cases of planetesimal evolution
in a 6 AU wide annulus centered at 35 AU.  Models without velocity 
evolution invariably produce several large bodies that accrete
practically all of the material in the annulus.
The runaway growth in these simulations occurs without dynamical 
friction or gas drag; it is a direct consequence of gravitational
focusing.  The time required to produce a runaway body in our models 
scales inversely with the initial mass of the annulus {\it and} 
with the initial radii and velocities of the planetesimals.  
For bodies with $r_0$ = 80--8000 m and $e_{init} = 10^{-3}$,
our simulations produce runaway growth in 100 Myr for 
annular masses of roughly 10--30 $M_{\earth}$.   The time scale for 
runaway growth increases to 700--2000 Myr for $e = 10^{-2}$.  
A minimum mass solar nebula with $\Sigma \propto R^{-3/2}$ contains 
7--15 \mearth~in a 6 AU wide annulus centered at 35 AU.  These models 
thus reach runaway growth in a minimum mass nebula on time 
scales comparable to the maximum formation time scale for 
Neptune ($\sim$ 50-100 Myr, \cite{lis96}; \cite{pol96}).  
Runaway growth on much shorter time scales, $\sim$ 10 Myr, 
requires annular masses that far exceed the minimum mass 
solar nebula, $\sim$ 100 $M_{\earth}$ in a 6 AU wide annulus 
for $e_{init} = 10^{-2}$ to $10^{-3}$.

Models with velocity evolution produce runaway growth on a much
wider range of time scales compared to constant velocity calculations.
First, dynamical friction and viscous stirring dominate the evolution 
of models with relatively large (8 km) initial bodies.  The velocities 
of these bodies thus increase as collisions produce more massive objects.
This velocity increase delays runaway growth by factors of
2 or more compared to constant velocity evolution.  The delay in
the runaway growth time increases with increasing $e_{init}$.
In contrast, collisional damping dominates the evolution of models 
with smaller (80--800 m) initial bodies.  These bodies `cool' until 
the largest objects have radii of 10--20 km.  Dynamical friction and 
viscous stirring then `heat up' the small bodies but this heating is 
small compared to the velocity increases of the $r_0$ = 8 km models.
For $e_{init}$ = $10^{-3}$, collisional damping enhances collision 
rates and decreases the time scale to runaway growth by factors of 
4--12 compared to constant velocity calculations.  Our simulations of 
minimum mass solar nebulae with $e_{init} = 10^{-3}$ reach 
runaway growth on time scales of 20--40 Myr for 80 m initial bodies, 
50--100 Myr for 800 m bodies, and 75--250 Myr for 8 km bodies.
These time scales increase by factors of 2--4 for $e_{init} = 10^{-2}$.

The formation of runaway bodies in constant velocity simulations
is surprising.  Previous simulations demonstrate a need for 
dynamical friction, which decreases the velocities of the most 
massive objects (\cite{wet89}, 1993; \cite{bar90}, 1991; \cite{oht92};
\cite{id92a}, 1992b; \cite{kok96}).  Gravitational focusing then
allows these bodies to sweep up lower mass bodies very rapidly.
Kokubo \& Ida (1996) summarize necessary and sufficient conditions
for runaway growth and show that the ratio of the maximum mass of
planetesimals to their mean mass increases dramatically during
runaway growth.  Our constant velocity simulations satisfy the
``necessary'' condition for runaway growth, 

\begin{equation}
2 R_{H,ij} \Omega \simless V_{ij} \simless V_{e,ij}
\end{equation}

\noindent
but do not meet the ``sufficient'' condition,

\begin{equation}
\frac{dV}{dm} < 0 ~ ,
\end{equation}

\noindent 
because our velocities are constant with mass, $dV/dm$ = 0.
Nevertheless, the maximum mass, $M_{max}$, of each constant 
velocity simulation increases much more rapidly than the 
mean mass, $<m_i>$ (Figure 14).  In Figure 14(a), the ratio 
$M_{max} / <m_i>$ increases slowly during the orderly growth 
phase and then increases rapidly when several isolated bodies 
begin to accrete most of the mass in the annulus.  This runaway 
growth is not as extreme as that seen in models with velocity
evolution.  We derive $M_{max} / <m_i> ~ \approx 10^6$ in constant
velocity models compared to $M_{max} / <m_i> ~ \approx 10^8$--$10^{10}$
in models with velocity evolution (Figure 14(b)).  Dynamical friction
is responsible for the larger increase in $M_{max} / <m_i>$ in
models with velocity evolution.

Although the constant velocity simulations are artificial, they are 
a useful guide for planetesimal growth in the outer solar system.
Our results show that collisional damping by small bodies can 
overcome viscous stirring and keep particle velocities low at $a$ = 
35 AU.  The relative efficiency of collisional damping should 
increase with $a$, because the collision rates decrease more
rapidly with $a$ than do the damping rates ($\tau_{coll} \propto
a^{-5/2}$ vs. $\tau_{damp} \propto a^{-3/2}$).  The situation in 
the outer solar system differs markedly from conditions at small $a$.  
In our simulations at 1 AU, bodies with $r_i \approx$ 80--800 m 
grow to 10 km objects in $\sim$ 1000 yr.  As in the 35 AU simulations
described above, collisional damping reduces the velocities of the small
bodies by a factor of $\sim$ 2 in 1000 yr.  In contrast to our 35 AU
models, viscous stirring and dynamical friction act quickly to increase 
velocities once larger bodies are produced at 1 AU.  Aside from the 
longer time scale to reach runaway growth, this evolution then 
follows closely the simulations in Figures 1--2.

With these considerations in mind, we suggest a modest modification 
to the sufficient condition for runaway growth,

\begin{equation} 
\frac{dV}{dm} \le 0 ~ ,
\end{equation}

\noindent
instead of the condition in Eqn. (3).  This condition maintains the 
relative growth rate necessary for runaway in a three-dimensional
system (see \cite{kok96})

\begin{equation}
\frac{1}{M} \frac{dm}{dt} \propto M^{1/3} ~ ,
\end{equation}

\noindent
when the gravitational focusing factor is large
(e.g., $V_{ij} \simless V_{e,ij}$).  

In addition to this new criterion for runaway growth, our
simulations show that Neptune and Pluto can grow in parallel
at $a \approx$ 30--35 AU.  For a minimum mass solar nebula,
previous calculations indicate that Neptune reaches its 
current size in no longer than 50--100 Myr (\cite{ip89}; 
\cite{lis96}; \cite{pol96}).  In our models of minimum mass 
solar nebulae at 35 AU, 800 m objects grow to Pluto-sized planets 
on similar time scales.  However, recent observations indicate
a much shorter lifetime, $\sim$ 5--10 Myr, for gaseous disks
surrounding nearby pre-main sequence stars (\cite{sar93}; 
\cite{str93}; \cite{rus96}; see also \cite{fer97}).  If our 
solar system evolved on a similar time scale, the formation of 
the gas-rich outer planets requires solar nebulae with masses 
2--5 times larger than the minimum mass solar nebula, $M_{min}$ 
(\cite{lis96}; \cite{pol96}).  Neptune then attains its current 
size in 5--10 Myr.  At 35 AU, 
objects reach 1000 km sizes in 10--20 Myr for $M_0 \approx$ 2--3 
$M_{min}$ and in 5--10 Myr for $M_0 \approx$ 5 $M_{min}$ assuming 
the initial bodies have small masses and eccentricities.  
Fragmentation should not change these conclusions unless 
collisional erosion can prevent the formation of 1 km bodies 
from smaller building blocks.

Our Kuiper Belt simulation produce Pluto-sized objects on
reasonable time scales for other plausible solar nebula models.
In particular, Cameron's (1995 and references therein) detailed 
disk models for the protosolar nebula have a shallower density
density distribution, $\Sigma = \Sigma_0 (a/a_0)^{-1}$, than
the minimum mass model we considered above to derive our success 
criteria (see also \cite{rud86}; \cite{rud91}).  
Cameron's model contains $\sim$ 100 \mearth~in our 6 AU annulus.  
The growth time for 1000+ km objects scales simply from results with 
smaller disk masses.  For models with velocity evolution, we estimate
$\tau_r \approx 3$ Myr for $r_0$ = 80 m,
$\tau_r \approx 8$ Myr for $r_0$ = 800 m, and
$\tau_r \approx 15$ Myr for $r_0$ = 8 km.
Although all of these models can produce Pluto-sized objects
before Neptune reaches its final mass, models with $r_0$ = 8 km 
produce more KBOs with $r_i \sim$ 100--300 km as outlined in
Sec. 3.3.

These results -- together with recent dynamical calculations of 
Malhotra (1993, 1995, 1996) -- suggest a self-consistent picture 
for the formation of Pluto-Charon in the outer solar system.
In this picture,  Pluto and Charon begin as $\sim$ 1 km planetesimals 
at $a \sim$ 35--40 AU and grow to their present sizes on a time scale 
of 10--100 Myr.  Both objects are runaway bodies more massive than 
the bulk of the planetesimal mass distribution and have low velocities 
due to collisional damping and dynamical friction.  At a somewhat 
smaller semi-major axis, $a \approx$ 25--30 AU, Neptune accretes 
its current mass in 10--100 Myr and migrates radially outward through 
the protosolar disk during the late stages of giant planet formation
(\cite{mal93}; \cite{pol96}; \cite{fer96}; see also \cite{ipa89}, 1991).  
During this outward migration, Neptune captures Pluto-Charon 
in the 3:2 resonance (\cite{mal95}).  This capture should 
effectively end further growth of Pluto-Charon, because the orbital 
elements increase to $e \sim$ 0.2 and $i \sim 10^{\circ}$ on a 
short time scale, $\sim$ 10 Myr, inside the resonance (\cite{mal95}).
High velocity collisions within the resonance should also hinder 
growth as in our large $e$
models.  Neptune also captures other KBOs at $a \approx$ 35--40 AU into 
the 3:2 and other resonances (\cite{mal96}; see also \cite{jew96}).  
Further growth of these objects is also slowed due to rapidly 
increasing velocities (\cite{mal95}, 1996; \cite{mor95}; \cite{dav97}).  
As long as collisional erosion does not decrease significantly 
the radii of captured KBOs, this sequence of events accounts for 
the general aspects of the mass distribution and orbital elements 
of observed KBOs in a simple way.

This picture for the formation of Neptune, Pluto-Charon, and KBOs 
differs from those of Stern (1995, 1996) and Stern \& Colwell (1997a,b), 
who studied KBO evolution at 40--70 AU.  Stern \& Colwell (1997)
concluded that the formation of Neptune and KBOs, including Pluto-Charon,
requires $\sim$ 100--1000 Myr due to the low collision rates at
$a \approx$ 30--50 AU.  Stern \& Colwell's (1997a; also \cite{ste95},
1996) time scale for KBO formation is a factor of 10--20 longer than 
our time scale.  Their results also conflict with the more detailed
gas dynamic calculations of Pollack \etal (1996). Although the exact 
origin of this discrepancy is unclear, we suspect mass spacing, 
collision rates, and (less probably) fragmentation may be responsible.
Our mass spacing, $\delta$ = 1.4, should produce more accurate
estimates for the growth time than $\delta$ = 2 (\cite{st97a})
or $\delta$ = 4 (\cite{ste96}).  We estimate 10\%--20\% delays 
in runaway growth for $\delta$ = 2 and expect very long delays
for $\delta$ = 4.  In addition, our collision rates include a 
factor, $\beta_{coll}$, that accounts for the gaussian distribution 
of impact velocities (\cite{grz92}; see Eqn. A14).  Neglecting this 
factor delays runaway growth by a factor of 3 (see also \cite{ws93}).  
Stern (1995) does not include this factor in his cross-section (see 
his Eqn. 5) and thus derives much longer growth times.  As noted above, 
our neglect of fragmentation encourages runaway growth.  We do not think 
that including this process should delay runaway growth by another factor 
of 3, for reasons outlined earlier, and plan to test this suspicion in 
our next paper.

However the theoretical issues may be resolved, several consequences
of our accretion models can be tested with additional observations.
All models that reach runaway growth produce 10--100 Pluto-sized objects 
with radii $r_{max} \approx$ 1000--2000 km and a roughly power-law 
mass distribution with a maximum radius at $\sim$ 0.1--0.2$r_{max}$
(see Figures 8--10).  If Pluto and Charon are runaway bodies produced
at $a \approx$ 35--40 AU, there should be several additional ``Plutos''
with similar orbital elements.  Malhotra (1995) reached a similar 
conclusion and noted that recent searches have not excluded the 
possibility of several additional Plutos in 3:2 orbital resonance
with Neptune.  The accretion models also predict a cumulative power law 
distribution, $N_C \propto r^{-3}$, for objects with $r \lesssim$ 
100--200 km.  By analogy with WS93, this shape is fairly independent 
of fragmentation as long as collisions produce overall growth 
instead of disruption.  The best fit to the observed distribution 
is shallower than expected, $N_C \propto r^{-2}$, but the data are not
yet accurate enough to preclude our model prediction (\cite{jew96}).
Larger surveys will provide a better test of this prediction (\cite{luu97}).

Finally, our results suggest that KBO formation is likely in other 
solar systems.  KBOs can grow in the dusty disks that surround many
nearby main-sequence stars, if the disk masses are within an order
of magnitude of the ``maximum'' disk masses for $\alpha$ Lyr, 
$\alpha$ PsA, and $\beta$ Pic (\cite{bac93}).  Stars with disk masses 
near the minimum dust masses of these well-studied A-type stars 
probably have few, if any, KBOs, but could have many objects with 
$r_i \sim$ 1--10 km.  Observations of nearby pre-main sequence stars 
also indicate substantial masses, $M_d \gtrsim 10^3~\mearth$, in 
circumstellar disks with radii of 100--1000 AU
(e.g., \cite{sar93}; see also \cite{bec96}).  These data imply $M_0$
$\sim$ 10--100 \mearth~in the Kuiper Belt.  With formation time scales 
of 1--10 Myr, KBOs can grow in massive circumstellar disks during the 
pre-main sequence contraction phase of a low mass star (see \cite{kh95}
and references therein).  In a less massive disk,  KBOs grow while the 
central star is on the main sequence.  We expect no KBO formation in
circumstellar disks with very low masses, $M_d \lesssim$ 10 \mearth.  
These disks can produce 1--10 km objects at $a \gtrsim$ 30 AU, unless 
fragmentation prevents growth of icy bodies.  We will explore this 
possibility in a subsequent paper.

\vskip 6ex 

We thank B. Bromley for making it possible to run our code on the 
JPL Cray T3D and for a generous allotment of computer time through
funding from the NASA Offices of Mission to Planet Earth, Aeronautics, 
and Space Science.  K. Kirby and J. Babb also graciously allocated us 
computing time on the CfA Dec Alpha computer.  Comments from A. Cameron, 
F. Franklin, M. Geller, M. Holman, J. Wood and an anonymous referee greatly 
improved our presentation.  We acknowledge G. Stewart for clarifying 
many details of the WS93 calculations.  We also thank S. Stewart for
designing and implementing a preliminary version of our computer
code as part of her Harvard Senior Thesis.  Finally, S.K. thanks 
S. Starrfield for teaching the tools for solving time-dependent 
differential equations.

\appendix

\section{APPENDIX}

\subsection{Overview}

Our evolution model follows procedures developed for other planet 
formation calculations, including Safronov (1969), Greenberg \etal 
(1978, 1984) and WS93.  We assume planetesimals are a statistical 
ensemble of masses with a distribution of horizontal and vertical 
velocities about a single Keplerian orbit.  
We consider a cylindrical annulus of width, $\Delta a$, and height, $H$, 
centered at a radius $a$ from the Sun.  Particles in the annulus have 
a horizontal velocity, $h_i(t)$, and vertical velocity, $v_i(t)$,
relative to an orbit with mean Keplerian velocity, $V_K$ (see
\cite{lis93}).  These velocities are related to the 
eccentricity, $e_i$, and inclination, $i_i$, through

\begin{equation}
V_i^2 = (\frac{5}{8} e_i^2 + \frac{1}{2} {\rm sin^2} i_i) V_K^2,
\end{equation}

\noindent
with

\begin{equation}
h_i^2 = \frac{5}{8} e_i^2 V_K^2
\end{equation}

\noindent
and

\begin{equation} 
v_i^2 = \frac{1}{2} {\rm sin^2} i_i   V_K^2
\end{equation}

\noindent
We approximate the continuous distribution of particle masses with 
discrete batches having an integral number of particles, $n_i(t)$, 
and total masses, $M_i(t)$ (\cite{ws93}).  
The average mass of a batch, $m_i(t)$ = 
$M_i(t) / n_i(t)$, evolves with time as collisions add and remove 
bodies from the batch.  This procedure naturally conserves mass and 
allows a coarser grid than simulations with fixed mass bins 
(\cite{wet90}; see also \cite{oht88}; \cite{kol92}).

To follow the evolution of particle number, we solve the coagulation
equations for all mass bins, $k$, during a time step, $\delta t$,

\begin{equation}
\delta n_k = \delta t \left [ \epsilon_{ij} A_{ij} n_i n_j ~ - ~ n_k A_{ik} n_i \right ] ~ - ~ \delta n_{k,gd}
\end{equation}

\begin{equation}
\delta M_k = \delta t \left [ \epsilon_{ij} A_{ij} n_i n_j m_k ~ - ~ n_k A_{ik} n_i m_k \right ] ~ - ~ m_k \delta n_{k,gd}
\end{equation}

\noindent
where $A_{ij}$ is the cross-section,
$\epsilon_{ij} = 1/2$ for $i = j$ and $\epsilon_{ij} = 1$ for $i \ne j$.
The three terms in A4--A5 represent (i) mergers of $m_i$ and $m_j$ into 
a body of mass $m_k = m_i + m_j$, (ii) loss of $m_k$ through mergers
with other bodies, and (iii) loss of $m_k$ by gas drag.
This treatment assumes (i) that each body can collide with every 
other body and (ii) that bodies do not fragment during collisions.
Assumption (i) is correct for all but the very largest bodies,
which become isolated from one another as their orbits circularize
due to dynamical friction (see below).  
We correct equations A4--A5 for this effect by calculating the 
``gravitational range'' of the largest bodies --
$R_{g,i} = K_1 a R_{H,ii_{mid}} + 2 a e_i$ (\cite{ws93}) --
where $K_1 = 2 \sqrt{3}$ and $R_{H,ij} = [(m_i + m_j)/3 \msun]^{1/3}$
is the mutual Hill radius.  As in WS93, the isolated bodies are the 
$N$ largest bodies that satisfy the summation, 
$ \sum_{i_{min}}^{i_{max}} ~ n_i R_{g,i} \ge \Delta a$.
Assumption (ii) is rarely correct, because all collisions produce some
debris unless the relative velocity of the two particles is very low
(see, for example, \cite{ws93}).  In this paper, we concentrate on
planetesimal growth and assume that all collisions result in mergers.
We will consider the effects of fragmentation in a separate paper.

To calculate the appropriate index $k$ for a specific collision
between batches $i$ and $j$, we first calculate a fixed grid of
masses, $m_l$, for $l$ = 1 to $N_{max}$ and $\delta$ = $m_{l+1}/m_l$.
The mass spacing, $\delta$, is constant throughout a calculation;
$N_{max}$ increases with time as more batches fill with particles.  
When a collision produces $n_k$ bodies with $m_k$, 
we augment either batch $l$ when $m_k \le \sqrt{m_l m_{l+1}}$ or
batch $l+1$ when $m_k > \sqrt{m_l m_{l+1}}$.  A complete cycle
through all mass batches produces new values for $n_k$ and $M_k$,
which yields new values for the average mass per bin, $m_k = M_k/n_k$.
This process conserves mass and provides a good description of
coagulation when $\delta$ is small (see below).

Besides collisions, several processes contribute to the velocity 
evolution of growing planetesimals, including dynamical friction, 
gas drag, and viscous stirring.  We assume that all collisions 
between mass batches conserve the horizontal and vertical components 
of kinetic energy, $E_{h,i} = m_i h_i^2$/2 and $E_{v,i} = m_i v_i^2$/2.
The change in the two components of kinetic energy due to collisions is

\begin{equation}
\delta E_{h,k}^{co} = - 1/2 ~ \left [ \delta n_k (m_k h_k^2) = \delta n_i (m_i h_i^2) + \delta n_j (m_j h_j^2) \right ]
\end{equation} 

\begin{equation}
\delta E_{v,k}^{co} = - 1/2 ~ \left [ \delta n_k (m_k v_k^2) = \delta n_i (m_i v_i^2) + \delta n_j (m_j v_j^2) \right ]
\end{equation} 

\noindent
for each pair of collisions between $m_i$ and $m_j$.  In these expressions,
$\delta n_i \le 0 $ represents the change in $n_i$ due to collisions with 
particles in batch $j$.  Batch $k$ loses kinetic energy due to collisions
with other batches (e.g., $\delta n_k >$ 0).  We also calculate 
the evolution of $h_i$ and $v_i$ due to gas drag (\cite{ada76}) 
and collective interactions, such as dynamical friction and 
viscous stirring, using a statistical treatment of the 
appropriate Boltzmann and Fokker-Planck equations (\cite{hor85}; 
see section A.3 below).  The complete change in the horizontal and vertical
kinetic energies is thus:

\begin{equation}
\delta E_{h,k} = \delta E_{h,k}^{co} + \delta E_{h,k}^{gd} + \delta E_{h,k}^{in} + \delta E_{h,k}^{lr}
\end{equation}
 
\begin{equation}
\delta E_{h,k} = \delta E_{v,k}^{co} + \delta E_{h,k}^{gd} + \delta E_{h,k}^{in} + \delta E_{h,k}^{lr}
\end{equation}

\noindent
where the superscripts ``gd'' (gas drag), ``in'' (inelastic), and
``lr'' (long range elastic collisions, such as viscous stirring and 
dynamical friction) refer to a specific type of velocity evolution 
outlined below (see also \cite{bar90}, 1991; \cite{ws93}).

We solve the complete set of evolution equations, A4, A5, A8, 
and A9, using an explicit method that automatically prevents 
large changes ($>$ 0.1\%) in the dynamical variables -- $n_i$, 
$M_i$, $h_i$, and $v_i$ -- by limiting the time step.  
As in WS93, we require integer values for $n_i$ and $\delta n_i$.
Section A.4 compares our numerical procedures with analytic results
from Wetherill (1990; see also \cite{oht88}; \cite{oht90}).
Section 2 of the main text compares calculations at 1 AU with
results from WS93.  In both cases, our procedures reproduce 
the expected results.  Before describing the analytic results,
we first describe in detail our treatment of the collision rates
(section A.2) and the velocity evolution (section A.3).

\subsection{The Collision Rate}

Approximations to the collision rates between planetesimals are
in the spirit of kinetic
theory, where the number of collisions is the product of the local 
density, the relative velocity, and a cross-section.  WS93 express
the number of collisions between a single body, $m_i$, and all of 
the bodies, $m_j$, as

\begin{equation}
n_{c,ij} = \alpha_{coll}~\left ( \frac{n_j}{4~H~a~\Delta a} \right ) ~V_{ij}~F_{g,ij}~(r_i + r_j)^2~\delta t
\end{equation}

\noindent
where $r_i$ and $r_j$ are the radii of the two bodies;
$V_{ij}^2 = V_i^2 + V_j^2$ is the relative velocity; and
$F_{g,ij}$ is the gravitational focusing factor.
The constant factor $\alpha_{coll}$ accounts for the gaussian distribution
of particle velocities and the difference between the collision
frequency of particles on Keplerian orbits and those in a box
(\cite{grz92}; \cite{ws93}).
The relative velocities and scale height depend on the 
individual particle velocities (\cite{ws93}):

\begin{equation}
H = \frac{\sqrt{2}}{\Omega} (v_i^2 + v_j^2)^{1/2}
\end{equation}

\begin{equation}
V_{ij} = (h_i^2 + v_i^2 + h_j^2 + v_j^2)^{1/2}
\end{equation}

\noindent
where $\Omega$ is the Keplerian angular frequency.
The total number of collisions for $m_i$ is simply $n_i n_c$;
the cross-section appropriate for equations A4--A5 is then:

\begin{equation}
A_{ij} = \alpha_{coll}~\left ( \frac{1}{4~H~a~\Delta a} \right ) ~V_{ij}~F_{g,ij}~(r_i + r_j)^2 ~ .
\end{equation}

We consider two approaches to compute the gravitational focusing 
factor, $F_{g,ij}$.  In the first case, we follow WS93 and set

\begin{equation}
F_{g,ij} = F_{WS,ij} = E_{ij} \left ( 1 + \beta_{coll} \frac{V_{e,ij}^2}{V_{ij}^2} \right ) ,
\end{equation}

\noindent
where $V_{e,ij}^2 = 2 G (m_i + m_j)/(r_i + r_j)$ is the mutual escape velocity.
The extra factors account for the gaussian distribution of impact 
velocities ($\beta_{coll}$, \cite{grz92}) and the deviations 
from two-body focusing at low relative velocities ($E_{ij}$, 
\cite{grz90}).  
We adopt WS93's prescription for the variation of $\beta_{coll}$ as
a function of the relative velocity in Hill units, 
$V_{H,ij} = V_{ij}/(R_{H,ij} V_K / a) $:

\begin{equation}
\beta_{coll} = \left\{ \begin{array}{l l l}
        2.7 & \hspace{5mm} & V_{H,ij} > 2 \\
        1.0 + 1.7 (V_{H,ij} - 1) & \hspace{5mm} & 1 \le V_{H,ij} \le 2 \\
        1.0 & \hspace{5mm} & V_{H,ij} < 1 \\
        \end{array}
        \right .
\end{equation}

\noindent
and we set 

\begin{equation}
E_{ij} = \left\{ \begin{array}{l l l}
         1 & \hspace{5mm} & V_{ij} > 0.13 V_{e,ij} \\
         \frac{4 \sqrt{ e^2 + sin^2 i} ~ E_k}{\pi^2 ~ sin i} & \hspace{5mm} & V_{ij} \le 0.13 V_{e,ij} \\
         \end{array}
         \right .
\end{equation}
\noindent

\noindent
where $E_k = \int_0^{\pi/2} \sqrt{1 - k^2 {\rm sin^2} \theta} ~ d \theta$
for $k^2 = 3/[4(1 + {\rm sin}~i/e)]$ (\cite{grz90}; 
\cite{gre91}).

At very low velocities ($V_{H,ij} <$ 2.3), we adopt the two-body 
collisional cross-sections of Greenberg \etal (1991):

\begin{equation}
F_{2B,ij} = \left\{ \begin{array}{l l l}
            \left ( 1 + V_{e,ij}^2/V_T^2 \right ) 
            \left ( \frac{V_T}{V_{ij}} \right ) \left ( \frac{R_H}{R_T} \right )
             & \hspace{5mm} & V_{H,ij} < 2.3 \\

            \hspace{5mm} \\

            0.5 (1 + V_{e,ij}^2/V_T^2)^{1/2}
            \left ( \frac{V_T}{V_{ij}} \right ) \left ( \frac{R_H}{R_T} \right )
            \left ( \frac{H}{r_i + r_j} \right )
             & \hspace{5mm} & V_{H,ij} < 2.3, v_{H,ij} < v_{H,crit} \\
            \end{array}
            \right .
\end{equation}

\noindent
where $R_T = a [(m_i + m_j)/1~\msun]^{2/5}$ is the Tisserand radius,
$V_T = 1.1 \Omega \Delta a_T$ is the Tisserand velocity,
$\Delta a_T = 2.5 R_H$ is the half width of the feeding zone,
$v_{H,ij} = v_{ij}/V_H$ is the relative vertical velocity in Hill units,
and $v_{H,crit} < 0.7~{\rm sin}~(0.9 [(r_i+r_j)/R_{H,ij}]^{1/2})$.

In our second approach to gravitational focusing, we modify the
piecewise analytical approximation of Spaute \etal (1991):

\begin{equation}
F_{S,ij} = \left\{ \begin{array}{l l l}
         1 + \beta_{coll} \left ( \frac{V_{e,ij}}{V_{ij}} \right )^2 
            & \hspace{5mm} & V_{ij} > 0.032 V_{e,ij} \\
         42.4042 \left ( \frac{V_{e,ij}}{V_{ij}} \right )^{1.2} 
            & \hspace{5mm} & 0.01 < V_{ij}/V_{e,ij} \le 0.032 \\
         10706.916 & \hspace{5mm} & V_{ij}/V_{e,ij} \le 0.01 \\
            \end{array}
            \right .  
\end{equation}

\noindent
with $\beta_{coll}$ as defined above.  These expressions are continuous and 
serve as a check on the more detailed expressions for $F_{WS,ij}$.  
For very low velocity encounters, we use the two body cross-sections 
for $F_{2B}$ defined above.

In these approximations to the cross-section, the transition from
$F_{WS}$ or $F_S$ to $F_{2B}$ at $V_{H,ij} \approx$ 2.3 is not smooth.  
To affect a smooth transition, we set 

\begin{equation}
F_{g,ij} = \left\{ \begin{array}{l}
           (1 - x_{2B,ij})~F_{WS,ij} + x_{2B,ij}~F_{2B,ij} \\
           (1 - x_{2B,ij})~F_{S,ij} + x_{2B,ij}~F_{2B,ij} \\
         \end{array}
         \right .
\end{equation}
\noindent
where

\begin{equation}
x_{2B,ij} = \left\{ \begin{array}{l l l}
         0 & \hspace{5mm} & V_{H,ij} > 3.3 \\
         0.5 (V_{H,ij} - 2.3) & \hspace{5mm} & 1.3 < V_{H,ij} < 3.3 \\
         1 & \hspace{5mm} & V_{H,ij} < 1.3 \\
         \end{array}
         \right . 
\end{equation}
\noindent

\subsection{Velocity Evolution}

As noted above, kinetic models approximate planetesimal orbital 
elements as a mean square random velocity, $V_i^2$ (\cite{saf69};
\cite{ste88}; \cite{ws93} and references therein).  We divide this
velocity into horizontal, $h_i$, and vertical, $v_i$, components
that are related to $V_i$, $e$, and $i$ (see equations A1--A3).
Hornung \etal (1985) derive analytic expressions for the time
evolution of planetesimal velocities using a kinetic approximation to 
average over the velocity distribution function.  WS93 reformulate
some of these results in terms of the eccentricity and inclination, 
which we adopt here for simplicity (see also \cite{ste88}). 
We calculate velocity changes due to 
(i) gas drag, which decreases particle velocities and causes particles
to spiral in through the disk;
(ii) dynamical friction from elastic collisions, which transfers
kinetic energy from larger to smaller bodies;
(iii) viscous stirring from elastic collisions, which taps the
solar gravitational field to increase the velocities of all bodies; and
(iv) collisional damping from inelastic collisions
(see also \cite{hor85}; \cite{bar90}, 1991; \cite{oht92}).

The time evolution of the eccentricity and inclination for
long-range, elastic encounters is (\cite{ws93}, Appendix C)

\begin{equation}
\frac{de_{vs,i}^2}{dt} = \sum_{j=1}^{j=N} \frac{C_{lr}}{4}~(m_i + m_j)~e_i^2~
(J_r + 4J_\theta)
\end {equation}

\begin{equation}
\frac{di_{vs,i}^2}{dt} = \sum_{j=1}^{j=N} \frac{C_{lr}}{2 \beta_{ij}^2}~(m_i + m_j)~i_i^2~J_z
\end{equation}

\noindent
for viscous stirring and

\begin{equation}
\frac{de_{df,i}^2}{dt} = \sum_{j=1}^{j=N} \frac{C_{lr}}{2}~(m_j e_j^2 - m_ie_i^2)~(K_r + 4K_\theta)
\end{equation}

\begin{equation}
\frac{di_{df,i}^2}{dt} = \sum_{j=1}^{j=N} \frac{C_{lr}}{2 \beta_{ij}^2}~(m_ji_j^2 - m_ii_i^2)~K_z
\end{equation}

\noindent
for dynamical friction.  In these expressions, 
$e_i$, $e_j$, $i_i$, and $i_j$ are the eccentricity and inclination
of each body, $\beta_{ij}^2 = (i_i^2 + i_j^2)/(e_i^2 + e_j^2)$
is the ratio of inclination to eccentricity, and $C_{lr} = 
16 G^2 \rho_j ({\rm ln} \Lambda + 0.55 ) / V_K^3(e_i^2 + e_j^2)^{3/2}$ 
is a function of the density of particles in batch $j$ and 
the relative horizontal velocity of the mass batches (\cite{ws93}). 
The functions $J_r$, $J_\theta$, $J_z$, $K_r$, $K_\theta$, and $K_z$ 
are definite integrals that are functions only of $\beta_{ij}$ 
(\cite{hor85}; \cite{bar90}, 1991; \cite{ws93}; \cite{oht92}
describes a similar approach to velocity evolution).

The time evolution of $e$ and $i$ due to collisional damping is

\begin{equation}
\frac{de_{in,i}^2}{dt} = \sum_{j=0}^{j=i} \frac{C_{in}}{2}~(m_j e_j^2 - m_i e_i^2 - (m_i + m_j) e_i^2)~(I_r + 4I_\theta)
\end{equation}
 
\begin{equation}
\frac{di_{in,i}^2}{dt} = \sum_{j=0}^{j=i} \frac{C_{in}}{\beta_{ij}^2}~(m_j i_j^2 - m_i i_i^2 - (m_i + m_j) i_i^2)~I_z 
\end{equation}

\noindent
where $C_{in} = \alpha_{coll} \epsilon_{ij} \rho_j V_{ij}  F_{g,ij} (r_i + r_j)^2$ 
and $\rho_j$ is the mass density (\cite{hor85}; see also \cite{oht92}).
We include terms from the collision rate, 
$\alpha_{coll}$, $\epsilon_{ij}$, and $F_{g,ij}$, for consistency.  
The integrals, $I_r(\beta_{ij})$ $I_\theta(\beta_{ij})$, and $I_z(\beta_{ij})$, 
are listed in the Appendices of Hornung \etal (1985).  
We integrate these expressions numerically.  The $I_z$ integral 
often diverges; we set $I_z = I - (I_r + I_\theta) $ to avoid 
these divergences.

In addition to dynamical friction and viscous stirring, we also consider
velocity evolution due to gas drag.  Gas drag reduces the velocities 
of all mass batches and also removes material from each mass match.
The inward drift of material is (\cite{ada76}):

\begin{equation}
\frac{\Delta a}{a} = 2 (0.97 e + 0.64 i + \eta/V_K) 
                    \left ( \frac{\eta}{V_K} \right )
                    \frac{\delta t}{\tau_0} ~ ,
\end{equation}

\noindent
where $\eta$ is the gas velocity relative to the local Keplerian
velocity, $V_K$.  We adopt $\eta$ = 60 m s$^{-1}$ for calculations 
at 1 AU (\cite{ws93}) and $\eta$ = 30 m s$^{-1}$ for calculations
at 35 AU (\cite{ada76}).  The characteristic drift time is

\begin{equation}
\tau_0 = \frac{365}{C_D}~\left ( \frac{m_i}{\rm 10^{21}~g} \right )^{1/3}
         \left ( \frac{\rm 1~AU}{a} \right ) 
         \left ( \frac{10^{-9}~\rm g~cm^{-3}}{\rho_g} \right ) T_K,
\end{equation}

\noindent
where $C_D$ = 0.5 is the drag coefficient, 
$\rho_g$ = $1.18 \times 10^{-9} ~ (a/{\rm 1~AU})^{-11/4}$ 
is the gas density (Nakagawa \etal 1983),
and $T_K$ is the orbital period (see \cite{ada76}, \cite{ws93}).
To simulate the disappearance of gas in the protosolar nebula,
we decrease the gas density with time:

\begin{equation}
\rho_g (t) = \rho_{g,0}~e^{-t/\tau_g}
\end{equation}

\noindent
with $\rho_{g,0} = 1.18~\times~10^{-9}$ g cm$^{-3}$ (a/1 AU)$^{-11/4} ~ 
(M_0/M_{min}) $ (\cite{rud86}; \cite{rud91}; \cite{ws93}).  The radial 
decrease of the gas density follows models for minimum mass solar nebulae; 
the mass dependence allows the density to scale with the mass of the annulus.

The number of bodies lost from the calculation at each time step
depends on their effectiveness at crossing $\Delta a$. We set the 
number of bodies lost from a batch as:

\begin{equation}
\frac{\delta n_{gd,i}}{n_i} = \left ( \frac{\Delta a}{a} \right )
                         \left ( \frac{\delta t}{\tau_0} \right ) .
\end{equation}

\noindent
This expression is used in equations A4--A5.

Finally, we adopt Wetherill \& Stewart's (1989) expression for
velocity damping due to gas drag:

\begin{equation}
\frac{dV_i}{dt} = \frac{-\pi C_D}{2m_i} \rho_g V_g^2 r_i^2 ,
\end{equation}

\noindent
where $C_D$ = 0.5 is the drag coefficient and
$V_g = (V_i (V_i + \eta))^{1/2}$ is the mean relative velocity of the gas.

We convert the differential equations, A21--A26 and A31, into 
a kinetic energy form in two steps.
We use $\beta_{i}$ to derive the appropriate horizontal and 
vertical components of the velocity, $V_i$, in Eqn A31.
Equations A2--A3 similarly yield $\delta h_i$ and $\delta v_i$
in terms of $\delta e_i$ and $\delta i_i$, where

\begin{equation}
\delta e_i^2 = \delta t \left [ \frac{d e^2}{dt} \right ]
\end{equation}

\noindent{and}
\begin{equation}
\delta i_i^2 = \delta t \left [ \frac{d i^2}{dt} \right ] ~ .
\end{equation}

\noindent
These substitutions yield:

\begin{equation}
(\delta h_i^{gd})^2 = - ~ \delta t \left [ 
\frac{\pi C_D \rho_g V_g^2 r_i^2}{2m_i (1 + 0.8 \beta_i^2)} \right ]
\end{equation}

\begin{equation}
(\delta v_i^{gd})^2 = 0.8~\beta_i^2~\delta h_i^2
\end{equation}

\begin{equation}
(\delta h_i^{in})^2 = \frac{5}{8}~V_K^2~\delta t \left [ \frac{d e_{in,i}^2}{dt} \right ]
\end{equation}

\begin{equation}
(\delta v_i^{in})^2 = \frac{1}{2}~{\rm sin}^2 i~V_K^2~\delta t \left [ \frac{d i_{in,i}^2}{dt} \right ]
\end{equation}

\begin{equation}
(\delta h_i^{lr})^2 = \frac{5}{8}~V_K^2~\delta t \left [ 
\frac{d e_{vs,i}^2}{dt} + \frac{d e_{df,i}^2}{dt} \right ]
\end{equation}

\begin{equation}
(\delta v_i^{lr})^2 = \frac{1}{2}~{\rm sin}^2 i~V_K^2~\delta t \left [ 
\frac{d i_{vs,i}^2}{dt} + \frac{d i_{df,i}^2}{dt} \right ]
\end{equation}

\noindent
We multiply these relations by $m_i$ for substitution into equations
A8--A9.

\subsection{Tests of the evolution code}

To test the validity of our numerical techniques, we compare 
our results with several test cases (see \cite{oht88}; \cite{oht90}; 
\cite{wet90}).  The coagulation equations, A4--A5, have analytic 
solutions for three simple forms of the cross-section, $A_{ij}$. 
Smoluchowski (1916) first solved the coagulation equation for 
$A_{ij} = \alpha_c$ = constant.  Trubnikov (1971) described solutions 
for $A_{ij} = \beta_c (m_i + m_j)$ and $A_{ij} = \gamma_c m_i m_j$.
Wetherill (1990) identified an inconsistency in Trubnikov's
results for $A_{ij} = \gamma_c m_i m_j$ and showed that this
cross-section produces runaway growth.
Tanaka \& Nakazawa (1994) verified Wetherill's new solution 
and placed limits on the validity of the coagulation equation
during runaway growth.

The analytic solutions to the coagulation equation provide
rigorous tests of numerical methods.  Two simple cases, 
$A_{ij}$ = $\alpha_c$ and $A_{ij} = \beta_c (m_i + m_j)$,
do not lead to runaway growth, but they test the ability of
numerical codes to reach a target mass at a specified time.
They also yield estimates for mass conservation over many 
time steps.  Numerical solutions for $A_{ij} = \gamma_c m_i m_j$
are challenging, because runaway growth requires a careful, 
automatic procedure for changing the time step.  
In all three cases, the time lag between the analytic and 
numerical solutions depends on the mass ratio between 
consecutive batches, $\delta$ (\cite{wet90}).
These tests thus yield a quantitative measure of the
largest allowed value for $\delta$ (\cite{oht88};
\cite{oht90}; \cite{wet90}).

To compare our numerical results with analytic solutions,
we follow conventions established by Ohtsuki \etal (1988) 
and Wetherill (1990).  
For $A_{ij} = \alpha_c$ and $A_{ij} = \beta_c (m_i + m_j)$, 
we plot log $N_im_i^2$ as a function of log $m_i$.
We evaluate log $N_k$ as a function of log $m_k$
and the fractional mass in the small body swarm and 
the runaway body as a function of a dimensionless time 
for $A_{ij} = \gamma_c m_i m_j$.  
The numerical calculations do not have regular mass intervals,
so we calculate $N = \delta N / \delta m$, where

\begin{equation}
\delta N = N_k + (N_{k+1} + N_{k-1})/2
\end{equation}

\begin{equation}
\delta m = m_{k+1} - m_{k-1}.
\end{equation}

Wetherill (1990) describes analytic solutions for each 
cross-section in detail.  The $A_{ij} = \alpha_c$ case has 
the simplest solution. If $n_0$ is the initial number of 
particles with mass $m_i$, the number of bodies with mass 
$m_k = km_i$ at a time, $t$, is

\begin{equation}
n_k = n_0 f^2 (1-f)^{k-1}
\end{equation}

\noindent
where $f = 1/(1 + \eta_1/2)$ and $\eta_1 = \alpha_c n_0 t$ is the
dimensionless time (see also \cite{sil79}; \cite{oht88}).  
The solution for $A_{ij}$ = $\beta_c (m_i + m_j)$ has a similar form:

\begin{equation}
n_k = n_0 \frac{k^{k-1}}{k!} f(1-f)^{k-1} e^{-k(1-f)}
\end{equation}

\noindent
where $f = e^{-\eta_2}$ and $\eta_2 = \beta_c n_0 t$. In both of these
expressions, $f$ is the fraction of bodies with $m_i$ that
have yet to undergo a collision at time $t$.

Models with $A_{ij} = \gamma_c m_i m_j$ lead to runaway growth 
when $\eta_3 = \gamma n_0 t$ = 1 (\cite{wet90}; \cite{bar90}; 
\cite{tan94}).  The number distribution for $\eta_3 \le 1$ is

\begin{equation}
n_k = n_0 \frac{(2k)^{k-1}}{k! k} (\eta_3/2)^{k-1} e^{-k \eta_3}
\end{equation}

\noindent
This solution fails to conserve mass for $\eta_3 > 1$;
a single runaway body then contains most of the total mass.  
The mass of the runaway body for $\eta_3 > 1$ is (\cite{wet90};
\cite{tan93}, 1994):

\begin{equation}
m_R = n_0 e^{-\int \sum_{k=1}^{N} k^2 (n_k/n_0) d \eta^{\prime}} ~ .
\end{equation}

Figure 15 compares our results for $\delta$ = 1.25 with the 
analytic solution for $A_{ij} = \alpha_c$.  The agreement is good
and again improves as $\delta$ decreases.  Our results for $\delta $
= 1.4--1.6 are consistent with the analytic solution,
although we have too few mass batches to make reliable
comparisons when $\eta_1 < 10-20$.  We did not attempt numerical
models for $\delta$ = 1.6--2 (the maximum allowed), but we
expect that these will produce satisfactory results for
$\eta > 100$.

Figure 16 shows results for $A_{ij} = \beta_c (m_i + m_j)$ and
$\delta = 1.25$.  The agreement between our calculation and
the analytic solution is quite good and improves as $\delta$ 
increases.  We find a slight excess of low mass bodies in
our numerical results compared to the analytic solution.
Wetherill's Figure 4 contains a similar excess. 
The peak of our normalized number distribution lags the
analytic result by 1.4\%.  This lag decreases with $\delta$
and is $<1\%$ for $\delta$ = 1.10.

Figures 17 and 18 summarize our results for $A_{ij} = \gamma_c m_i m_j$
and $\delta$ = 1.25 for $\eta \le$ 1.  The numerical solution follows the
analytic model very closely for $\eta_3 < 0.95$ and then begins
to diverge at large masses as $\eta$ approaches unity (Figure 17).  
The numerical model begins runaway growth at $\eta_3$ = 1.012
and lags the analytic model by 1.2\%.  The numerical runaway 
begins much closer to the predicted result, $\eta_3 = 1.005$,
for $\delta$ = 1.08.  Larger values for $\delta$ produce runaways that
are delayed by much longer factors.  The lag is 2.7\% for 
$\delta$ = 1.4 and 8.7\% for $\delta$ = 2.
Wetherill (1990) quotes similar results for his numerical models
with $\delta$ = 1.07 and $\delta$ = 1.25.

Figure 18 describes the evolution of the runaway growth model for
$\delta$ = 1.08 and $\eta_3 >$ 1.  The calculated mass distribution 
initially lags the analytic result by less than 1\% for $\eta_3$
marginally larger than 1 (see also \cite{ws93}) but matches the 
analytic result almost exactly at $\eta_3$ = 1.05 (Figure 18;
left panel).  The calculation continues to match the analytic 
result until $\eta_3 \approx$ 5.  The right panel of Figure 18 plots
the mass of the runaway body for $\eta_3 \ge$ 1.  The calculated
mass agrees with the analytic prediction, Eqn. A45, to 1\% or better
for all $\eta_3 \ge$ 1.  Models with $\delta$ = 1.25 have greater
difficulty reaching large $\eta_3$ due to their poorer mass resolution.
These models have larger $n_k$ at high masses, which reduces the 
time step considerably compared to models with small $\delta$.
The calculation then requires a significant amount of computer time 
and does not agree as well with the analytic predictions.  
Our models with $\delta \ge$ 1.4 fail to reach $\eta_3 \approx$ 1.1 
if we maintain our criterion of small $\delta n_k$ per time step.
Relaxing this criterion allows reasonable time steps but produces 
very poor agreement, $\gtrsim$ 20\%, with the analytic solution.  

These results confirm our limits on $\delta n_k$ for the Kuiper
Belt simulation described in the main text, $\lesssim$ 0.1\%
per time step, for $\delta$ = 1.08--1.4.  Models with larger 
$\delta$ fail to follow growth properly unless the time steps
are unreasonably small.

\vfill
\eject

\begin{center}
\begin{tabular}[t]{l c l l}
\multicolumn{4}{c}{{\sc Table 1.} Basic Model Parameters} \\
\\
\tableline
\tableline
Parameter & Symbol & 1 AU Models & 35 AU Models \\
\tableline
Width of annulus & $\delta a$ & ~~0.17 AU & ~~6 AU \\
Initial Velocity & $V_0$      & ~~4.7 m s$^{-1}$ & ~~4.5--45 m s$^{-1}$ \\
Particle mass density & $\rho_0$ & ~~3 g cm$^{-3}$ & ~~1.5 g cm$^{-3}$ \\
Relative gas velocity & $\eta$ & ~~60 m s$^{-1}$ & ~~30 m s$^{-1}$ \\
Time Step        & $\delta t$ & ~~0.5 yr & ~~5--250 yr \\
Number of mass bins & $N$     & ~~100--150 & ~~64--128 \\
Mass Spacing of bins& $\delta$ & ~~$\le$ 1.20 & ~~1.40 \\
\tableline
\end{tabular}
\end{center}

\clearpage

\begin{center}
\begin{tabular}[t]{l c c c c c c}
\multicolumn{7}{c}{{\sc Table 2.} Model Results at 1 AU$^{(a)}$} \\
\\
\tableline
\tableline
 & \multicolumn{3}{c}{$\delta = 1.25$} & \multicolumn{3}{c}{$\delta = 1.40$} \\
Time (yr) & $r_{max}$ (km) & $m(r_{max})$ (kg) & $N(r_{max}$) & $r_{max}$ (km) & $m(r_{max})$ (kg) & $N(r_{max}$) \\
\tableline
5.0 $\times 10^2$ &~~~19.8 & 9.7 $\times 10^{19}$ & 3 &~~~20.9 & 1.2 $\times 10^{20}$ & 1 \\
1.0 $\times 10^4$ &~~513.3 & 1.7 $\times 10^{24}$ & 3 &~~492.3 & 1.5 $\times 10^{24}$ & 1 \\
2.5 $\times 10^4$ & 1167.3 & 2.0 $\times 10^{25}$ & 1 & 1203.0 & 2.2 $\times 10^{25}$ & 1 \\
5.0 $\times 10^4$ & 1540.8 & 4.6 $\times 10^{25}$ & 3 & 1515.7 & 4.4 $\times 10^{25}$ & 1 \\
1.0 $\times 10^5$ & 1890.8 & 8.5 $\times 10^{25}$ & 2 & 1746.7 & 6.7 $\times 10^{25}$ & 3 \\
1.5 $\times 10^5$ & 1948.3 & 9.3 $\times 10^{25}$ & 5 & 2382.7 & 1.7 $\times 10^{26}$ & 1 \\
\tableline
\multicolumn{7}{l}{$^{(a)}$These results are for the WS93 prescription of
gravitational focusing, equations } \\
\multicolumn{7}{l}{A14--A16, as summarized in the main text.} \\
\end{tabular}
\end{center}

\clearpage

\begin{center}
\begin{tabular}[t]{l c c c c c c}
\multicolumn{7}{c}{{\sc Table 3.} Model Results at 1 AU$^{(a)}$} \\
\\
\tableline
\tableline
 & \multicolumn{3}{c}{$\delta = 1.25$} & \multicolumn{3}{c}{$\delta = 1.40$} \\
Time (yr) & $r_{max}$ (km) & $m(r_{max})$ (kg) & $N(r_{max}$) & $r_{max}$ (km) & $m(r_{max})$ (kg) & $N(r_{max}$) \\
\tableline
5.0 $\times 10^2$ &~~~19.8 & 9.7 $\times 10^{19}$ & 3 &~~~20.9 & 1.2 $\times 10^{20}$ & 1 \\
1.0 $\times 10^4$ &~~550.8 & 2.1 $\times 10^{24}$ & 2 &~~492.3 & 1.5 $\times 10^{24}$ & 1 \\
2.5 $\times 10^4$ & 1167.5 & 2.0 $\times 10^{25}$ & 1 & 1223.0 & 2.3 $\times 10^{25}$ & 1 \\
5.0 $\times 10^4$ & 1458.6 & 3.9 $\times 10^{25}$ & 4 & 1735.0 & 6.6 $\times 10^{25}$ & 2 \\
1.0 $\times 10^5$ & 1969.0 & 9.6 $\times 10^{25}$ & 1 & 2174.8 & 1.3 $\times 10^{26}$ & 1 \\
1.5 $\times 10^5$ & 2121.0 & 1.2 $\times 10^{26}$ & 1 & 2335.0 & 1.7 $\times 10^{26}$ & 1 \\
\tableline
\multicolumn{7}{l}{$^{(a)}$These results are for an adaptation of the 
Spaute et al. (1991) prescription of} \\
\multicolumn{7}{l}{gravitational focusing, equation A18, as summarized 
in the main text.} \\
\end{tabular}
\end{center}

\epsffile{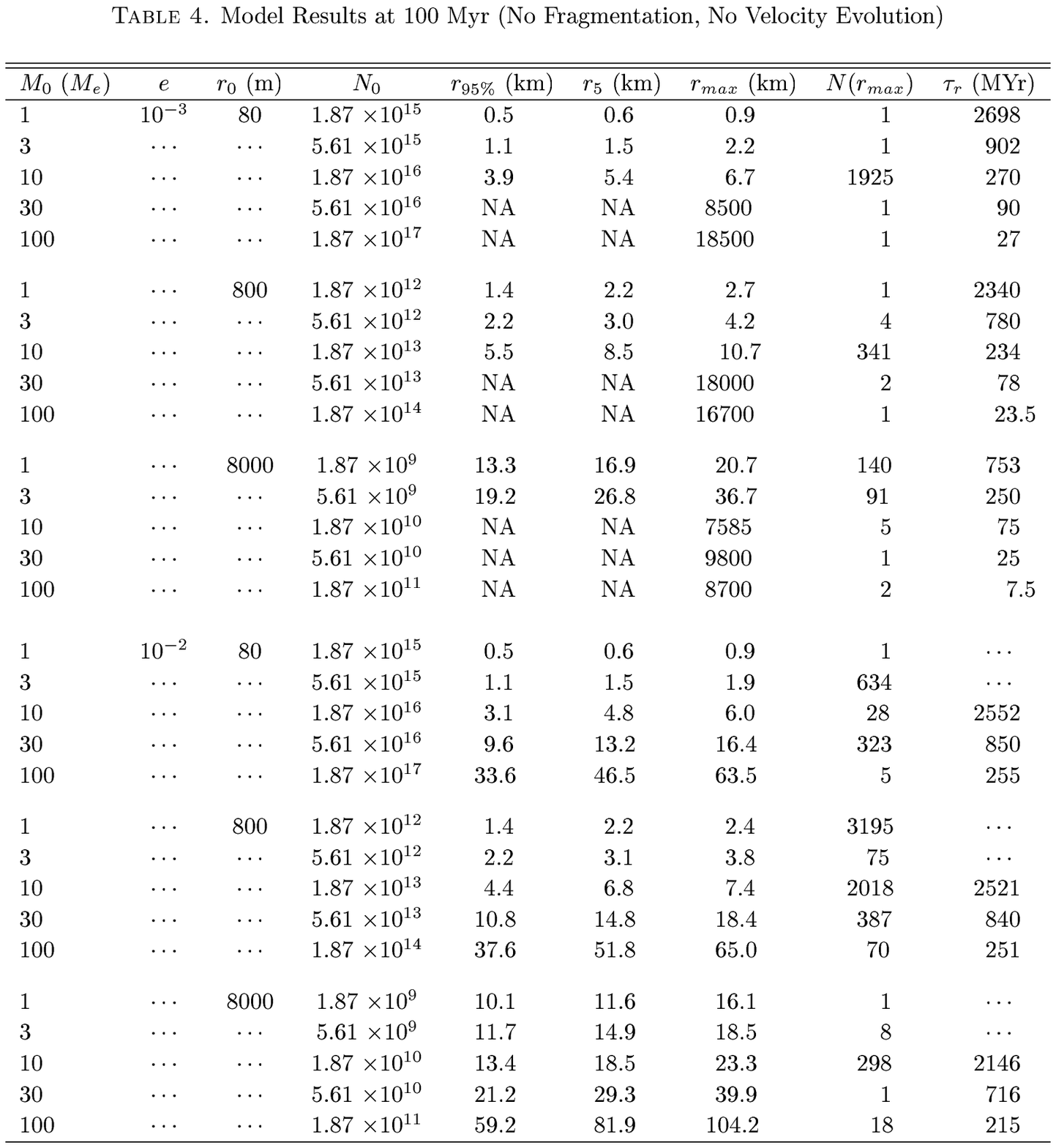}

\hskip -10ex
\vskip -20ex

\epsffile{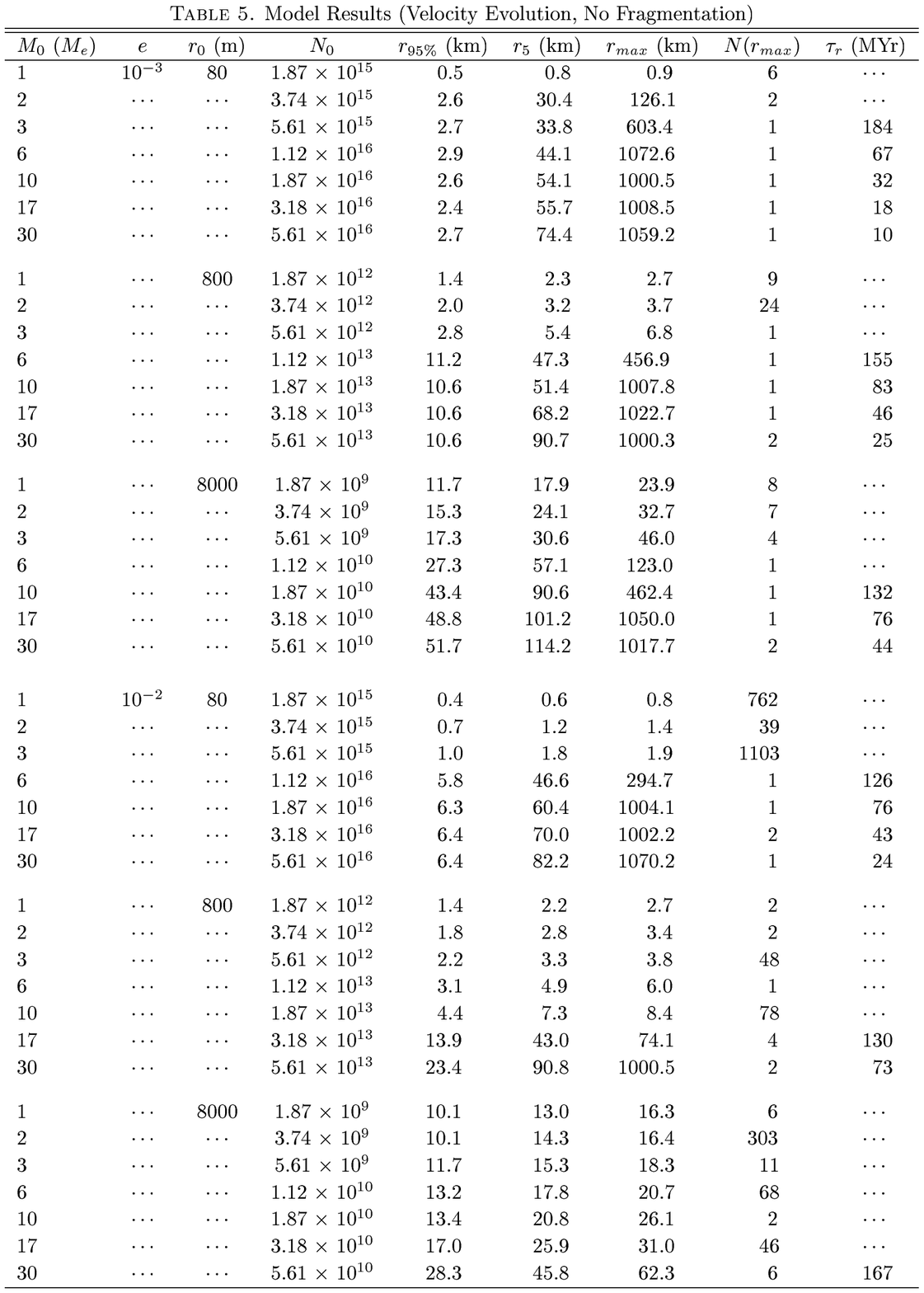}

\clearpage

\centerline{\bf FIGURE CAPTIONS}
\vskip 4ex

\figcaption[Kenyon.fig1.ps]
{Results at 1 AU for $M_0$ = 0.667 $M_{\earth}$ and $\delta$ = 1.25.
(a) Cumulative mass distribution at selected times.  
The ``runaway plateau'' forms at $\sim 2~\times~10^4$ yr;
it includes 17\% of the total mass at $5~\times~10^4$ yr,
18\% of the total mass at $10^5$ yr,
and 22\% at $1.5~\times~10^5$ yr.
(b) Horizontal velocity distribution.
Viscous stirring increases all velocities with time;
dynamical friction breaks the runaway bodies and 
increases velocities of the lowest mass bodies.}

\figcaption[Kenyon.fig2.ps]
{Results at 1 AU for $\delta$ = 1.4.  The runaway plateau
is more poorly resolved as the mass spacing increases.
(a) Cumulative mass distribution at selected times.  
The ``runaway plateau'' includes 12\% of the total mass 
at $5~\times~10^4$ yr; this component comprises 18\% of the 
total mass at $10^5$ yr and 22\% at $1.5~\times~10^5$ yr.
(b) Horizontal velocity distribution.  The horizontal velocities 
of low mass objects are 5\% larger with $\delta$ = 1.4 than with
$\delta$ = 1.25. Higher mass bodies have velocities 2--3 larger
than comparable masses in $\delta$ = 1.25 simulations.}

\figcaption[Kenyon.fig3.ps]
{Cumulative size distribution for $M_0$ = 10 \mearth, $r_0$ = 8 km,
and $e = 10^{-3}$.  The eccentricity of this model is constant
in time.  Collisional growth is quasi-linear for 50 Myr 
until the largest bodies have $r_{max}$ = 50 km.  Runaway growth
begins when $r_{max} \simgreat $ 100 km; these bodies then grow
to sizes of $10^3$ km to $10^4$ km in 20--30 Myr.}

\figcaption[Kenyon.fig4.ps]
{Radius evolution as a function of initial mass for constant
velocity models, with (a) $r_0$ = 8 km, (b) $r_0$ = 800 m, and (c)
$r_0$ = 80 m. The eccentricity is $e = 10^{-3}$.
The time to runaway growth scales inversely with initial mass,
$\tau_r \approx \tau_0 ~ (M_0 / 1 M_{\earth})^{-1}$; 
$\tau_0 \approx $ 753 Myr for $r_0$ = 8 km,
$\tau_0 \approx $ 2340 Myr for $r_0$ = 800 m, and
$\tau_0 \approx $ 2700 Myr for $r_0$ = 80 m.}

\figcaption[Kenyon.fig5.ps]
{Same as Fig. 4, but with  eccentricity $e = 10^{-2}$. (a) $r_0$ = 8
km, (b) $r_0$ = 800 m, and (c) $r_0$ = 80 m.}

\figcaption[Kenyon.fig6.ps]
{Cumulative size distribution for a constant velocity model with
$\delta$ = 1.1 for $M_0$ = 10 \mearth, $r_0$ = 8 km, and $e = 10^{-3}$. 
This model produces a cumulative size distribution with two
distinct power laws: $N_C \propto r^{-2.7}$ for 
8 km $\lesssim r_i \lesssim$ 300 km and
$N_C \propto r^{-1.7}$ for 
300 km $\lesssim r_i \lesssim$ 5000 km.}

\figcaption[Kenyon.fig7.ps]
{Time scales for viscous stirring (solid line) and collisional
damping (dashed line) as a function of mass.
(a) The two curves show the time scales for interactions 
only between particles of the same mass for a realistic 
cumulative mass distribution.  The two plots curve up at
high masses due to the small number of particles.
Collisional damping balances the velocity increase due to
viscous stirring at $\sim 10^{15}$ kg.
(b) Time scales integrated over all particles during the 
initial stages, $\sim$ 10 Myr, of a model with velocity 
evolution.  Collisional damping overcomes velocity increases
from viscous stirring only for $m_i \lesssim 10^{12}$ kg.
(c) As in (b) but for a late stage in the growth process,
15--16 Myr.  The integrated effects of viscous stirring now
increase the velocities of all particles.}

\figcaption[Kenyon.fig8.ps]
{Model with $M_0$ = 10 \mearth, $r_0$ = 8 km, and $e = 10^{-3}$, with
velocity evolution: (a) cumulative size distribution, and (b)
horizontal velocity as a function of time. Collisional growth is
quasi-linear until the largest bodies have $r_{max}$ = 50--100 km
at 50--60 Myr.  
Runaway growth begins when $r_{max} \simgreat $ 500 km at $\sim$ 
100 Myr; these bodies then grow to sizes of $10^3$ km in another 50--80 Myr.}

\figcaption[Kenyon.fig9.ps]
{Model with $M_0$ = 10 \mearth, $r_0$ = 800 m,
and $e = 10^{-3}$: (a) cumulative size distribution, and (b)
horizontal velocity as a function of time. 
Collisional growth is quasi-linear for 45--50 Myr until the 
largest bodies have $r_{max}$ = 50--100 km.  The transition
to runaway growth requires $\sim$ 10 Myr, when Charon-sized
objects form.  These bodies grow to sizes of 1000 km in 
another 20 Myr. The velocities of the smallest objects increase
with due due to viscous stirring.  Dynamical friction reduces
the velocities of the largest objects.  The velocity minimum 
at 3--5 km indicates the batches that contain the largest 
fraction of the total mass.}

\figcaption[Kenyon.fig10.ps]
{Model with $M_0$ = 10 \mearth, $r_0$ = 80 m, and $e = 10^{-3}$:
(a) cumulative size distribution, and (b) horizontal velocity as 
a function of time.  The largest bodies reach $r_{max}$ = 50--100 km
in 20 Myr and $r_{max}$ = 500 km in only 25 Myr.  Runaway growth 
begins at 25 Myr and the largest bodies achieve $r_{max} \sim $ 1000 km
after another 8 Myr. As in Figure 9, dynamical friction and viscous
stirring increase the velocities of the smallest objects at the
expense of the largest objects.  Dynamical friction produces a
velocity minimum in batches that contain the largest fraction of 
the total mass.}

\figcaption[Kenyon.fig11.ps]
{Evolution of the maximum radius with time for models with
different initial mass ($M_0$) and initial radius ($r_0$),
for low initial eccentricity ($e = 10^{-3}$).  The time scale
to reach runaway growth decreases with smaller $r_0$ and 
with larger $M_0$.} 

\figcaption[Kenyon.fig12.ps]
{Evolution of the maximum radius as in Figure 11, for models 
with large initial eccentricity ($e = 10^{-2}$).  Models with
high $e$ require 2--4 times more mass to reach runaway growth
on time scales similar to that of low $e$ models.}

\figcaption[Kenyon.fig13.ps]
{Summary of velocity evolution models for
(a) $e = 10^{-3}$ and (b) $e = 10^{-2}$.  
Filled circles indicate successful simulations that produce 
a few Pluto-sized objects and $\sim 10^5$ KBOs in 100 Myr or less;
open circles indicate simulations that produce no Plutos and too 
few KBOs in 100 Myr or less); filled circles within a larger open
circle indicate partially successful simulations that produce a few 
Pluto-sized objects but too few KBOs in 100 Myr or less). }

\figcaption[Kenyon.fig14.ps]
{The evolution of $M_{max}/\langle m \rangle$, as a function
of time. (a) constant velocity models, and (b) with velocity evolution.
The rapid increase in $M_{max}/\langle m \rangle$ at the latter stages 
of many simulations indicates runaway growth.}

\figcaption[Kenyon.fig15.ps]
{Evolution of the mass distribution for a constant cross-section,
$A_{ij} = \alpha_c$.  The solid lines plot analytic results for
four values of $\eta$; the symbols indicate results of the numerical
simulations.}

\figcaption[Kenyon.fig16.ps]
{Evolution of the mass distribution for 
$A_{ij} = \beta_c(M_i + M_j)$.  The solid lines plot analytic results 
for four values of $\eta$; the symbols indicate results of the numerical
simulations.}

\figcaption[Kenyon.fig17.ps]
{Evolution of the mass distribution for 
$A_{ij} = \gamma_c M_i M_j$.  The solid lines plot analytic results for
four values of $\eta$; the symbols indicate results of the numerical
simulations.}

\figcaption[Kenyon.fig18.ps]
{Runaway growth for $A_{ij} = \gamma_c M_i M_j$.
(a) Evolution of the residual mass distribution for 
four values of $\eta >$ 1.  The simulations lag the
analytic model for $\eta \approx$ 1 and then follow
it closely for larger $\eta$.
(b) Evolution of the mass of the runaway body for
the simulation (symbols) and the analytic model (solid curve)
as a function of $\eta$.}

\hskip -10ex
\epsffile{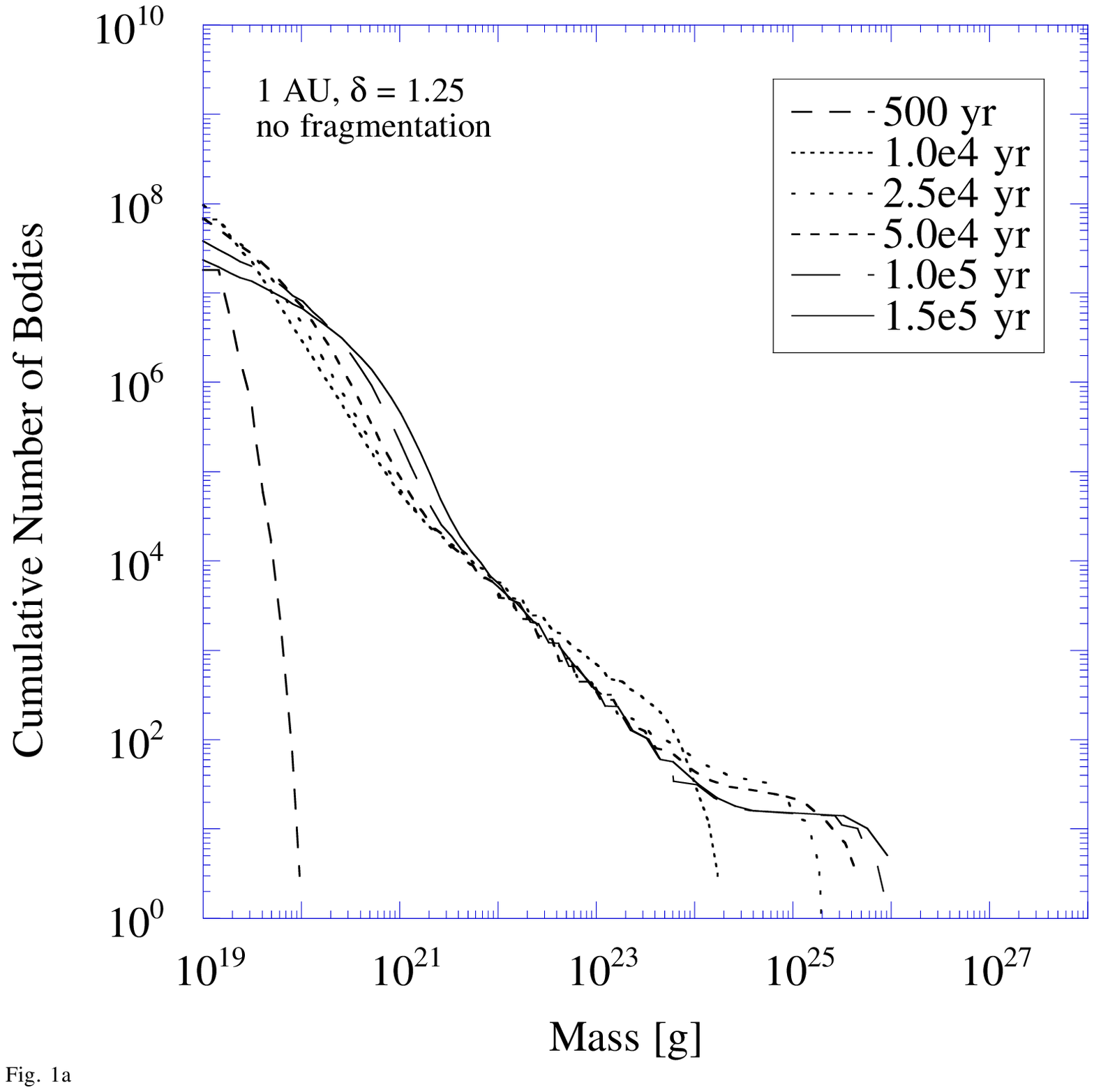}
 
\hskip -10ex
\epsffile{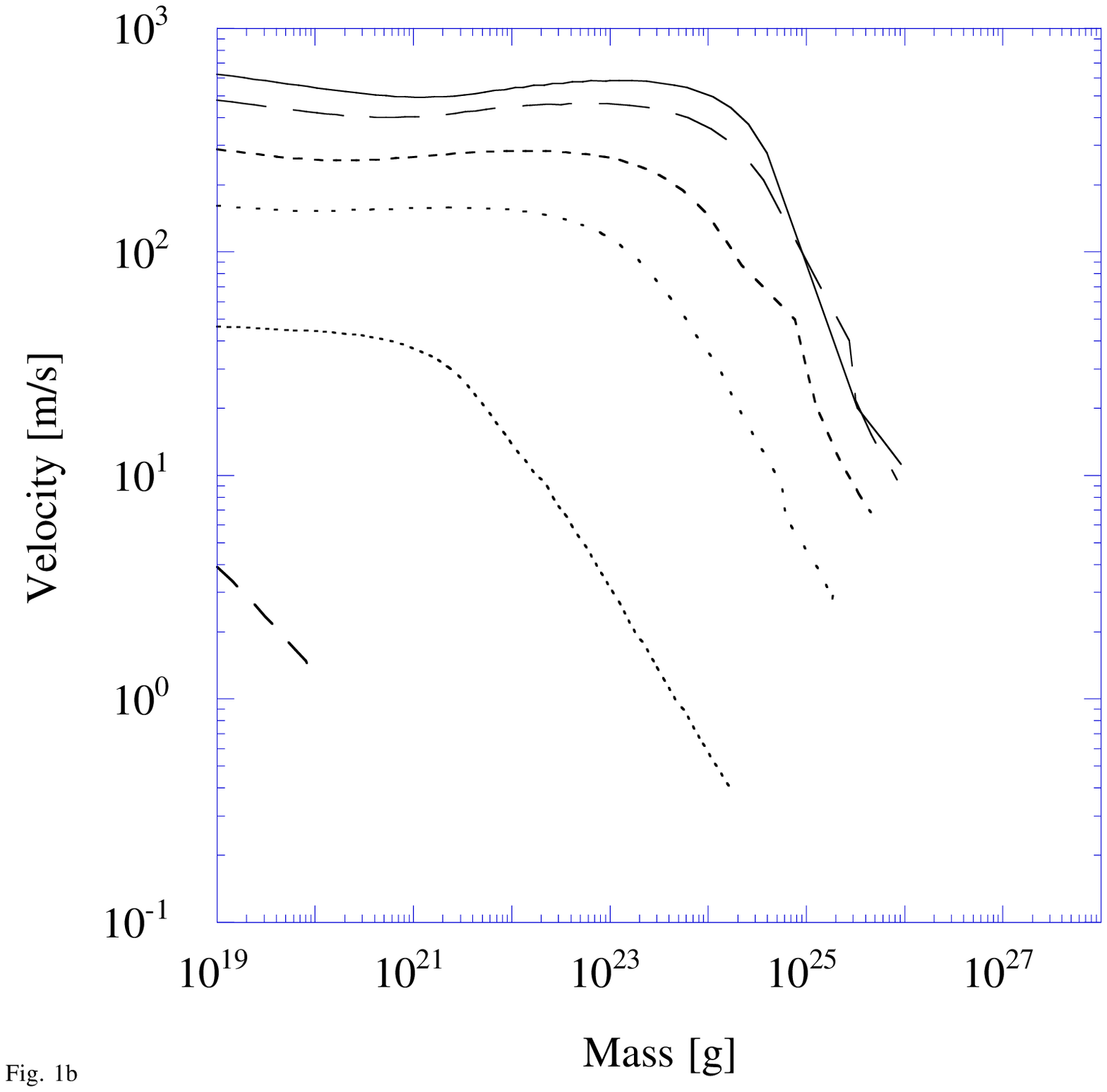}
 
\hskip -10ex
\epsffile{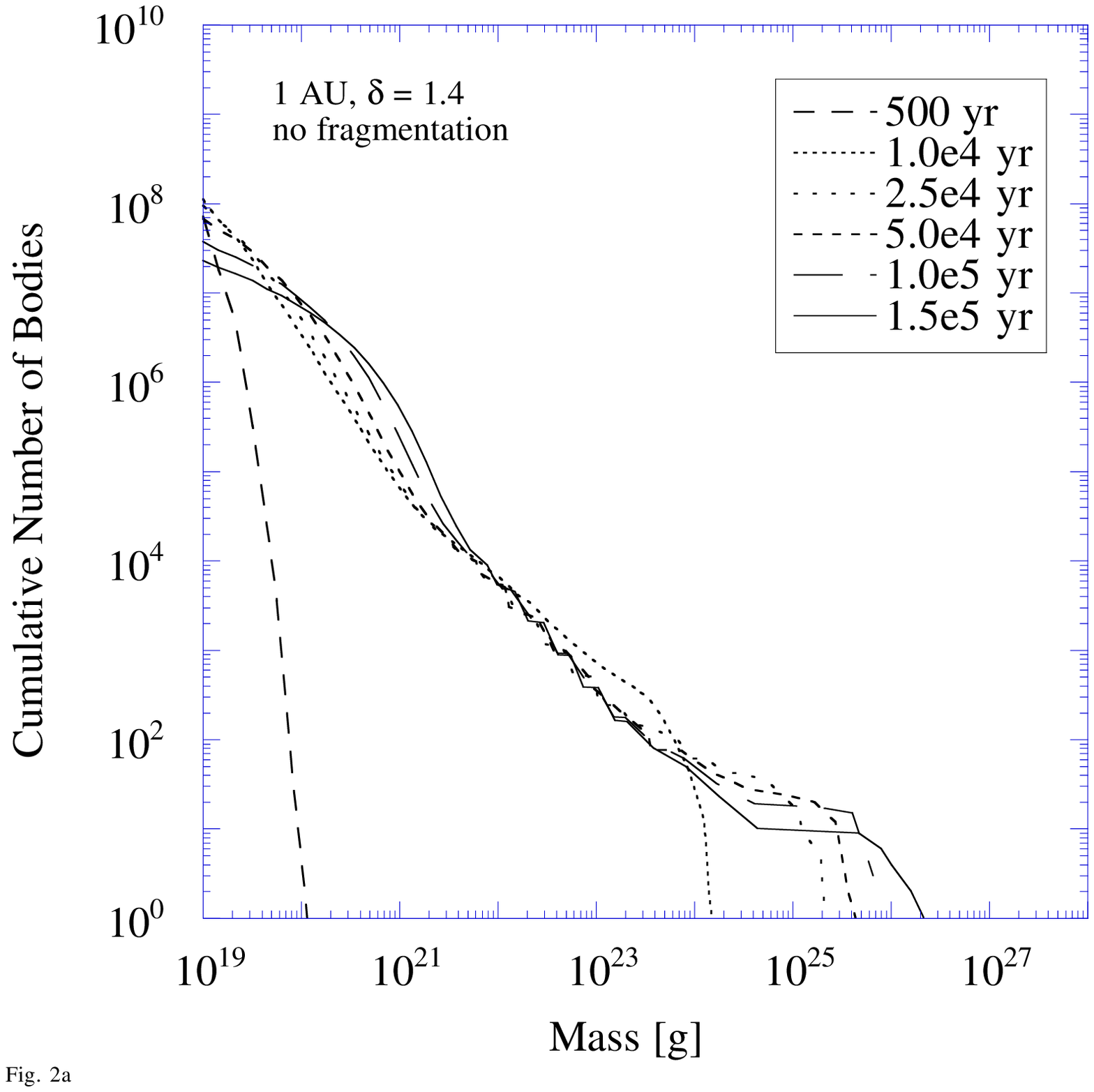}
 
\hskip -10ex
\epsffile{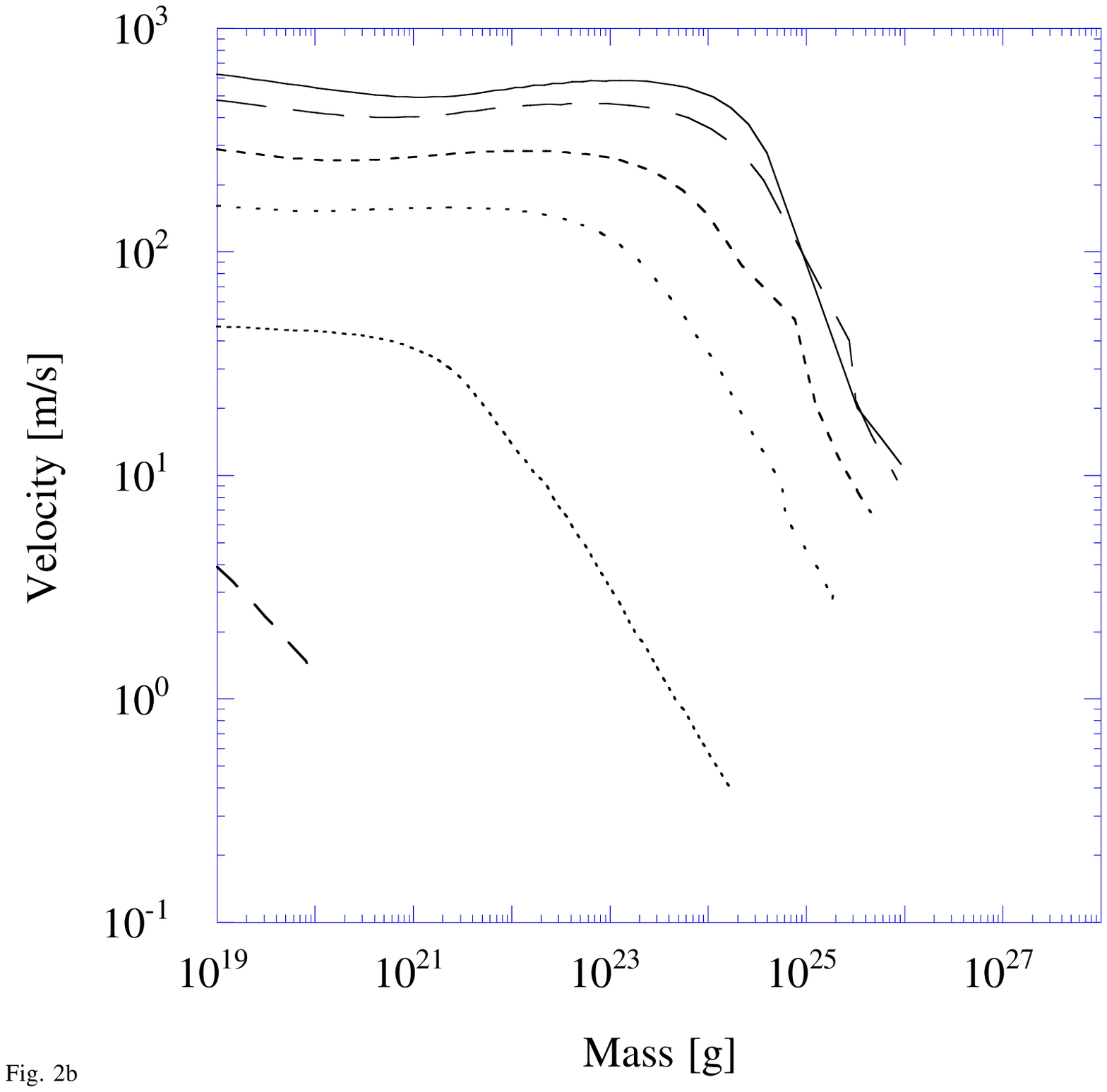}
 
\hskip -10ex
\epsffile{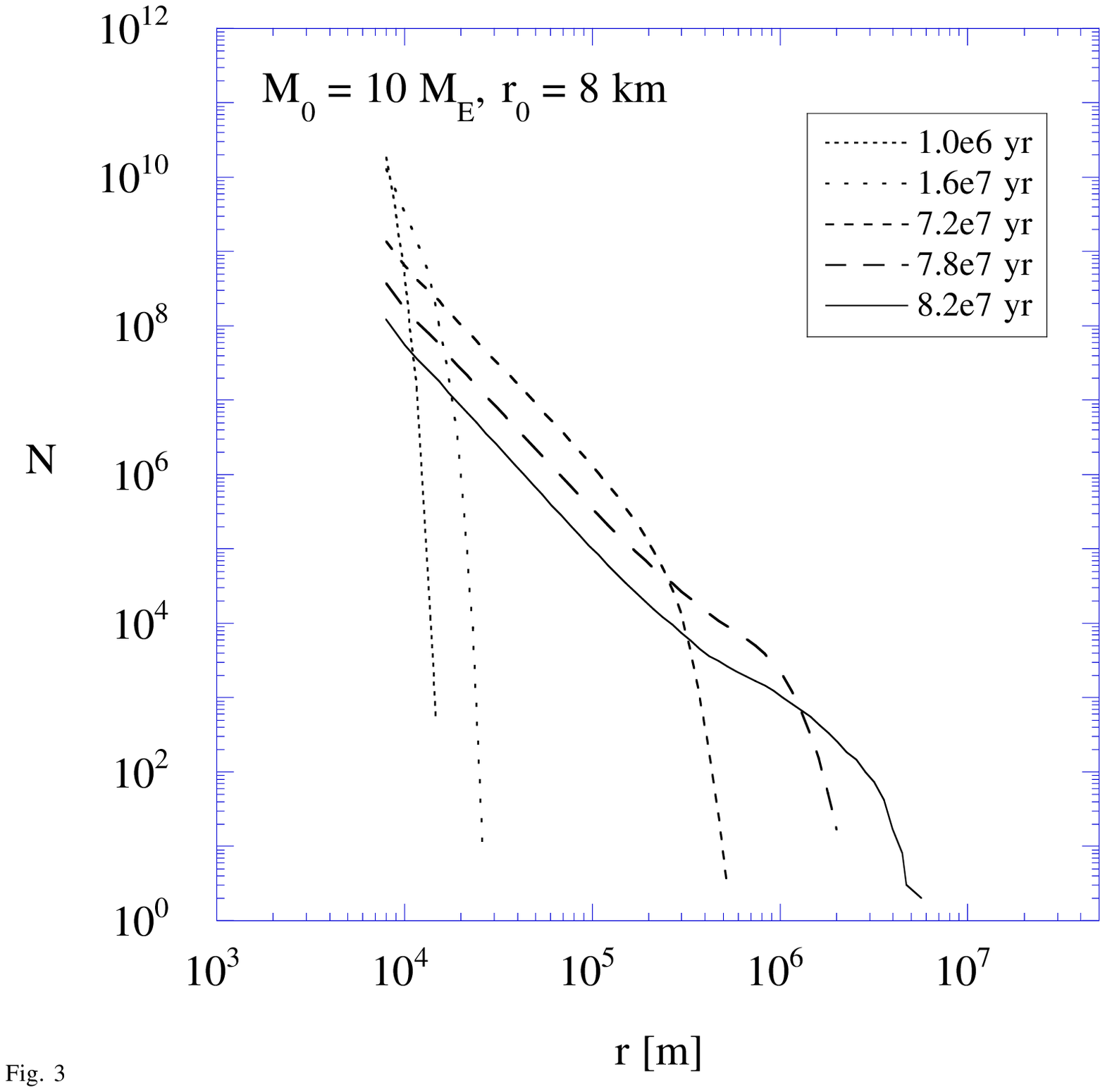}
 
\hskip -10ex
\epsffile{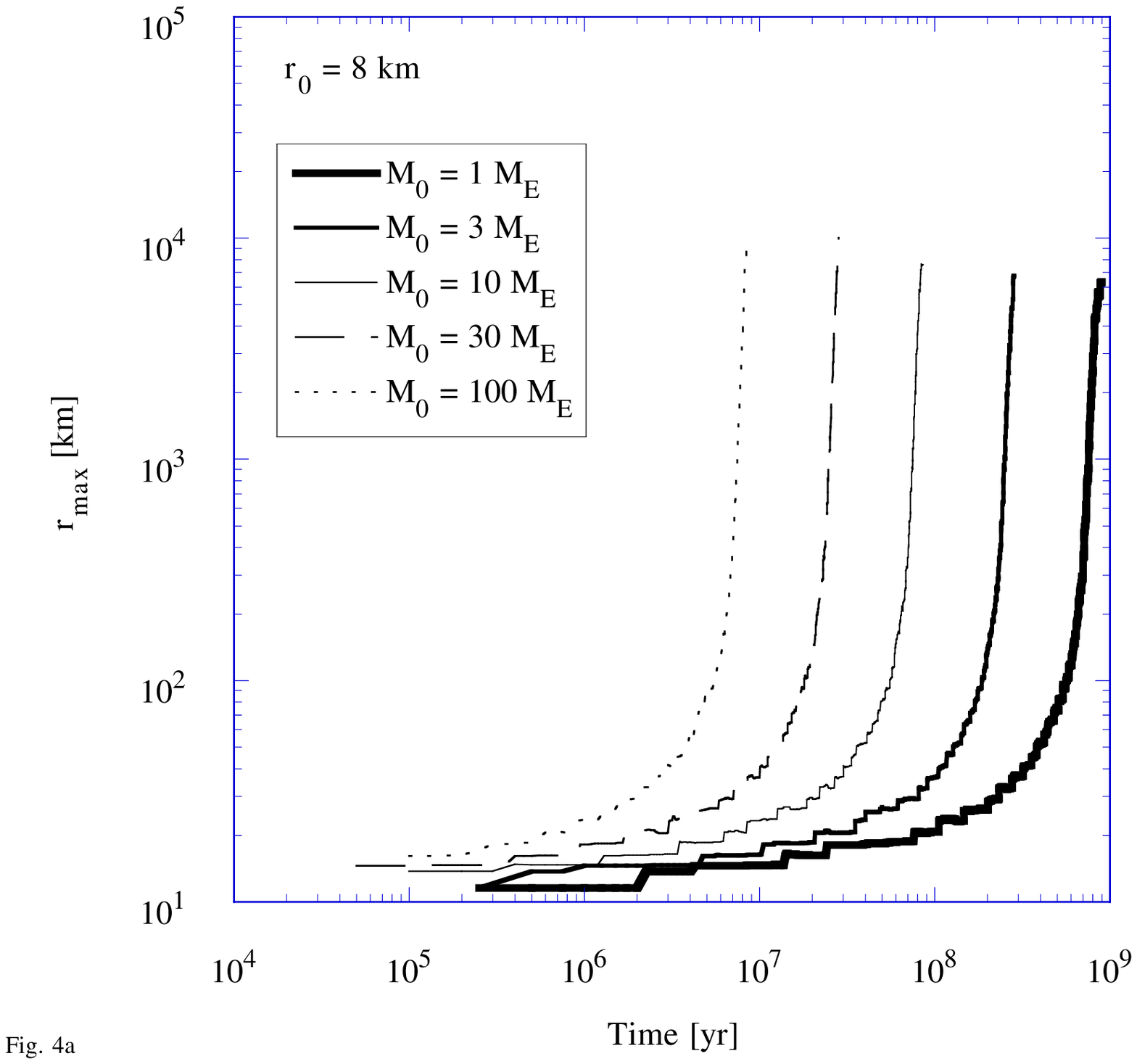}
 
\hskip -10ex
\epsffile{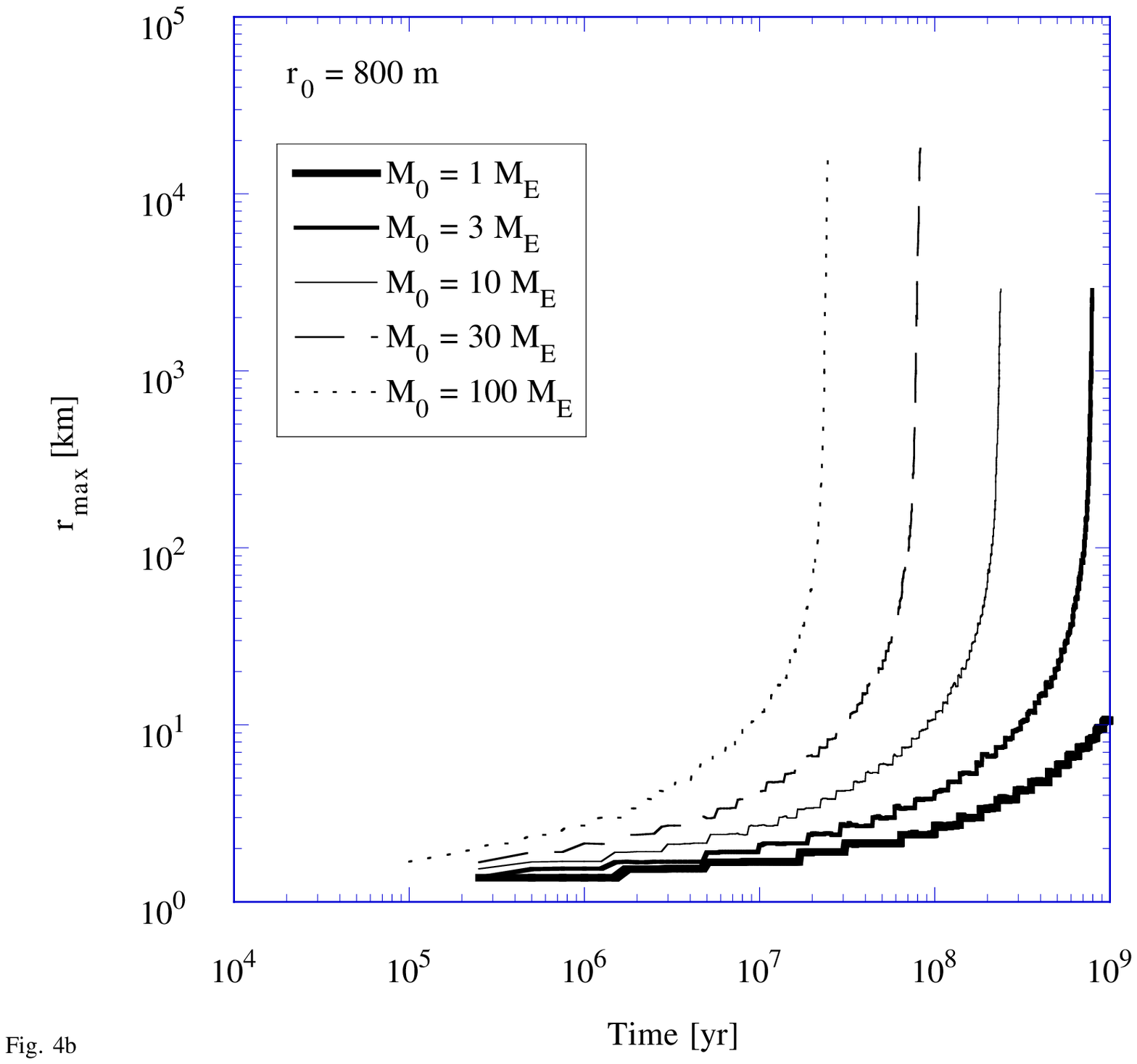}
 
\hskip -10ex
\epsffile{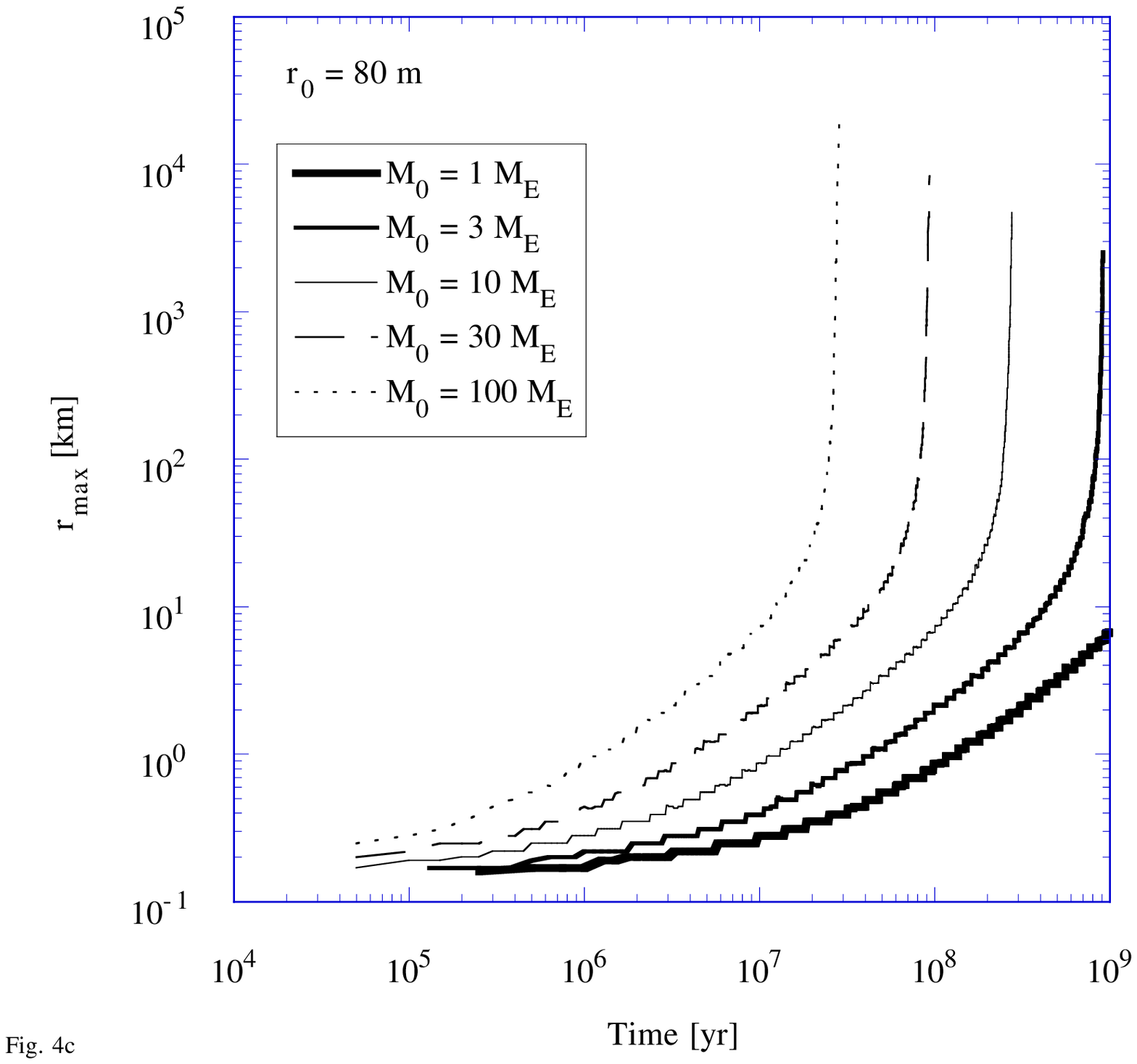}
 
\hskip -10ex
\epsffile{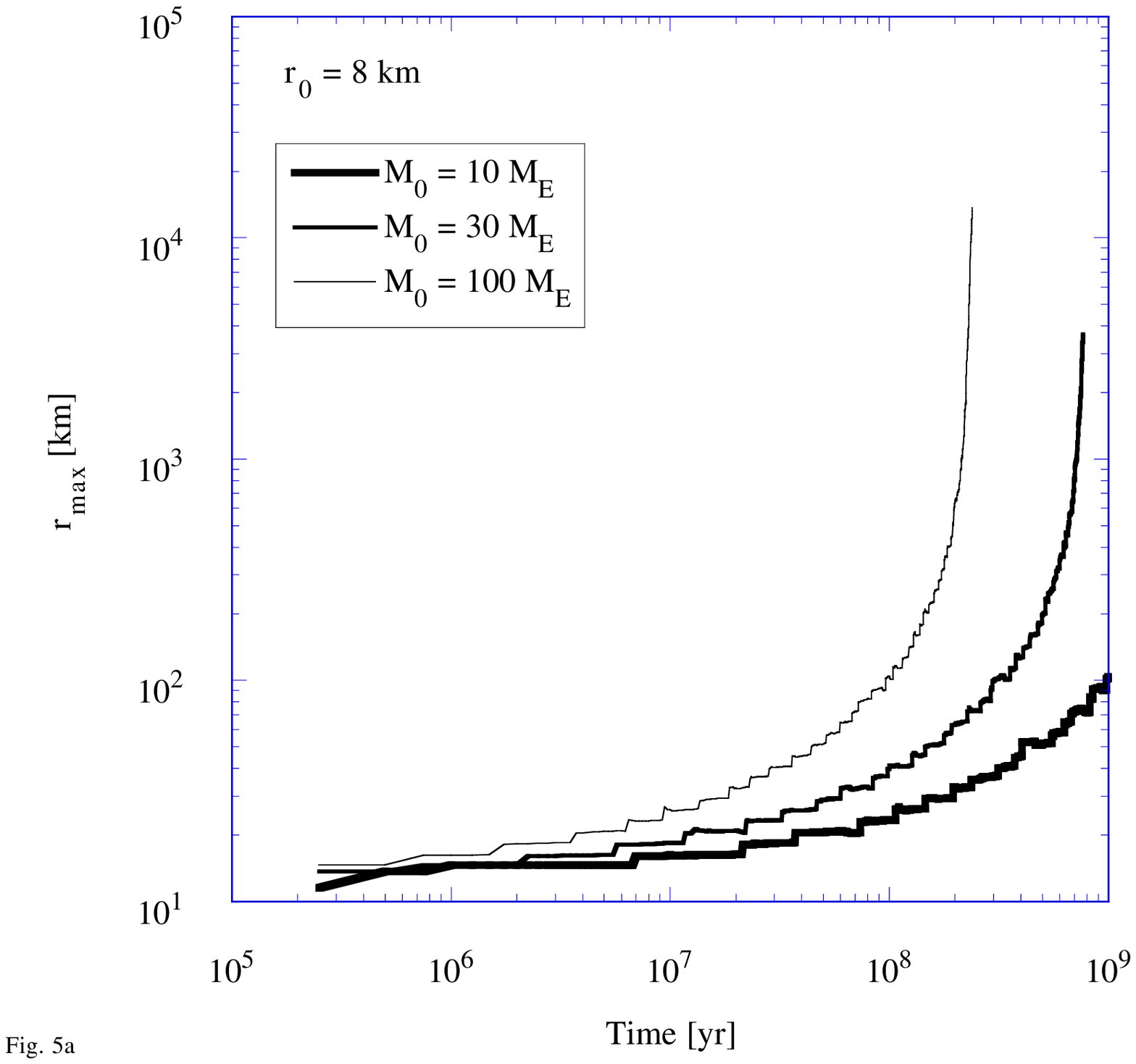}
 
\hskip -10ex
\epsffile{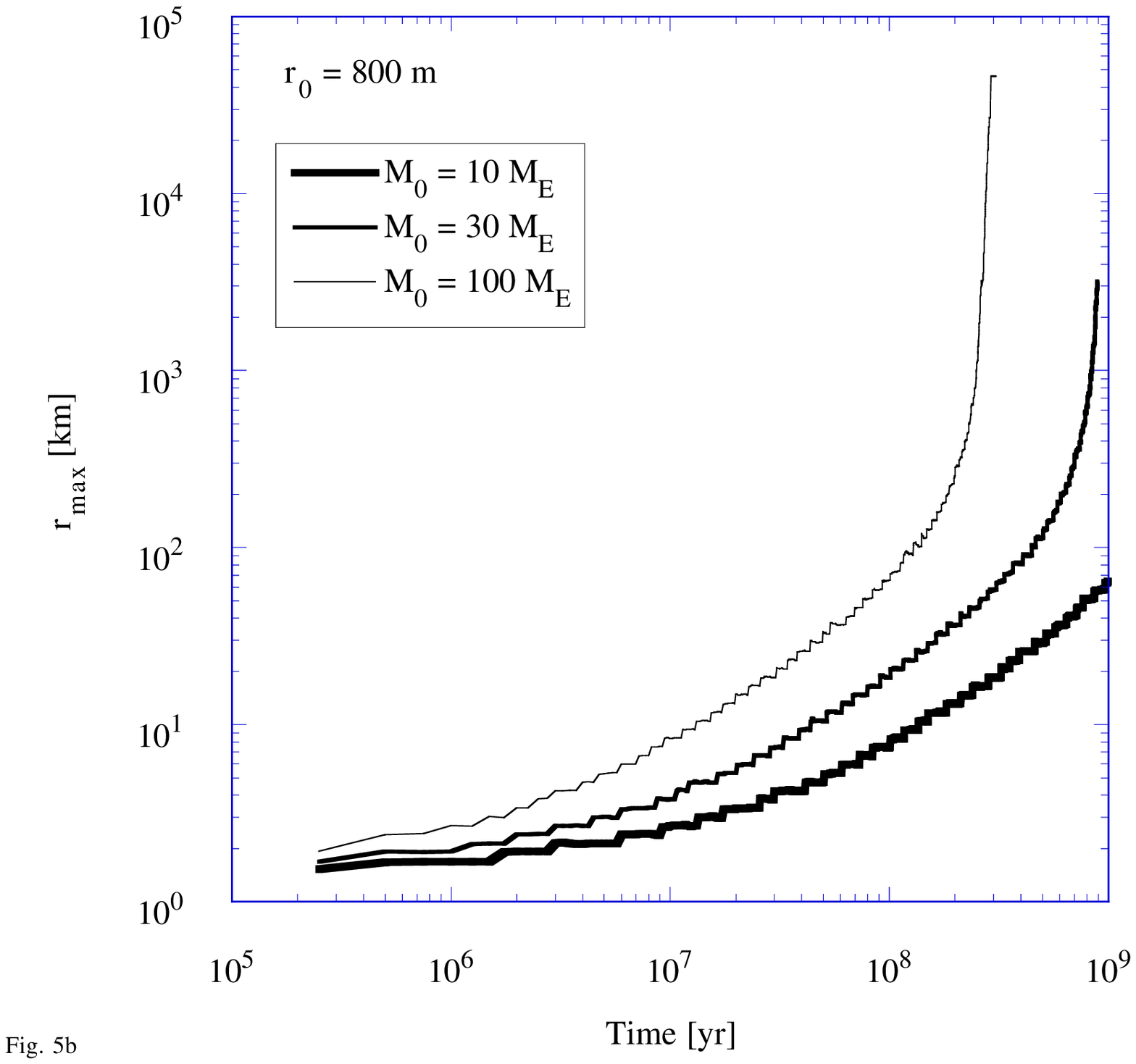}
 
\hskip -10ex
\epsffile{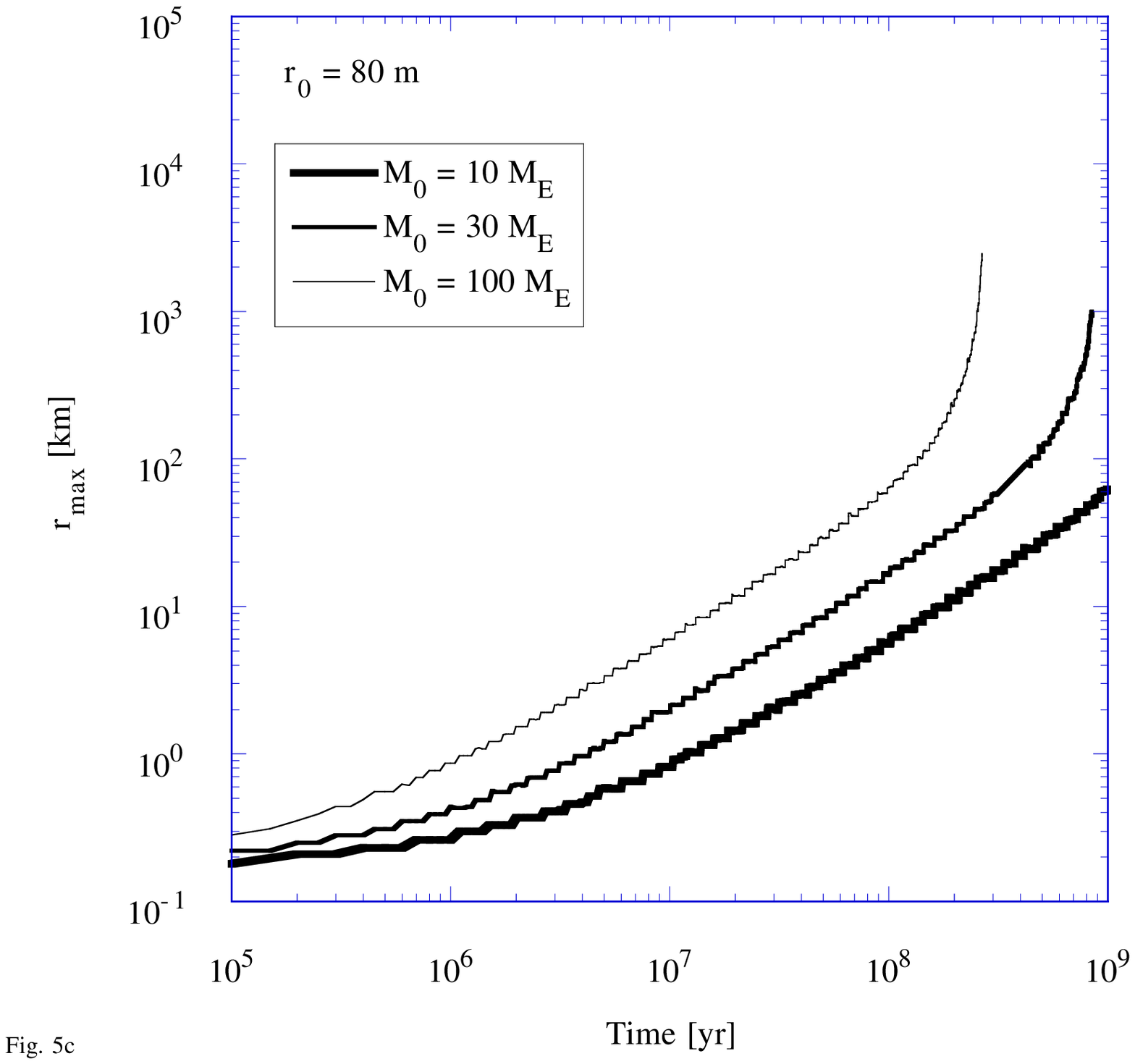}
 
\hskip -10ex
\epsffile{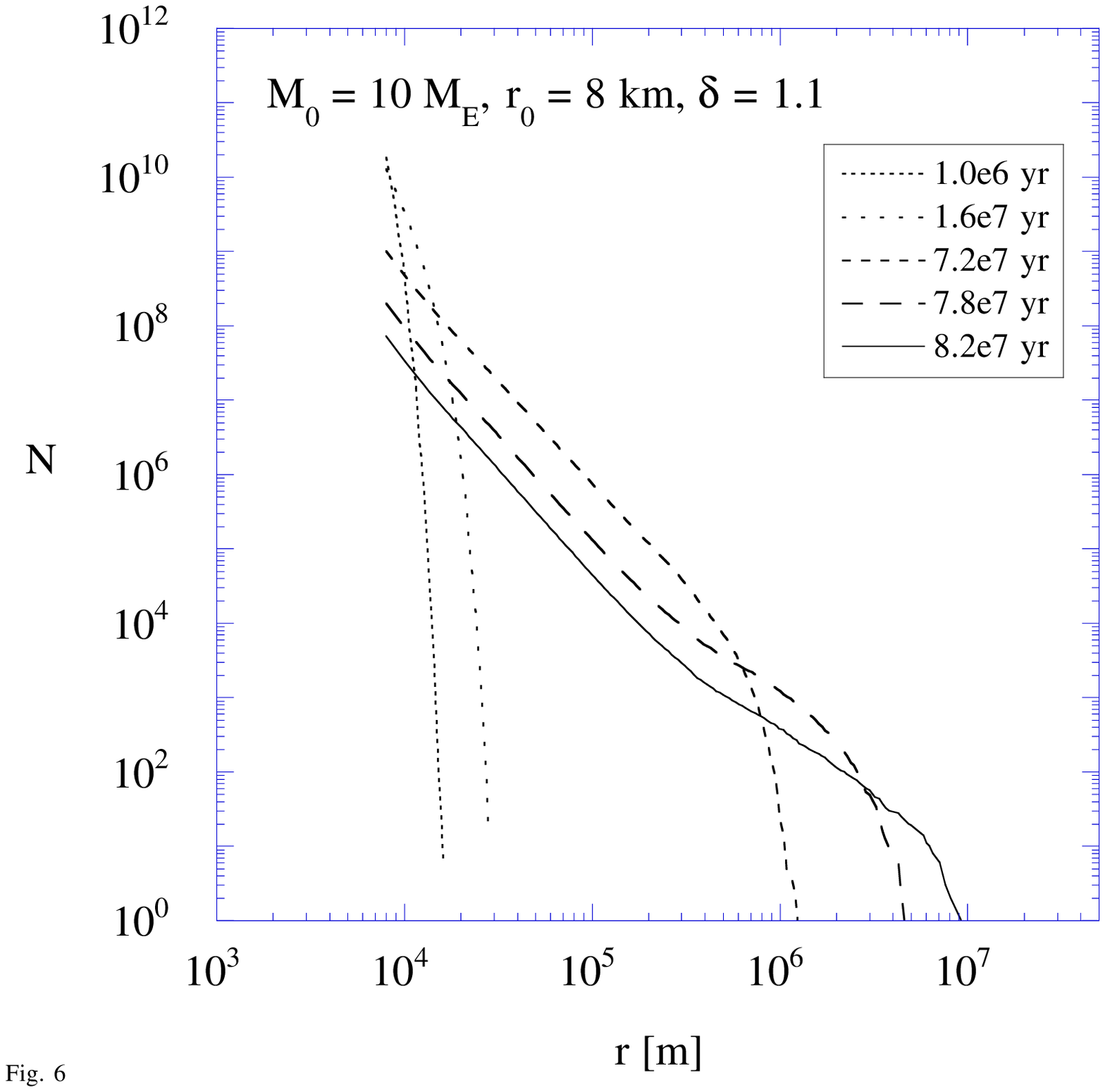}
 
\hskip -10ex
\epsffile{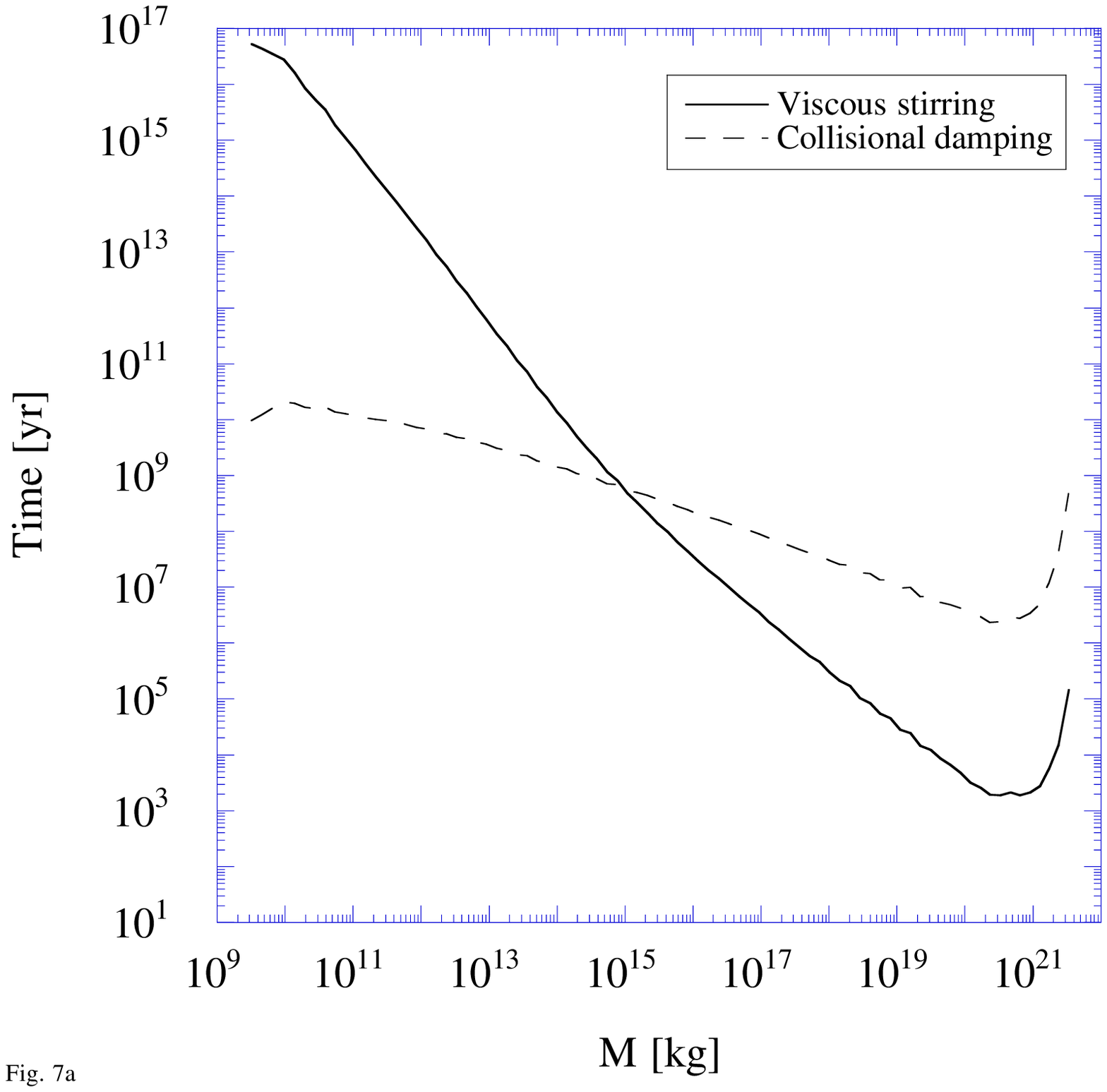}
 
\hskip -10ex
\epsffile{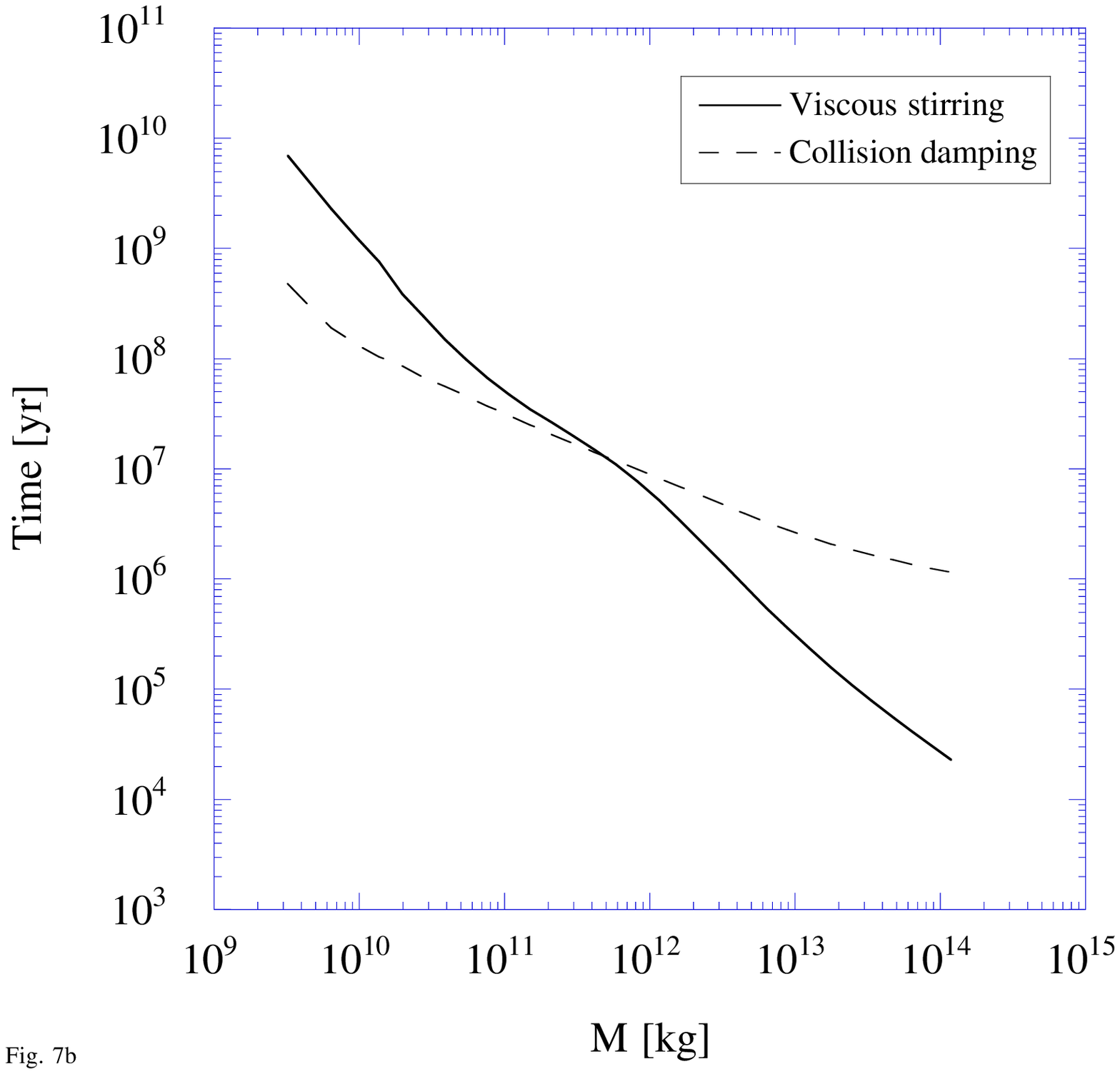}
 
\hskip -10ex
\epsffile{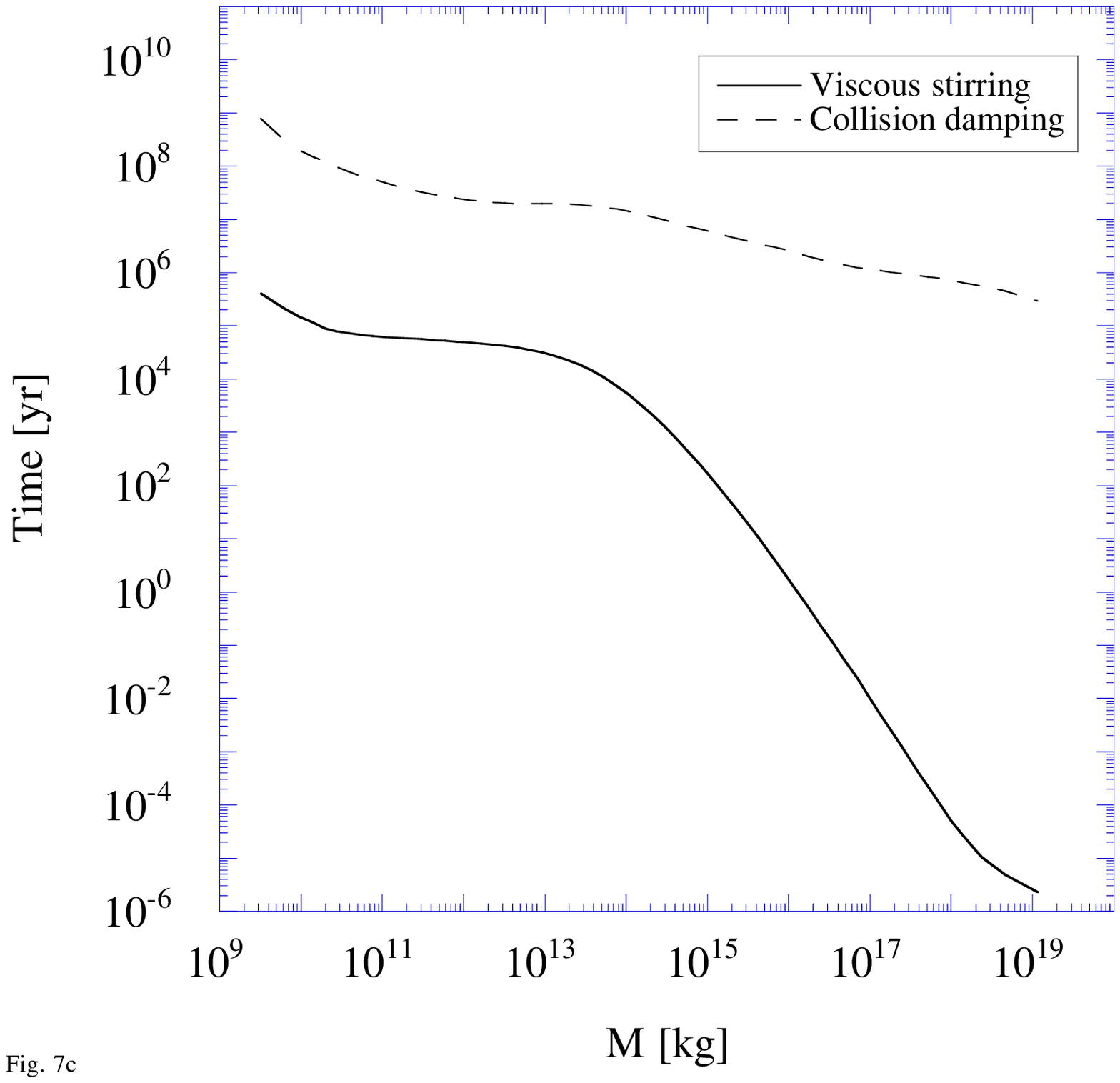}
 
\hskip -10ex
\epsffile{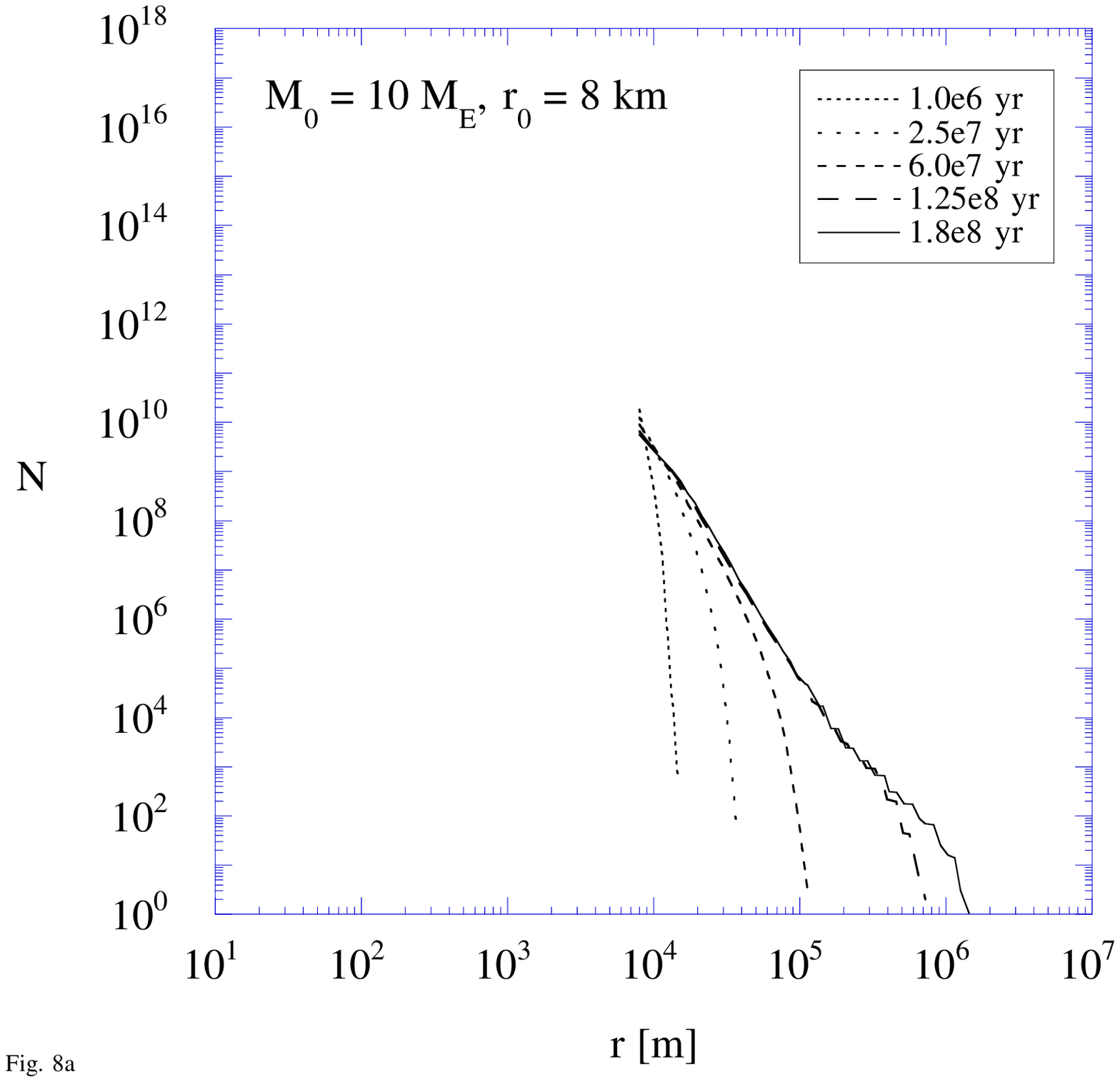}
 
\hskip -10ex
\epsffile{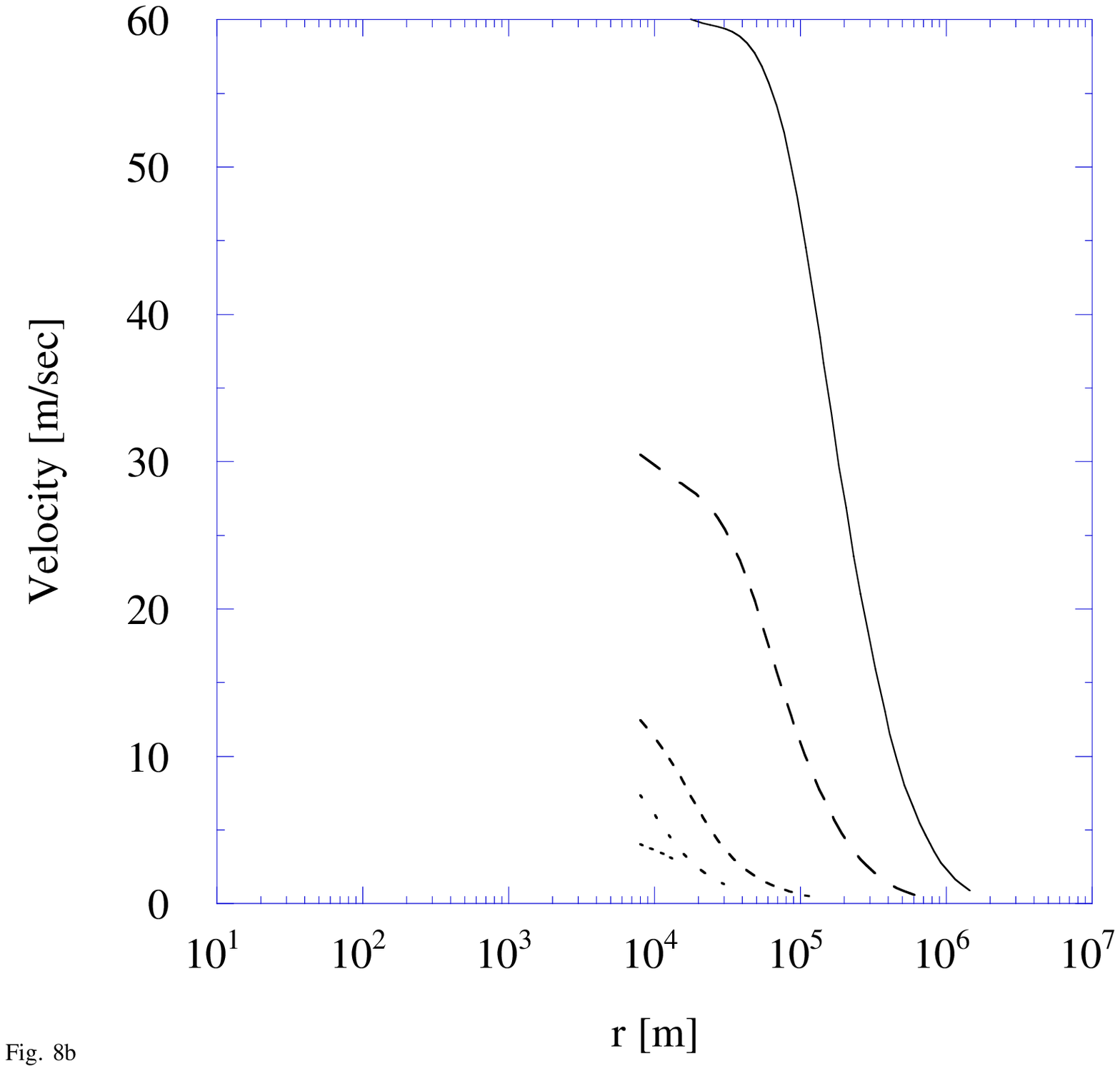}
 
\hskip -10ex
\epsffile{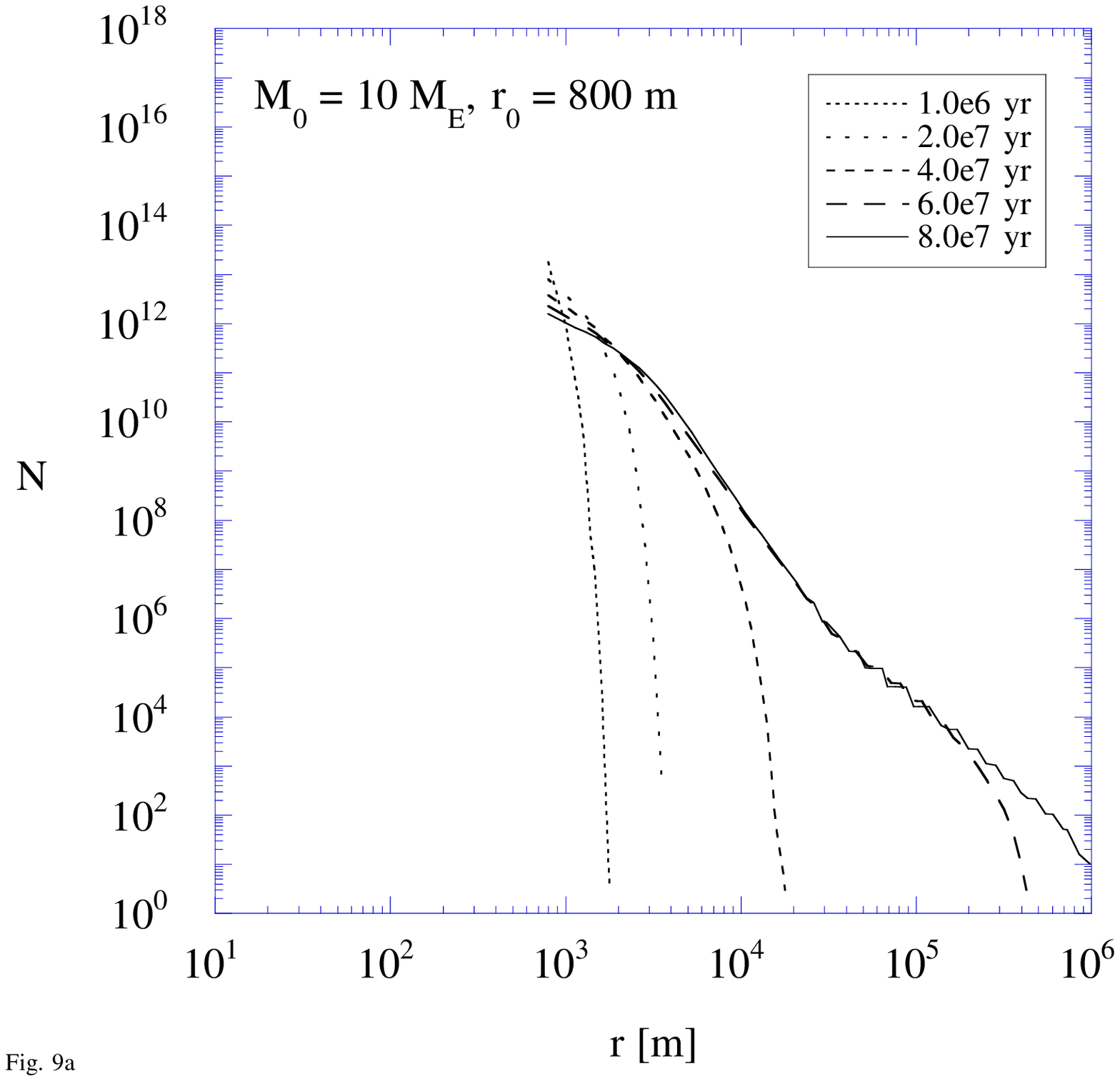}
 
\hskip -10ex
\epsffile{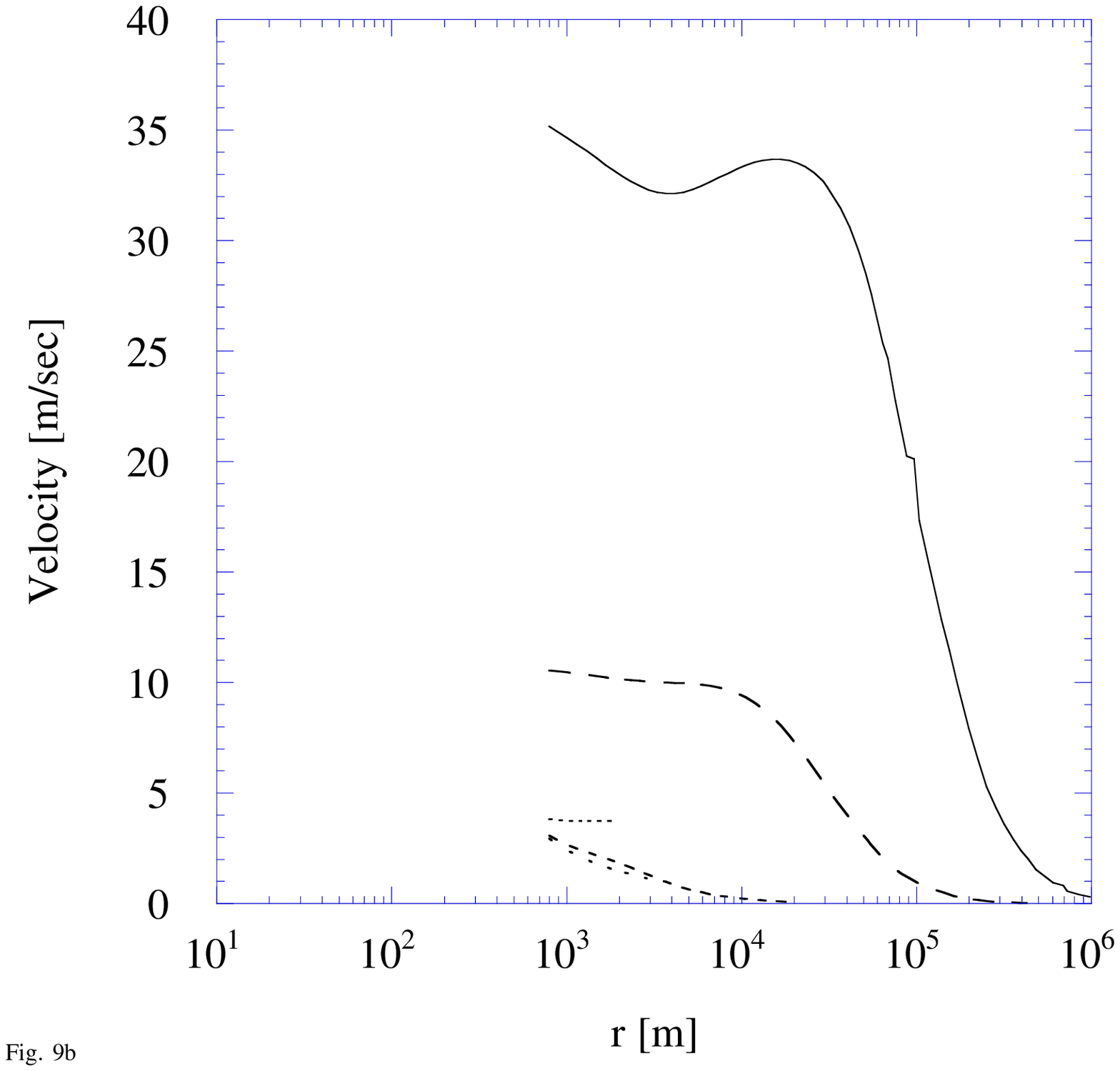}
 
\hskip -10ex
\epsffile{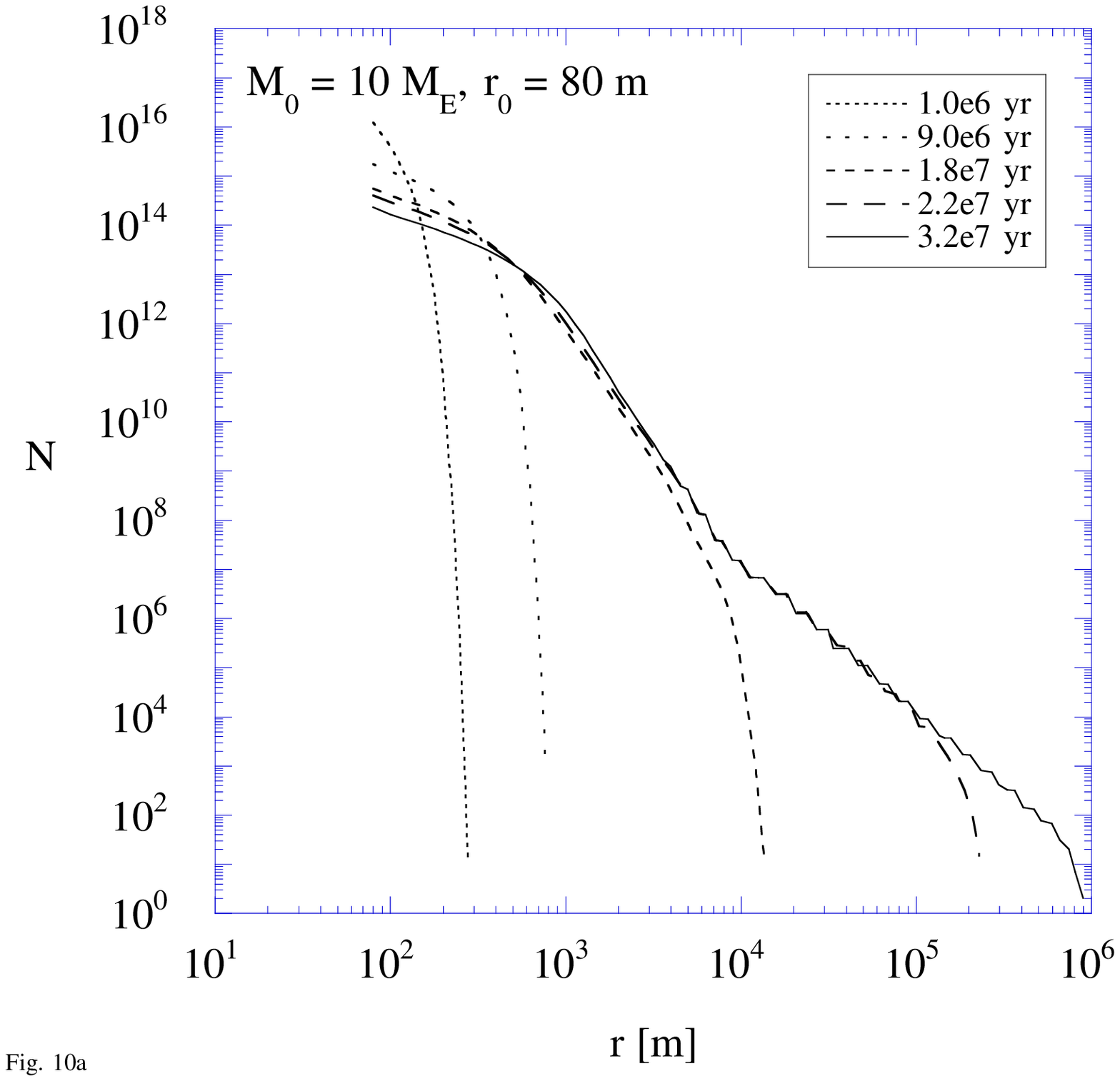}
 
\hskip -10ex
\epsffile{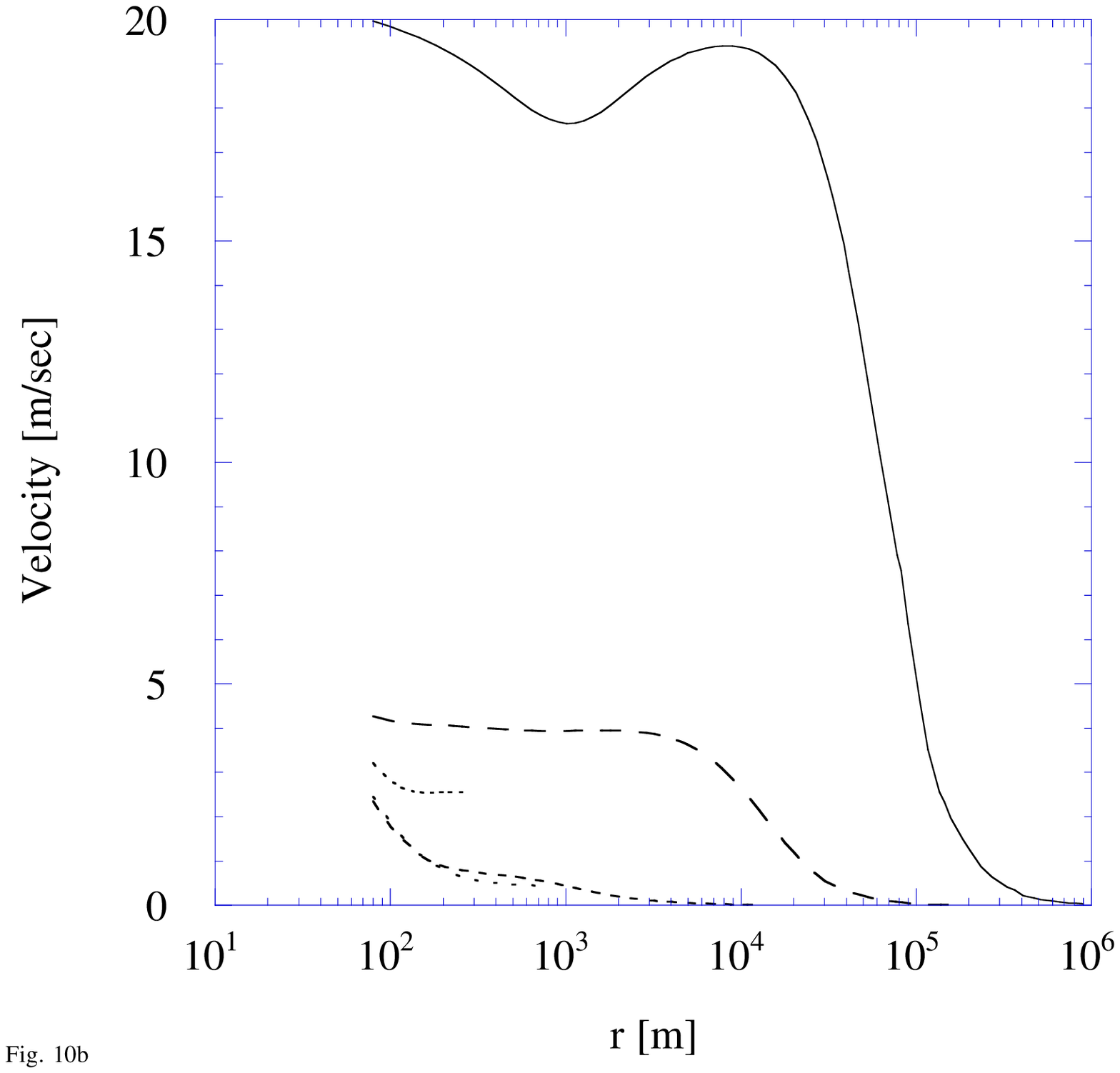}
 
\hskip -10ex
\epsffile{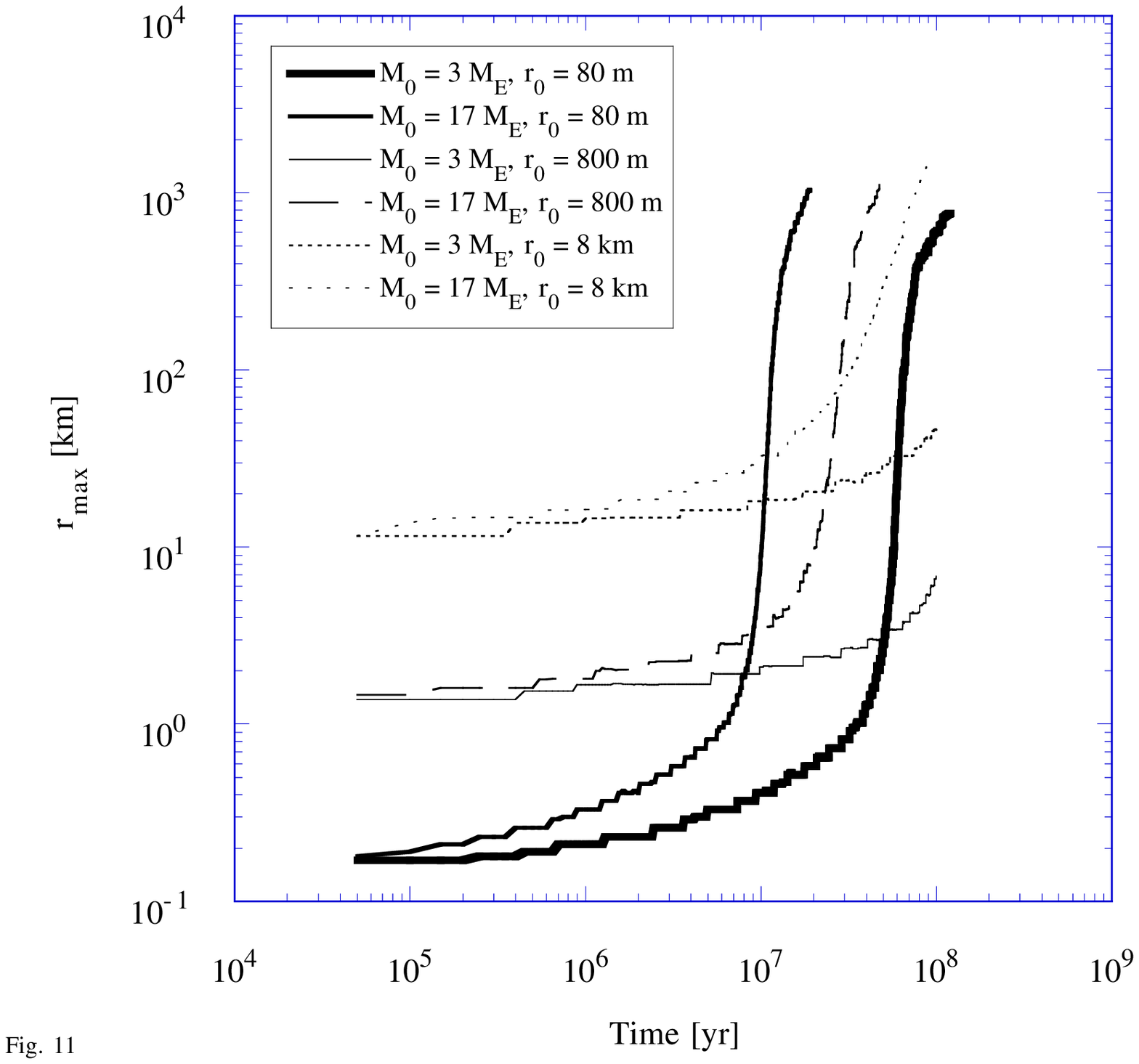}
 
\hskip -10ex
\epsffile{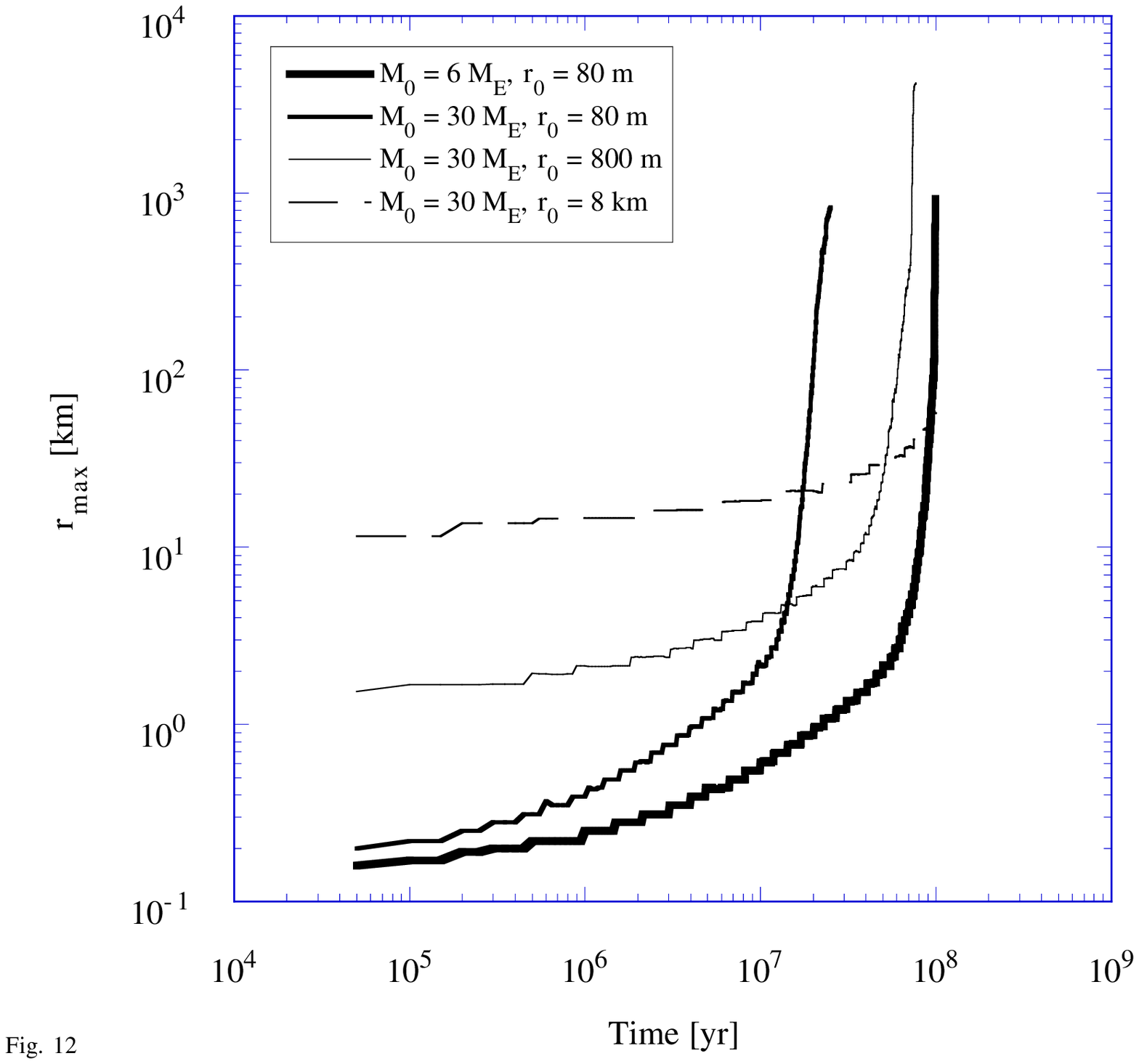}
 
\hskip -25ex
\epsfxsize=8.5in
\epsffile{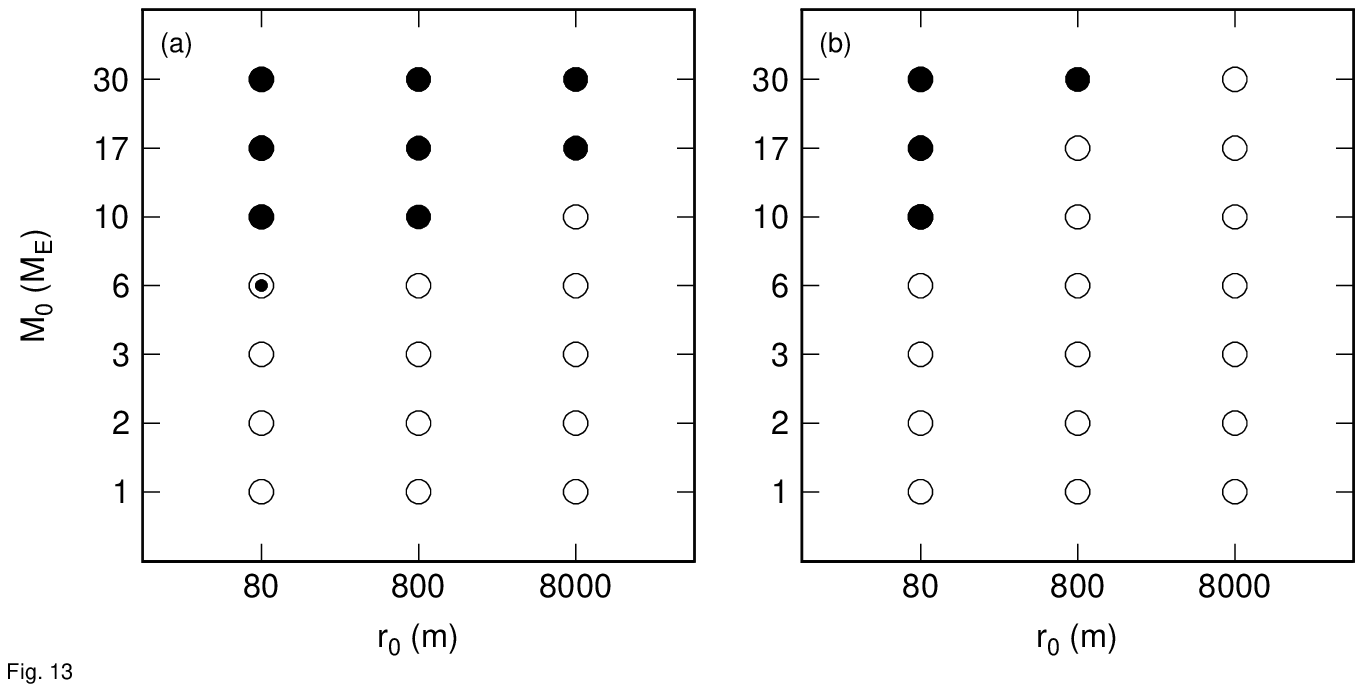}
 
\hskip -10ex
\epsffile{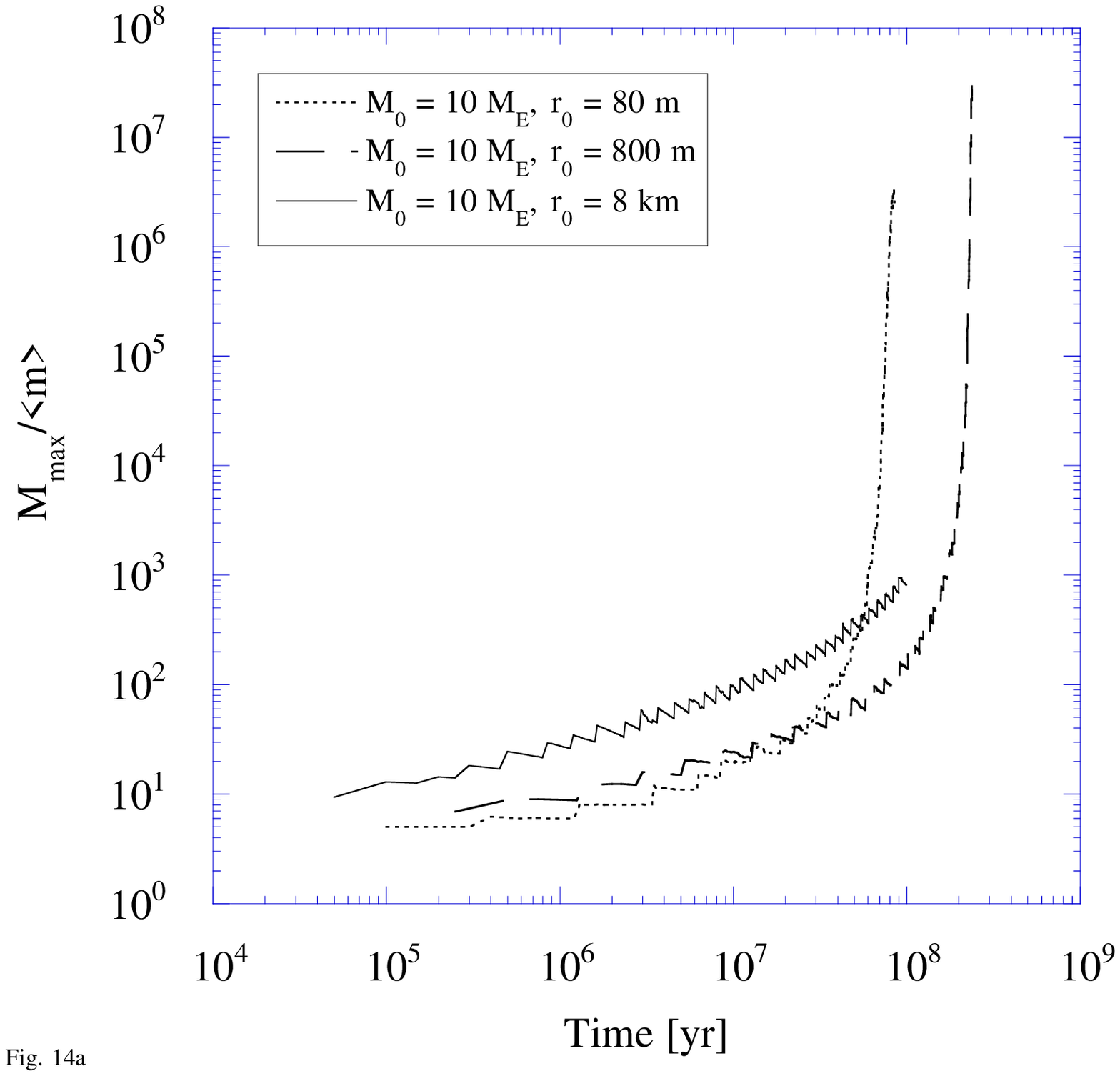}
 
\hskip -10ex
\epsffile{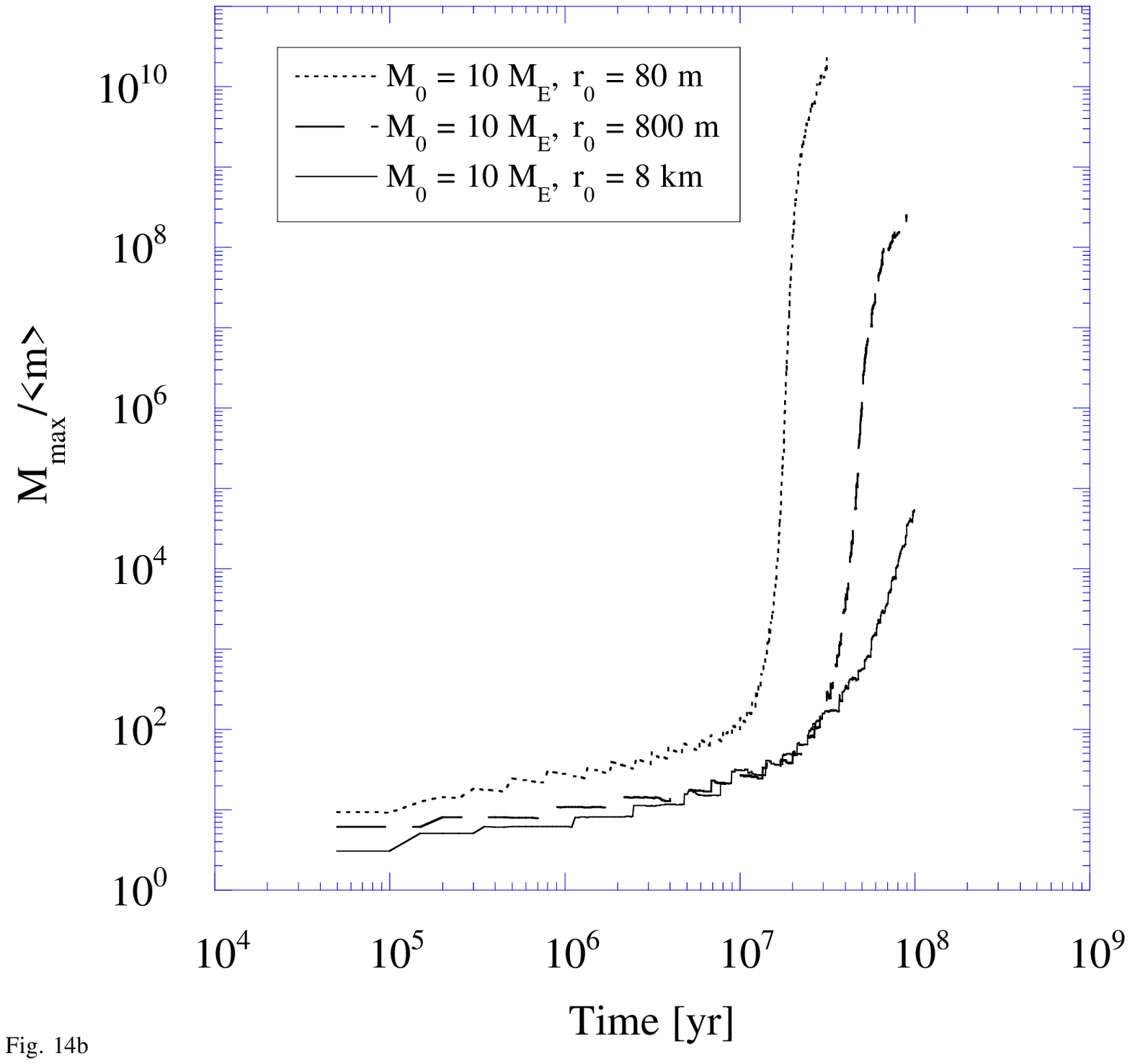}
 
\hskip -10ex
\epsffile{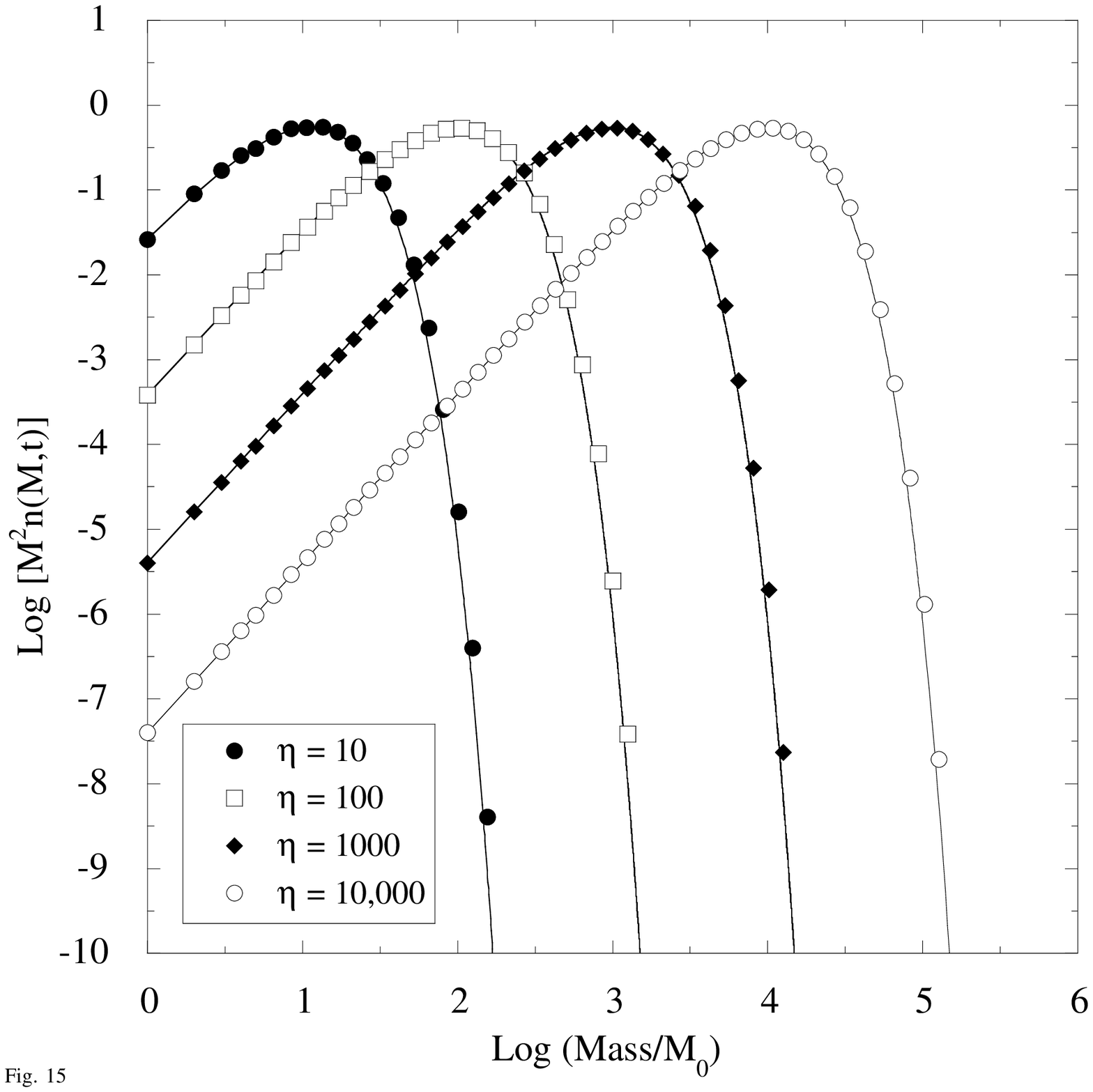}
 
\hskip -10ex
\epsffile{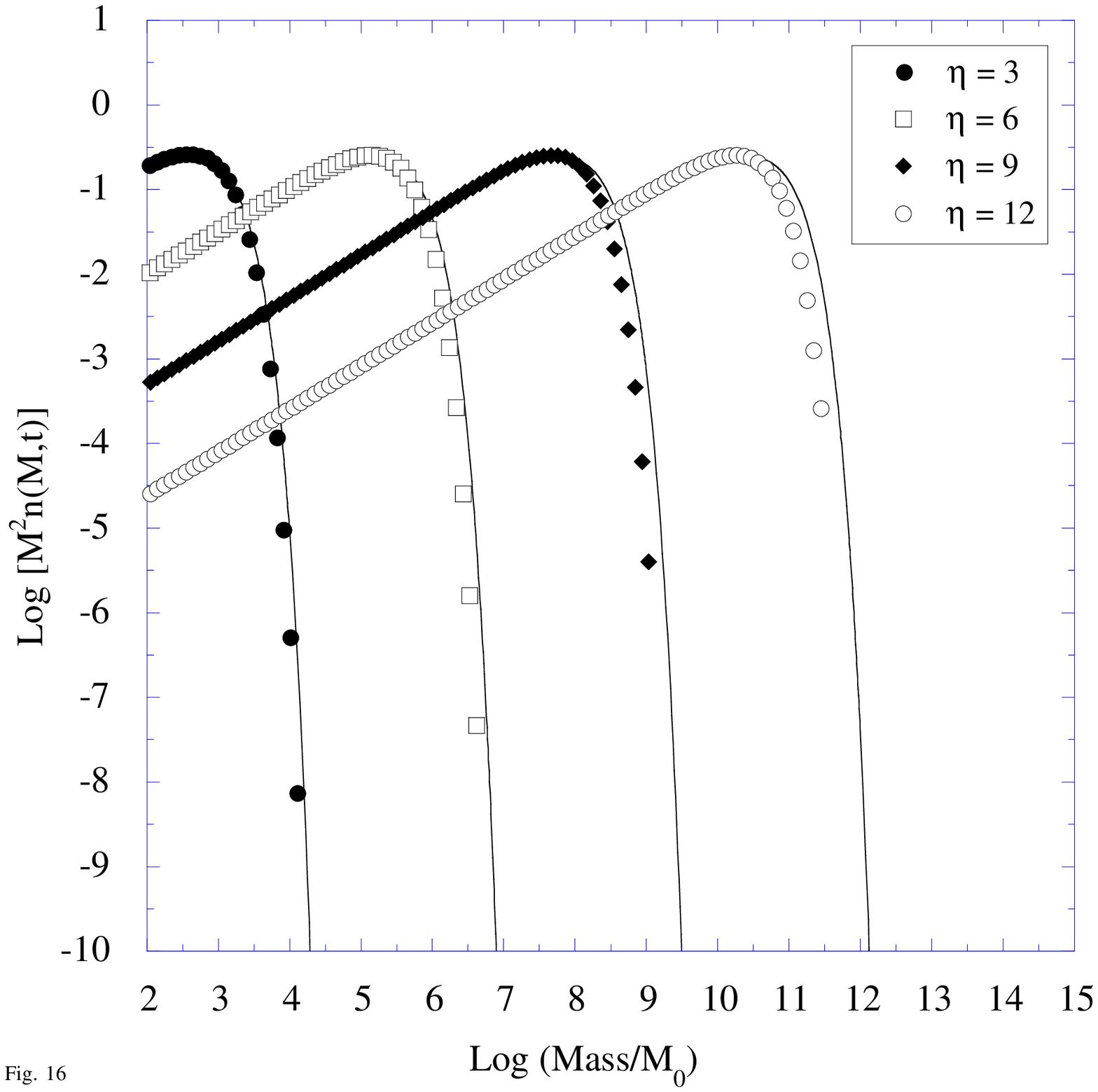}
 
\hskip -10ex
\epsffile{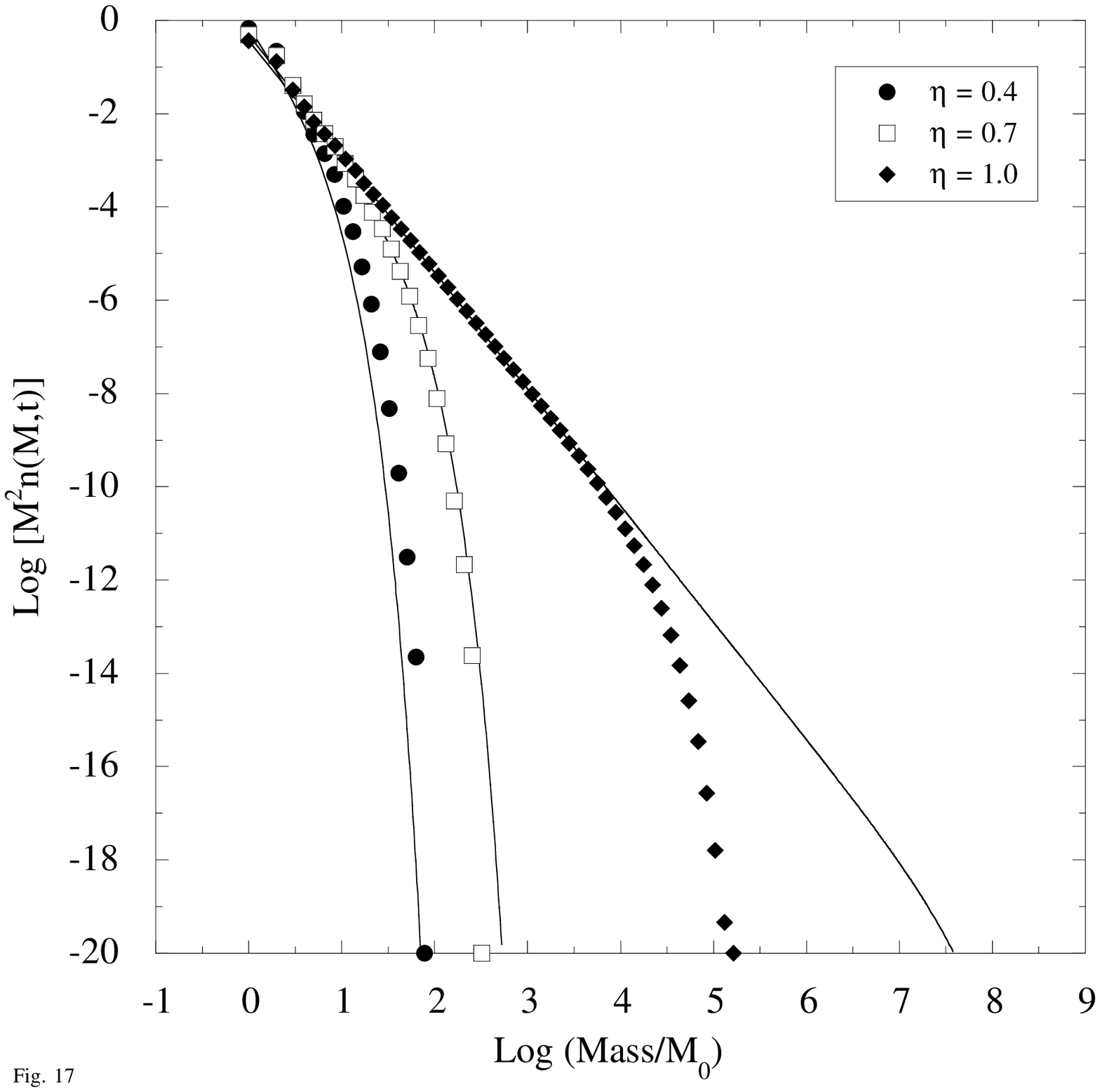}
 
\hskip -10ex
\epsffile{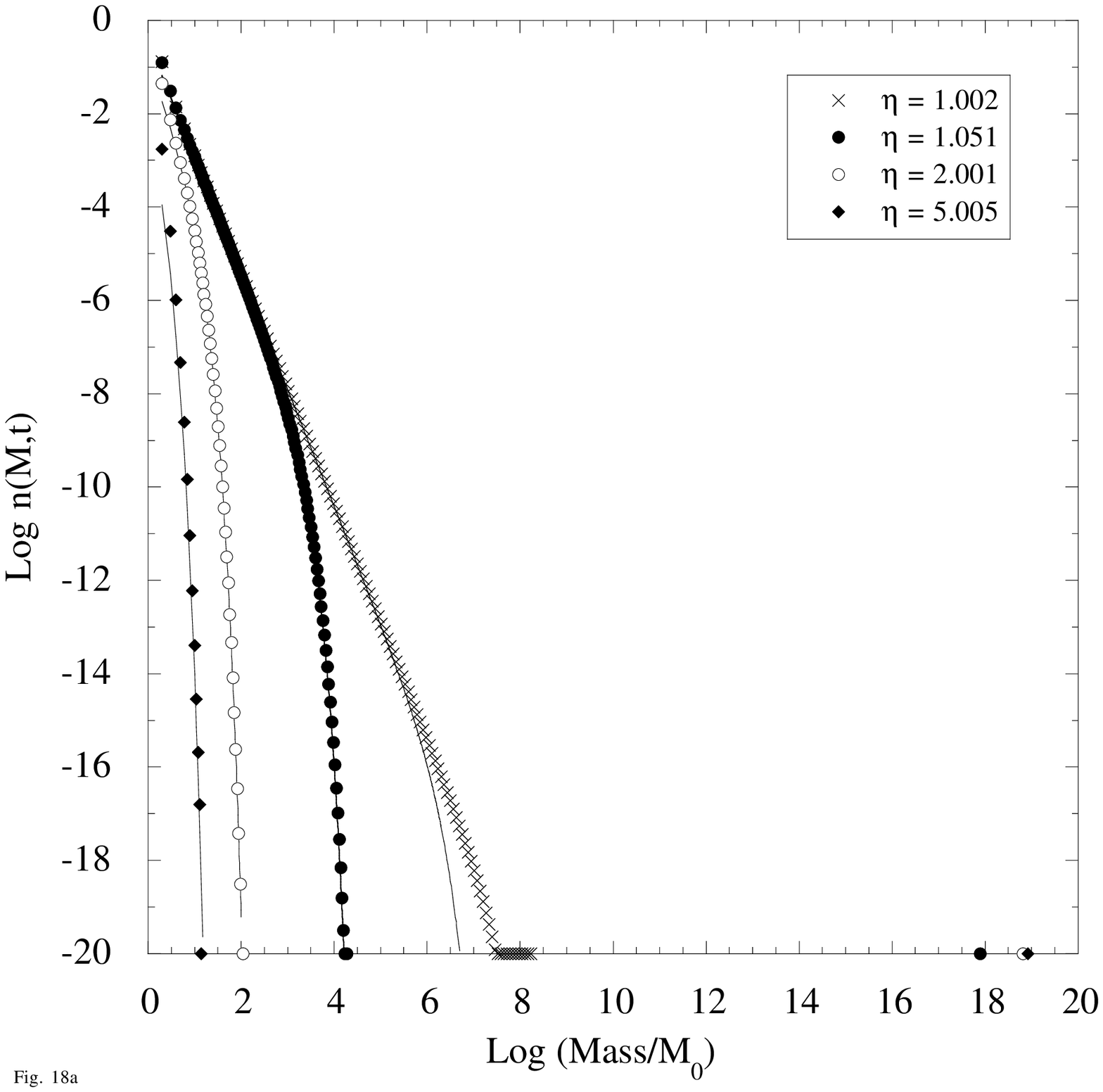}
 
\hskip -10ex
\epsffile{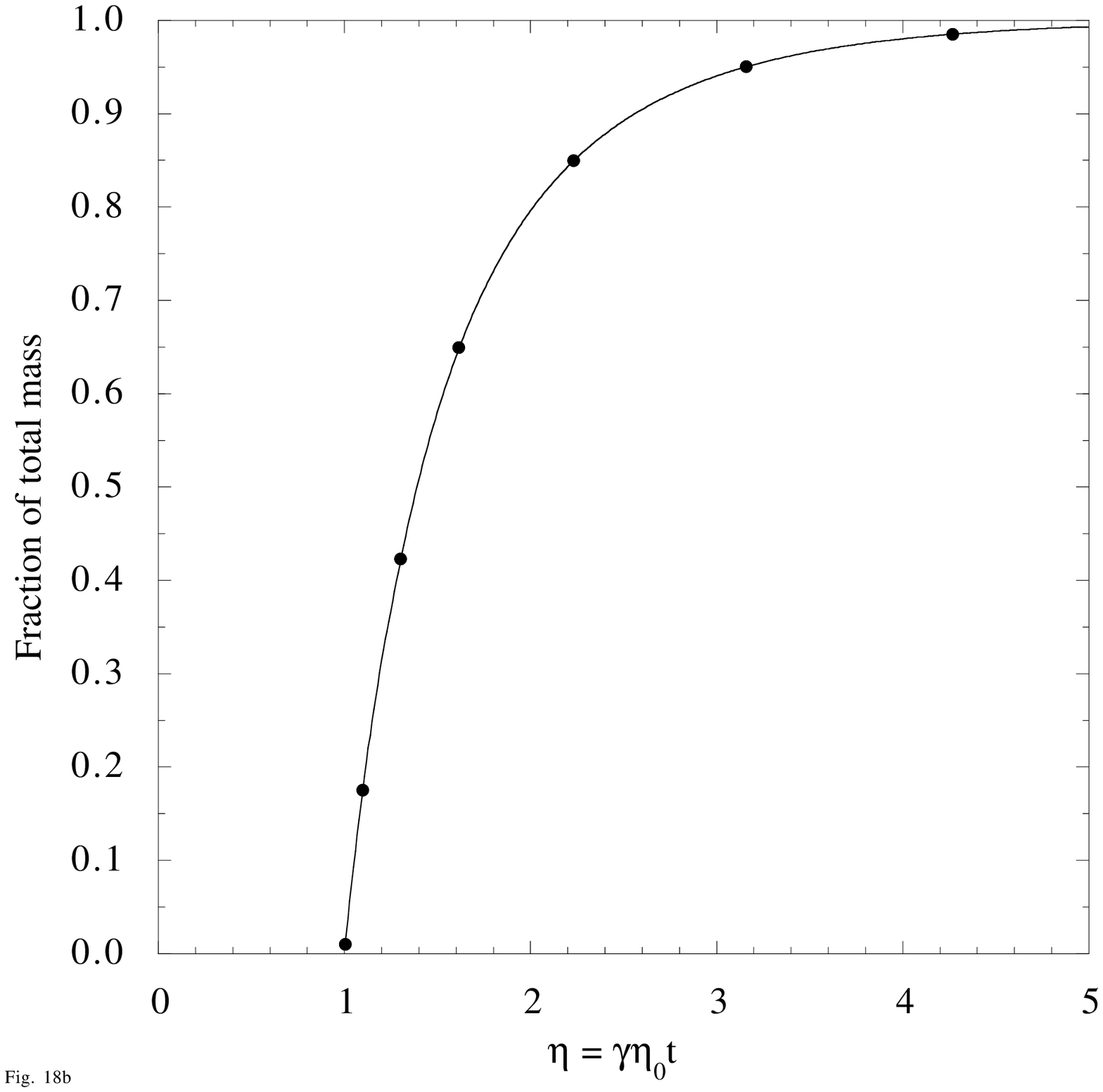}
 

\begin{thebibliography}{}

\bibitem[Aarseth \etal 1993]{aar93}
Aarseth, S. J., Lin, D. N. C., \& Palmer, P. L. 1993, ApJ, 403, 351

\bibitem[Adachi \etal 1976]{ada76} Adachi, I., 
Hayashi, C., Nakazawa, K. 1976, Progress
of Theoretical Physics 56, 1756 - 1771.

\bibitem[Bailey 1994]{bai94} Bailey, M. 1994,  
In Asteroids, Comets, Meteors 1993,
eds. A. Milani, M. DiMartino, and A. Cellino (Kluwer Academic
Publishers, Dordrecht), pp. 443 - 459.

\bibitem [Backman \& Paresce 1993]{bac93} 
Backman, D. E., \& Paresce, F. 1993, in
{\it Protostars and Planets III}, eds. E. H. Levy \& J. I. Lunine,
Tucson, Univ of Arizona, p. 1253

\bibitem[Barge \& Pellat 1990]{bar90} Barge, P., \& Pellat, R. 1990,
Icarus, 85, 481

\bibitem[Barge \& Pellat 1991]{bar91} Barge, P., \& Pellat, R. 1991,
Icarus, 93, 270

\bibitem[Barge \& Pellat 1993]{bar93} Barge, P., \& Pellat, R. 1993,
Icarus, 104, 79

\bibitem[Beckwith \& Sargent 1996]{bec96} Beckwith, S. V. W., \&
Sargent, A. I. 1996, Nature, 383, 139

\bibitem[Boss 1993]{bos93} Boss, A. P. 1993, ApJ, 417, 351


\bibitem[Davis \& Farinella 1997]{dav97} Davis, D. R., \& 
Farinella, P. 1997.  Icarus, 125, 50

\bibitem[Duncan \etal 1990]{dun95} Duncan, M. J., Levison, H. F., 
\& Budd, S. M. 1995,  AJ, 110, 3073


\bibitem[Edgeworth 1949]{edg49} Edgeworth, K. E. 1949, MNRAS, 109, 600

\bibitem[Fern\'andez 1997]{fer97} Fern\'andez, J. A. 1997, Icarus, 129, 106

\bibitem[Fern\'andez \& Ip 1981]{fer81} Fern\'andez, J. A., \& 
Ip, W.-H. 1981, Icarus, 47, 470

\bibitem[Fern\'andez \& Ip 1984]{fer84} Fern\'andez, J. A., \& 
Ip, W.-H. 1984, Icarus, 58, 109

\bibitem[Fern\'andez \& Ip 1996]{fer96} Fern\'andez, J. A., \& 
Ip, W.-H. 1996, Astrophys Sp Sci, 44, 431

\bibitem[Goldreich \& Ward 1973]{gol73} Goldreich, P., \& 
Ward, W. R. 1973,  ApJ, 183, 1051

\bibitem[Greenberg \etal 1978]{gre78}
Greenberg, R., Wacker, J. F., Hartmann, W. K., \& Chapman,
C. R. 1978, Icarus, 35, 1

\bibitem[Greenberg \etal 1991]{gre91} Greenberg, R., Bottke, W., 
Carusi, A., Valsecchi, G. B. 1991, Icarus, 94, 98

\bibitem[Greenzweig \& Lissauer 1990]{grz90}
Greenzweig, Y., \& Lissauer, J. J. 1990, Icarus, 87, 40

\bibitem[Greenzweig \& Lissauer 1992]{grz92}
Greenzweig, Y., \& Lissauer, J. J. 1992, Icarus, 100, 440

\bibitem[Hayashi 1981]{hay81} Hayashi, C. 1981, Prog Theor Phys Suppl, 70, 35

\bibitem[Hayashi \etal 1985]{hay85} Hayashi, C., Nakazawa, K., 
\& Nakagawa, Y. 1985,  In Protostars and Planets II, eds. 4 (U. of
Arizona Press, Tucson, pp. 1100 - 1153.

\bibitem[Holman \& Wisdom 1993]{hol93}
Holman, M. J., \& Wisdom, J. 1993,  AJ, 105, 1987

\bibitem[Hornung \etal 1985]{hor85} Hornung, P., Pellat, R., \& Barge, P.
1985, Icarus, 64, 295

\bibitem[Housen \etal 1991]{hou91} Housen, K. R., Schmidt, R. M., 
\& Holsapple, K. A. 1991, Icarus, 94, 180

\bibitem[Ida 1990]{ida90} Ida, S. 1990, Icarus, 88, 129

\bibitem[Ida \& Makino 1992a]{id92a} Ida, S., \& Makino, J. 
1992, Icarus, 96, 107

\bibitem[Ida \& Makino 1992b]{id92b} Ida, S., \& Makino, J. 
1992, Icarus, 98, 28

\bibitem[Ip 1989]{ip89} Ip, W.-H. 1989, Icarus, 80, 167

\bibitem[Ipatov 1989]{ipa89} Ipatov, S. I. 1989, Solar System Res, 23, 119

\bibitem[Ipatov 1991]{ipa91} Ipatov, S. I. 1991, Soviet Astron Lett, 17, 113

\bibitem[Jewitt \etal 1996]{jew96} Jewitt, D., Luu, J., \& Chen, J. 1996,
AJ, 112, 1225

\bibitem[Jewitt \& Luu 1995]{jew95} Jewitt, D., \& Luu, J. 1995, AJ, 109, 1867

\bibitem[Kenyon \& Hartmann 1995]{kh95} Kenyon, S. J., \& Hartmann, L. W.
1995, ApJS, 101, 177

\bibitem[Kokubo \& Ida 1996]{kok96} Kokubo, E., \& Ida, S. 1996, 
Icarus, 123, 180

\bibitem[Kolvoord \& Greenberg 1992]{kol92} Kolvoord, R. A. 
\& Greenberg, R. 1992, Icarus, 98, 2


\bibitem[Kuiper 1951]{kui51} Kuiper, G. P. 1951.  
"On the Origin of the Solar System."  In
Astrophysics: A Topical Symposium, ed. J. A. Hynek
(McGraw-Hill, New York, 357 - 424.

\bibitem[Levison \& Duncan 1993]{lev93} Levison, H. F., \& 
Duncan, M. J. 1993,  ApJ, 406, L35

\bibitem[Lissauer \& Stewart 1993]{lis93} Lissauer, J. J., 
Stewart, G. R. 1993,  Growth of Planets from Planetesimals.  In 
Protostars and Planets III, eds. E. H. Levy and J. I. Lunine 
(U. of Arizona Press, Tucson, 1061 - 1088. 

\bibitem[Lissauer \etal 1996]{lis96}
Lissauer, J. J., Pollack, J. B., Wetherill, G. W., \& Stevenson,
D. J. 1996.  "Formation of the Neptune System."  In Neptune and
Triton, eds. D. P. Cruikshank, M. S. Matthews, and A. M. Schumann
(U. of Arizona Press, Tucson, pp. 37 - 108.


\bibitem[Luu \etal 1997]{luu97} Luu, J. X., Marsden, B., Jewitt, D., 
Trujillo, C. A., Hergenother, C. W., Chen, J., \& Offutt, W. B. 1997,
Nature, in press

\bibitem[Malhotra 1993]{mal93} Malhotra, R. 1993, Nature, 365, 819

\bibitem[Malhotra 1995]{mal95} Malhotra, R. 1995, AJ, 110, 420

\bibitem[Malhotra 1996]{mal96} Malhotra, R. 1996,  AJ, 111, 504



\bibitem[Morbidelli \etal 1995]{mor95} Morbidelli, A., Thomas, F., 
\& Moons, M. 1995, Icarus, 118, 322

\bibitem[Nakagawa \etal 1983]{nak83} Nakagawa, Y., Hayashi, C., 
\& Nakazawa, K. 1983,  Icarus, 54, 361

\bibitem[Ohtsuki \& Nakagawa 1988]{oht88} Ohtsuki, K., \& 
Nakagawa, Y. 1988, Prog Theor Phys (Suppl), 96, 239

\bibitem[Ohtsuki \etal 1990]{oht90} Ohtsuki, K., Nakagawa, Y., \&
Nakazawa, K. 1990, Icarus, 83, 205

\bibitem[Ohtsuki 1992]{oht92} Ohtsuki, K. 1992, Icarus, 98, 20

\bibitem[Pollack \etal 1996]{pol96} Pollack, J. B., Hubickyj, O., 
Bodenheimer, P., Lissauer, J. J., Podolak, M., \& Greenzweig, Y. 1996,
Icarus, 124, 62

\bibitem[Ruden \& Lin 1986]{rud86} Ruden, S. P., \& Lin, D. N. C. 1986,
ApJ, 308, 883

\bibitem[Ruden \& Pollack 1991]{rud91} Ruden, S. P., \& Pollack, J. B. 1991,
ApJ, 375, 740

\bibitem[Russell \etal 1996]{rus96} Russell, S. S., Srinivasan, G., 
Huss, G. R., Wasserburg, G. J., \& Macpherson, G. J. 1996, Science, 273, 757

\bibitem[Safronov 1969]{saf69} Safronov, V. S. 1969, Evolution of
the Protoplanetary Cloud and Formation of the Earth and Planets,
Nauka, Moscow [Translation 1972, NASA TT F-677]

\bibitem[Sargent \& Beckwith 1993]{sar93}
Sargent, A. I., \& Beckwith, S. V. W. 1993, Phy Tod, 46, 22


\bibitem[Silk \& Takahashi 1979]{sil79} Silk, J., \& Takahashi, T. 1979,
ApJ, 229, 242

\bibitem[Spaute \etal 1991]{spa91} Spaute, D., Weidenschilling, S. J.,
Davis, D. R., \& Marzari, F. 1991, Icarus, 92, 147 


\bibitem[Stern 1995]{ste95} Stern, S. A. 1995, AJ, 110, 856

\bibitem[Stern 1996]{ste96} Stern, S. A. 1996, AJ, 112, 1203

\bibitem[Stern \& Colwell 1997]{st97a} Stern, S. A., \& 
Colwell, J. E. 1997a,  AJ, 114, 841

\bibitem[Stern \& Colwell 1997]{st97b} Stern, S. A., \& 
Colwell, J. E. 1997b,  ApJ, 490, 879

\bibitem[Stewart \& Wetherill 1988]{ste88} Stewart, G. R., \& 
Wetherill, G. W. 1988, Icarus, 75, 542

\bibitem[Strom \etal 1993]{str93} Strom, S. E., Edwards, S., \& 
Skrutskie, M. F. 1993, in {\it Protostars and Planets III}, 
eds. E. H. Levy \& J. I. Lunine, Tucson, Univ of Arizona, p. 837

\bibitem[Tanaka \& Nakazawa 1993]{tan93} Tanaka, H., \& Nakazawa, K. 
1993,  J Geomag Geoelec, 45, 361.

\bibitem[Tanaka \& Nakazawa 1994]{tan94} Tanaka, H., \& Nakazawa, K. 
1994, Icarus, 107, 404.


\bibitem[Trubnikov 1971]{tru71} Trubnikov, B. A. 1971, 
Doklady Acad. Nauk SSSR, 196, 1316





\bibitem[Weidenschilling 1977]{wei77} Weidenschilling, S. J. 1977,
Astrophys Sp Sci, 51, 153

\bibitem[Weidenschilling 1980]{wei80} Weidenschilling, S. J. 1980,
Icarus, 44, 172


\bibitem[Weidenschilling \& Cuzzi 1985]{wei93} Weidenschilling, S. J.,
\& Cuzzi, J. N., 1993, in {\it Protostars and Planets III},
edited by E. H. Levy \& J. I. Lunine, Tucson, Univ of Arizona, p. 1031


\bibitem[Weidenschilling \& Davis 1992]{wei92} Weidenschilling, 
S. J., \& Davis, D. R. 1992, Lun Planet Sci, XXIII, 1507


\bibitem[Wetherill 1990]{wet90} Wetherill, G. W. 1990,
Icarus, 88, 336

\bibitem[Wetherill \& Stewart 1989]{wet89}
Wetherill, G. W., \& Stewart, G. R. 1989.  Icarus 77, 300 - 357. 

\bibitem[WS93]{ws93} Wetherill, G. W., \& Stewart, G. R. 1993,  
Icarus, 106, 190 (WS93) 


\end{thebibliography}
\end{document}